\documentclass[final,5p,times,twocolumn]{elsarticle}

\usepackage{amssymb}
\usepackage{longtable} 
\usepackage{pdflscape} 
\usepackage{rotating}

\usepackage[T1]{fontenc} 
\usepackage[utf8]{inputenc}
\usepackage[english]{babel}

\usepackage{microtype}
\usepackage{graphicx}
\usepackage{caption}
\usepackage{subcaption}
\usepackage{adjustbox}
\usepackage{multirow}
\usepackage{algorithmicx}
\usepackage[ruled]{algorithm}
\usepackage{algpseudocode}
\usepackage[table,xcdraw]{xcolor}
\usepackage{multicol}
\usepackage{float}
\usepackage{bbm}
\usepackage{xcolor}
\usepackage{soul}
\usepackage{hyperref}
\usepackage{booktabs}
\usepackage{lscape}
\usepackage{longtable}
\hypersetup{
    colorlinks=true,
    linkcolor=blue,
    filecolor=magenta,      
    urlcolor=cyan
}

\journal{Elsevier Journal}

\begin{document}

\begin{frontmatter}

\title{Fatigue Monitoring Using Wearables and AI: Trends, Challenges, and Future Opportunities}

\author[inst1]{Kourosh Kakhi}
\affiliation[inst1]{organization={Institute for Intelligent Systems Research and Innovation (IISRI)}, 
            city={Geelong},
            postcode={3217}, 
            state={VIC},
            country={Australia}}

\author[inst2]{Senthil Kumar Jagatheesaperumal}
\affiliation[inst2]{organization={Department of Electronics and Communication Engineering, Mepco Schlenk Engineering College}, 
            city={Sivakasi},
            state={Tamil Nadu},
            country={India}}

\author[inst1]{Abbas Khosravi}
\author[inst1]{Roohallah Alizadehsani\textsuperscript{*,}}

\author[inst3]{U Rajendra Acharya}
\affiliation[inst3]{organization={School of Mathematics, Physics and Computing, University of Southern Queensland}, 
            city={Springfield},
            country={Australia}}

\begin{abstract}
\textbf{Background}: Monitoring fatigue is essential for improving safety, particularly for people who work long shifts or in high-demand workplaces. The development of wearable technologies, such as fitness trackers and smartwatches, has made it possible to continuously analyze physiological signals in real-time to determine a person’s level of exhaustion. This has allowed for timely insights into preventing hazards associated with fatigue.

\textbf{Methods}: This review focuses on wearable technology and artificial intelligence (AI) integration for tiredness detection, adhering to the PRISMA principles. Studies that used signal processing methods to extract pertinent aspects from physiological data, such as ECG, EMG, and EEG, among others, were analyzed as part of the systematic review process. Then,  to find patterns of weariness and indicators of impending fatigue, these features were examined using machine learning and deep learning models. 

\textbf{Results}: It was demonstrated that wearable technology and cutting-edge AI methods could accurately identify weariness through multi-modal data analysis. By merging data from several sources, information fusion techniques enhanced the precision and dependability of fatigue evaluation. Significant developments in AI-driven signal analysis were noted in the assessment, which
should improve real-time fatigue monitoring while requiring less interference.
Conclusion: Wearable solutions powered by AI and multi-source data fusion present a strong option for real-time tiredness monitoring in the workplace and other crucial environments. These developments open the door for more improvements in this field and offer useful tools for enhancing safety and reducing fatigue-related hazards .

\end{abstract}




\end{frontmatter}


\section{Introduction}\label{sec:intro}
Fatigue is usually related to decreased physical activity or difficulty performing a task. Most sportspeople and laborers who perform intensive physical activities are often subjected to fatigue issues. The psychological state of the affected persons subjected to their sensations usually leads to perceived fatigue, which directly impacts the central nervous system and affects the mood \cite{1}. Moreover, the central and peripheral contractions lead to performance fatigue, mainly related to muscle activation and contractile functions \cite{2}. Monitoring such fatigue symptoms among highly active individuals is often challenging because they are busy with their professional tasks. In this regard, wearable devices could assist in observing their physical activities and log the data for analysis using AI frameworks.
Currently, the most prominent related works can be observed in the literature. Lambay et al. \cite{3} looked at machine learning techniques when looking at how to find physical fatigue in construction and manufacturing. They emphasized how hard it is to see fatigue and how important it is to have strong, all-around methods to keep an eye on and deal with fatigue in environments where people do the same thing repeatedly. Zong et al. \cite{4} conducted a systematic review on fatigue in construction workers, identifying various causes and evaluation methods and emphasizing the importance of interventions to alleviate fatigue and improve occupational health and safety. Li et al. \cite{5} introduced a decentralized approach for monitoring construction equipment operators’ fatigue using facial images, highlighting the balance between technical efficiency and data privacy.

In the aviation industry, Ziakkas et al. \cite{6} talked about the role of AI in fatigue risk management systems (FRMS). They said that FRMS could use data from wearable devices like EEGs and smartwatches to detect better and predict real-time fatigue, making flights safer and saving money. Similarly, the author in \cite{7} provided further insights into AI-based strategies for detecting fatigue and sleep problems in aviation, emphasizing the advancements and challenges in using AI to enhance aviation safety by predicting fatigue-related incidents. Kong et al. \cite{8} proposed a non-invasive method for fatigue detection using a multi-modal fusion of heart rate and facial features, achieving high accuracy in detecting driver fatigue and contributing to traffic safety.

Taylor et al. \cite{9} demonstrated that a combination of wearable bracelets and RF sensing, analyzed through Random Forest and ResNet algorithms, achieved 100\% accuracy in detecting fatigue. The bracelet alone also yielded high accuracy, highlighting the potential of hybrid sensing approaches in fatigue detection.  Similarly, Biro et al. \cite{10} used IMUs to track athletes’ performance and employed AI to predict fatigue and stamina, achieving high predictive accuracy and facilitating individualized training adjustments to reduce the risk of overtraining. In \cite{11}, the authors discussed how wearable biosensors can be used for non-invasive fatigue diagnosis. They also discussed the progress and problems of using biochemical reactions and data communication modules in biosensors.
Alam et al. \cite{12} suggested an activity-aware recurrent neural network to measure cognitive and mental fatigue. This was much better than the current best methods. Goumopoulos et al. \cite{13} used wearable devices for HRV analysis to detect mental fatigue, demonstrating high accuracy and potential for continuous, unobtrusive monitoring in everyday life. Lambert et al. \cite{14} surveyed AI models for mental fatigue, highlighting the need for balanced parameter acquisition and model validation to improve accuracy and reliability. Lastly, Yaacob et al. \cite{15} conducted a systematic review of AI techniques in brain-computer interfaces for mental fatigue detection, identifying research gaps and future directions for automated neurofeedback.

With the help of real datasets, Mu et al. \cite{16} created a system for detecting fatigue based on ECG and HRV features. This system worked very well at monitoring exercise-induced fatigue in real-time. Similarly, Russell et al. \cite{17} used a single sensor and a deep learning model to measure mental and physical fatigue during long-lasting mountain events. This shows that field fatigue research can be used in real life. Liu et al. \cite{18} created a dynamic muscle fatigue classification model using sEMG and a better whale optimization algorithm. This model was very good at predicting muscle fatigue when conditions were changing.

Narteni et al. \cite{19} looked into explainable AI (XAI) for predicting physical fatigue. They stressed how important it is for AI models used in the workplace to be trustworthy and easy to understand. Giorgi et al. \cite{20} looked into neurophysiological parameters for assessing mental fatigue in self-driving cars and found sensitive indicators for real-time fatigue detection. Jiao et al. \cite{21} looked into how HRV and EDA features could be used to detect driver fatigue. They were very good at capturing the body’s responses to fatigue, which could improve driving and traffic safety.

El et al. \cite{22} looked at machine learning (ML) and deep learning (DL) methods for finding driver fatigue. They discussed each method’s usefulness and reliability and suggested areas for future research. Zhao et al. \cite{23} proposed an rPPG-based learning fatigue detection system, demonstrating high accuracy in early fatigue detection using facial videos. Chang et al. \cite{24} implemented a drowsiness-fatigue detection system using wearable smart glasses, significantly enhancing road safety by alerting drivers and following vehicles. 

Liu et al. \cite{25} developed ML models for detecting fatigue during repetitive physical tasks, achieving high accuracy and demonstrating the potential of wearable sensors in workplace safety. Finally, Maman et al. \cite{26} proposed a data analytic framework for managing physical fatigue using wearable sensors, establishing feature and ML algorithm selection criteria to improve worker safety. The table in Appendix \ref{tab:Apendix} shows the summary of fatigue monitoring related works using AI frameworks.

\subsection{The Rise of AI and Wearables}\label{subsubsec:RiseAIWearable}

\begin{figure}[h]
    \centering
    \includegraphics[scale=0.18]{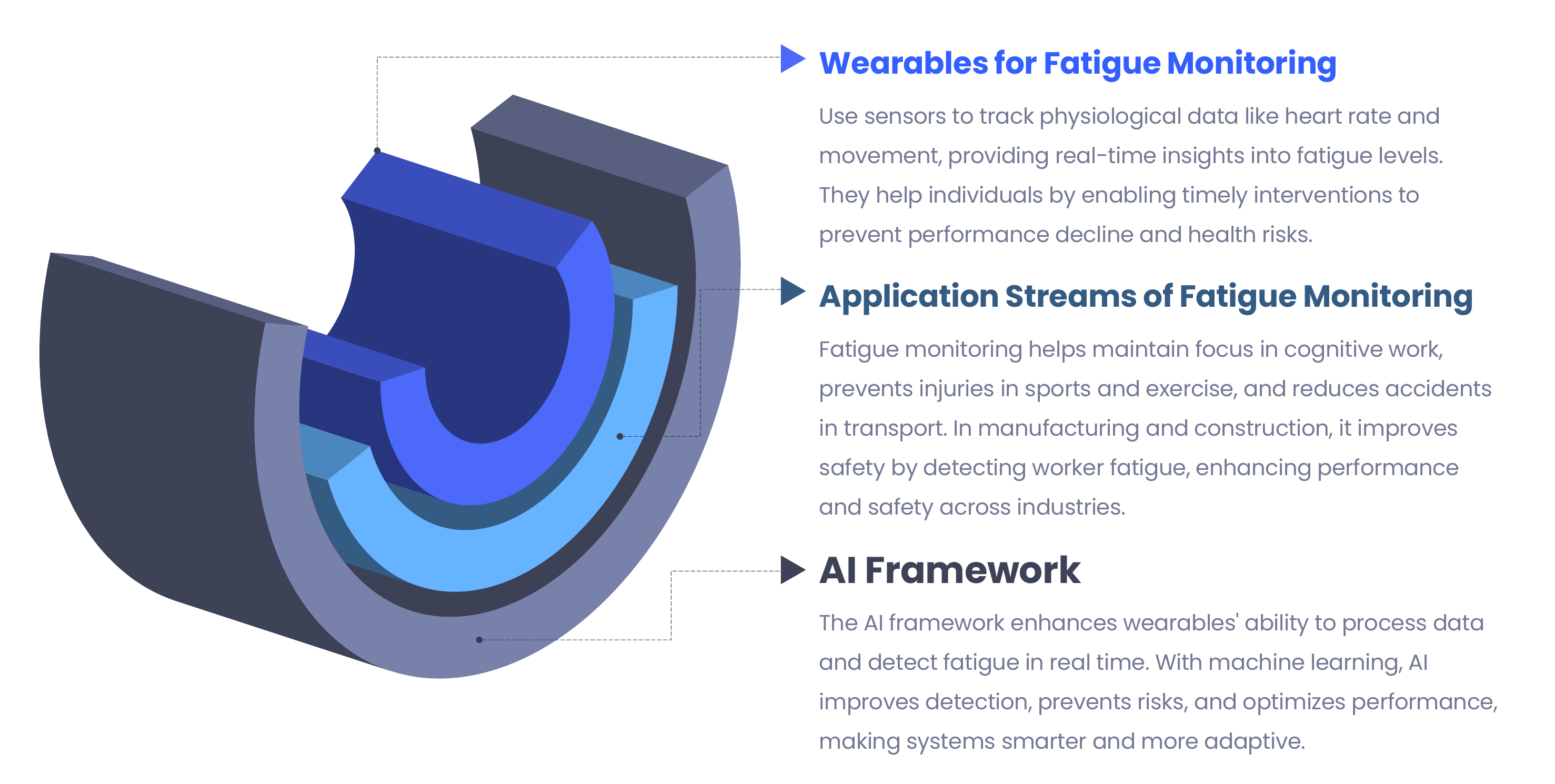}
    \caption{Wearable technology that interacts with people and fatigue monitoring powered by AI frameworks.}
    \label{fig:Fig. 1}
\end{figure}

All fields of science and engineering, including the human factor, are currently experiencing a revolution due to the integration of AI technology. Both traditional ML methods (e.g., DT, SVM, RF) and advanced AI techniques such as deep learning and transformers are used to analyze vast amounts of multi-source datasets, including images, time series, and demographic information. These techniques identify subtle patterns in the data, identify dependencies/relationships, and predict response variables. Figure \ref{fig:Fig. 1} shows the representation of wearable technology that interacts with people and fatigue monitoring solutions powered by AI frameworks.

Additionally, AI is being used to improve the efficiency of systems and boost their accessibility. Integrating big data and AI’s ability to process and analyze large amounts of data are key factors in this revolution, helping to make different systems and organizations more agile, efficient, and effective.

Also, wearable technology has been revolutionized in recent years. The availability, affordability, and unobtrusiveness of wearables have contributed to their popularity. Wearables like smartwatches and fitness trackers are becoming increasingly popular among consumers due to their ease of use and portability. They are now widely available and relatively inexpensive, making them accessible to many users. Moreover, these devices are typically passive, meaning they do not require active input from the user. They can automatically track movement, heart rate, sleep patterns, and other metrics, providing users with valuable insights about their health and wellness. Additionally, wearables are designed to be unobtrusive and worn throughout the day and night, allowing continuous monitoring. These factors have contributed to the rise of wearables, making them an increasingly popular choice for personal health and fitness tracking.

\subsection{Issues with Current Methods for Fatigue Monitoring}\label{subsubsec:FatigueMonitoringMethod}

There are several issues with the current methods and technologies used for fatigue monitoring. Self-reported fatigue can be unreliable as it relies on the individual’s perception and subjective interpretation of the symptom. People may have different thresholds for what they consider "fatigue," and other elements like mood, motivation, and cognitive biases may also play a role. Additionally, people may under report or over-report their fatigue levels due to social desirability bias, where they want to present themselves in a positive light. Furthermore, poor recall may affect self-reported fatigue, particularly if the individual is asked to report their fatigue levels retrospectively. Finally, self-reported fatigue may not always be accurate due to the complexity of the symptom, which various physical, psychological, and environmental factors can influence.

In addition to the issues related to self-reported fatigue, there are several other issues with monitoring fatigue. One of the major issues is the lack of a standard definition of fatigue, which makes it difficult to compare studies and results. Additionally, there is a lack of standardized and validated measures for fatigue, making it difficult to assess and monitor the symptoms accurately. Another problem is that there are numerous causes of fatigue, including illnesses, medications, lifestyle choices, and psychological factors, making it difficult to pinpoint the root cause.

Additionally, numerous physical, psychological, and environmental factors  can affect fatigue, making it difficult to monitor with a single measurement or device. Furthermore, other elements like stress, physical activity, or sleep quality may  impact physiological measures like heart rate, electroencephalogram, or actigraphy. Finally, monitoring fatigue in real-world settings outside of a laboratory can be challenging due to the need for continuous, discreet, and reliable data collection.

This work provides a comprehensive literature review of wearables and AI for fatigue monitoring. It covers various aspects of fatigue monitoring, such as the physiological signals used, the different types of wearables available, the methods used for data analysis, and the applications of fatigue monitoring in different industries. The literature review also looks at the current state of the technology, the challenges faced, and the future directions for the field. This work aims to provide a comprehensive overview of the current state of the research on fatigue monitoring using wearables and artificial intelligence and identify the gaps in the literature that need further research. Fatigue significantly impacts several industries, including construction, sports, aviation, healthcare, transportation, human health, and performance. Such fatigue may have negative impacts, resulting in accidents, injuries, and lower productivity. These effects include decreased awareness, diminished cognitive function, and decreased physical performance. The feasibility and efficacy of traditional fatigue detection techniques are limited because they frequently rely on subjective evaluations or intrusive procedures.	These challenges may be partly resolved by combining wearable technology with AI, making it possible to assess weariness continuously, non-invasively, and in real-time. To examine wearables and AI fatigue monitoring trends, challenges, and future potential. This article explains how these technologies might enhance performance, safety, and health. The key contributions of this article are listed as follows:

\begin{itemize}
    \item Thoroughly examined various fatigue measurement approaches, highlighting their strengths and limitations.
    
    \item Emphasis on physiological signal-based monitoring and assessing its implication over real-time fatigue monitoring using AI and deep learning.
    
    \item Critical analysis of wearables and AI through a literature review that covers the application of AI and wearables
across different physiological signals, providing insights into fatigue monitoring.

    \item Identify potential research challenges such as real-time data access and ergonomics while identifying gaps in predictive accuracy and explainable AI.
    
    \item Proposes future research opportunities, including new sensor development, edge computing integration, and improved multimodal fusion techniques.
    
\end{itemize}

The rest of the article is organized as follows: Section \ref{sec:FundamentalsandMethodology} begins with an introduction to fatigue and the role of AI and wearables, followed by a detailed exploration of fatigue measurement methods in Section \ref{Sec:FatigueMeasurementMethods}. Section \ref{sec:DBFatigueMonitoring} summarizes the popular fatigue datasets that the AI frameworks could use for appropriate predictions. The article then discusses the advantages of physiological signals for fatigue monitoring in Section \ref{sec:PhysiologicalSignalsFatigueMonitoring}. A comprehensive literature review on AI and wearable applications is presented in Section \ref{sec:FatigueMonitoringUsingWearablesAI}. The research challenges are presented in Section \ref{sec:ResearchChallenges}, and from the synthesized findings, identified research challenges and highlights of future opportunities are summarized in Section \ref{sec:ResearchGapsandFutureOpportunities}. The article concludes with a summary of contributions towards fatigue measurement approaches in Section \ref{sec:Conclusion}.

\section{Fundamentals and Methodology}\label{sec:FundamentalsandMethodology}
This section summarizes the fundamental ideas and context related to fatigue, its effects, and measurement methods. Subsequently, it reviews the systematic literature search approach to find pertinent studies on wearable and AI-based fatigue monitoring.

\subsection{Fatigue Background}\label{subsubsec:Fatigue Background}
\begin{figure}[h]
    \centering
    \includegraphics[scale=0.25]{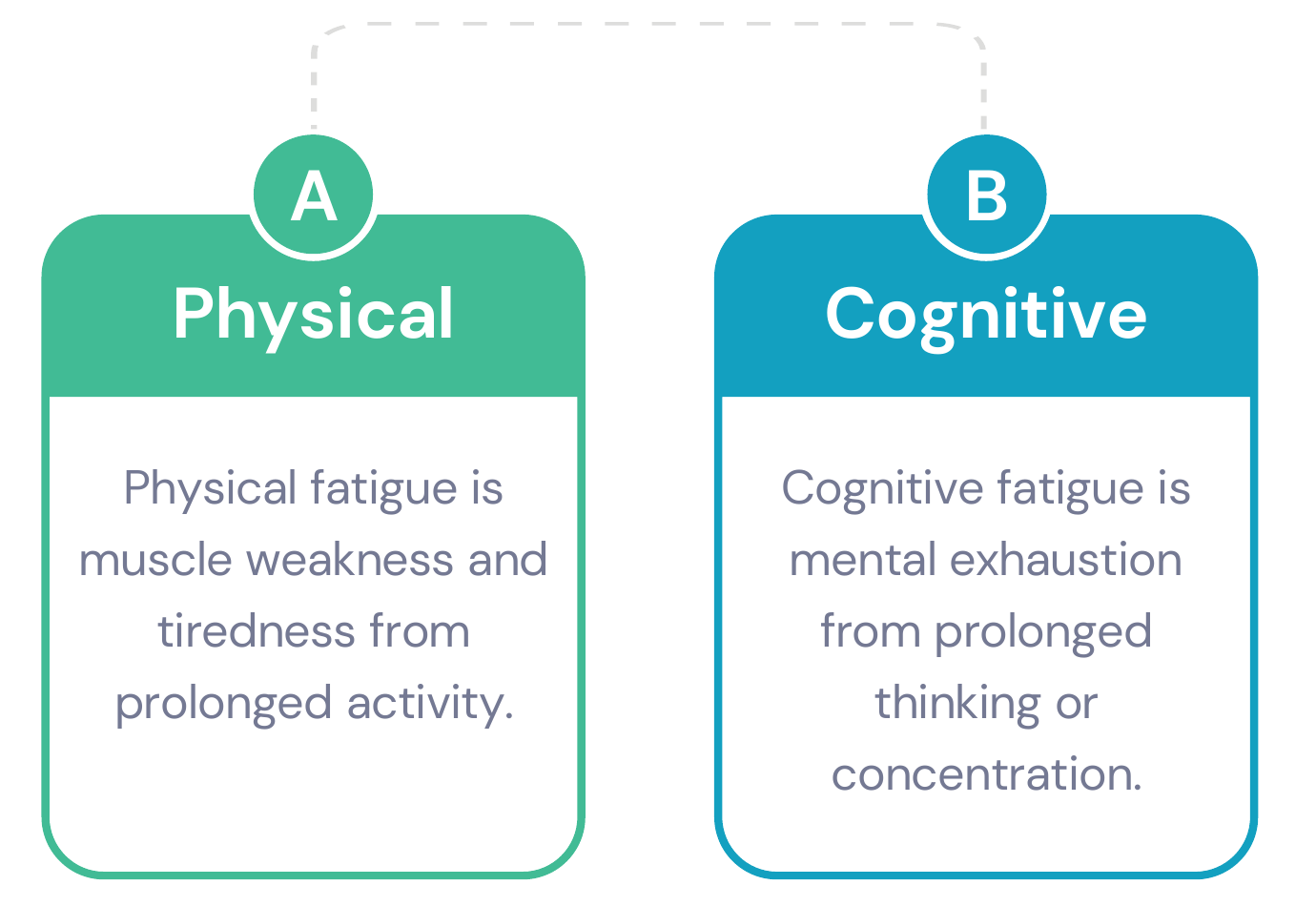}
    \caption{Identified two main types of fatigue.}
    \label{fig:Figure2}
\end{figure}

Fatigue is a multifaceted condition characterized by a decline in physical, cognitive, or mental performance due to prolonged activity or stress (see Figure \ref{fig:Figure2}) . This section explores the various types of fatigue and the impact of AI in detecting and managing fatigue, as well as reviews related studies that contribute to our understanding of this issue.

\begin{figure}[h]
    \centering
    \includegraphics[scale=0.16]{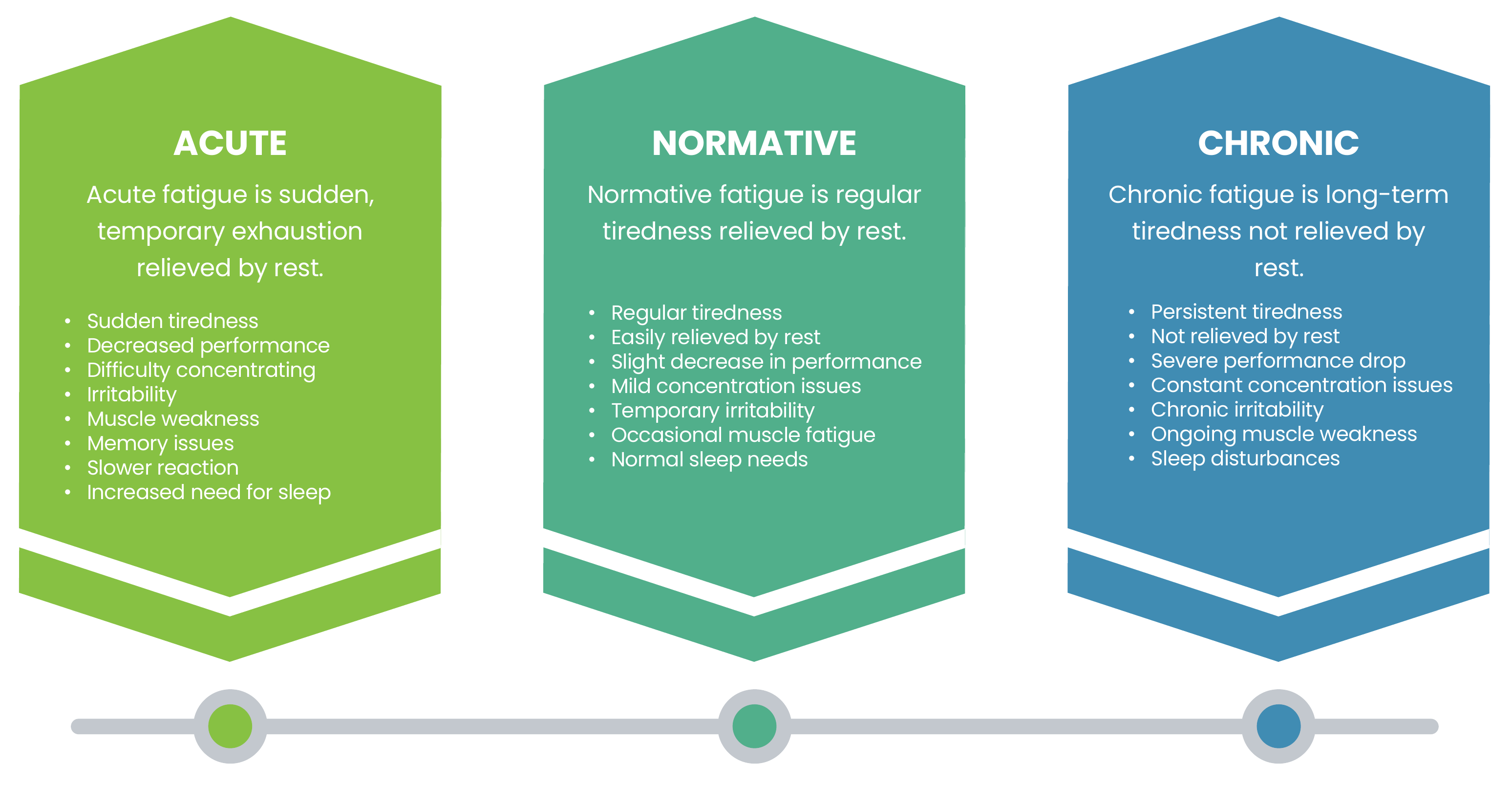}
    \caption{Three forms of fatigue are identified.}
    \label{fig:Figure3}
\end{figure}

Fatigue occurs when one does not get enough sleep, works for an extended period at a high mental or physical intensity, or does the same mundane duties repeatedly \cite{27}. A tired individual may not recognize their condition. Therefore, this could potentially create risk and has to be managed \cite{28}. Suppose the fatigue condition of an individual can be diagnosed in time, and a corresponding early warning can be supplied. In that case, \cite{28}, there may be a significant reduction in the incidence of an accident due to fatigue. For instance, the Transport Accident Commission in Australia \cite{29} reports that fatigue is a significant factor in 20\%-30\% of fatal road accidents in Australia. Also, research indicates that fatigue is directly responsible for 30\% of all fatal accidents in the manufacturing sector \cite{30}. Workers who are very drowsy or exhausted are 70\% more likely to be engaged in industrial mishaps than attentive and well-rested workers \cite{31}. There are two main types of fatigue physical and cognitive \cite{32}. Each type could be then divided into three levels according to the severity of fatigue: acute, normative, and chronic, as shown in Figure \ref{fig:Figure3}. Acute fatigue is a normal response to physical or mental activity and typically lasts for a short period, for example, after a long day or a hard workout.
Rest, sleep, and other forms of relaxation can help with acute fatigue.

Chronic fatigue is a more severe and long-term type of fatigue. It is characterized by persistent feelings of exhaustion that are not relieved by rest and sleep. Underlying medical conditions, such as anemia, diabetes, thyroid disorders, and certain infections, can cause chronic fatigue. Lifestyle factors, such as poor nutrition, lack of exercise, and chronic stress, can also cause it. Normative fatigue is a type of fatigue that is considered normal and acceptable. It is a normal part of aging, familiar in healthy people with demanding jobs, children, or other responsibilities. Normative fatigue is a state of tiredness that can be managed with good sleep hygiene and healthy lifestyle habits, but it doesn’t require medical attention.

\begin{figure}[h]
    \centering
    \includegraphics[scale=0.4]{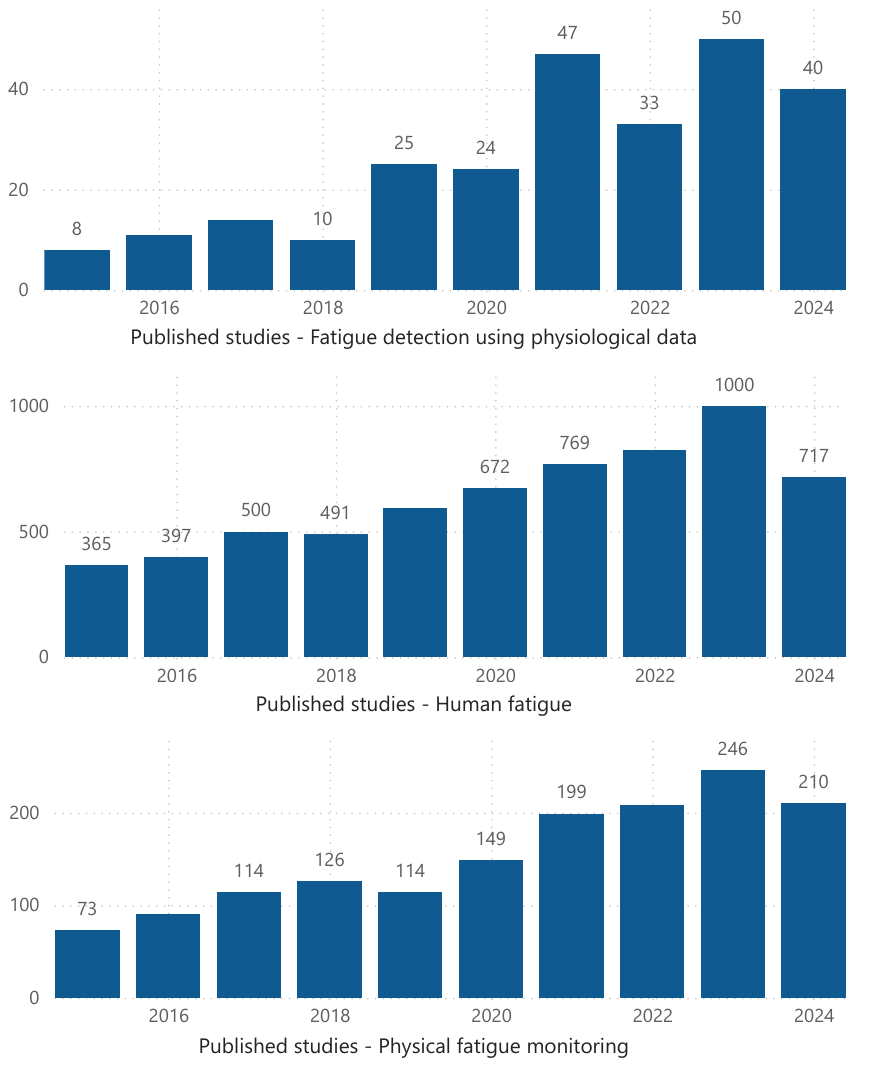}
    \caption{Published studies between 2015 and 2024 related to fatigue detection using physiological data (as per a search conducted on Engineering Village©).}
    \label{fig:Figure4}
\end{figure}

There is a growing interest in detecting and monitoring mental and physical fatigue using wearables and artificial intelligence \cite{28} \cite{33} \cite{34} \cite{35} \cite{36} \cite{37}. According to a recent search of Engineering Village ©, more than 8,000 articles have been published between 2015 and 2024 on the topic of
fatigue measurement and detection based on human medical or physiological data (see Fig. \ref{fig:Figure4}). Fatigue is subjective to be reliably measured; however, it depends on many other physical, mental, and environmental multidimensional factors \cite{38}. A growing body of evidence suggests that lifestyle, social, psychological, and general wellness difficulties, rather than a pre-existing medical illness, are primary contributors to fatigue. It is essential to differentiate mental fatigue from neuromuscular fatigue (muscle fatigue). Muscle or neuromuscular fatigue results from repeated muscle actions \cite{39}. In contrast, mental fatigue is the inability to execute mental activities involving self-motivation and internal signals without physical reason.

Human fatigue may be better understood by using physiological signals. These signals may aid in early diagnostics and therapeutics by estimating essential biomarkers, which can reveal the fatigue state formed in the human body. To create a fully automated detection system, studies of physiological signals have undergone significant evolution in recent years \cite{40}. There are a variety of techniques available for identifying signs of fatigue.

Recent studies have shown that methods for identifying static muscle fatigue are relatively advanced \cite{37}. Sensing devices are often used for signal collection. The subjective and objective levels of human fatigue are used to determine the muscular exhaustion state, and various classification techniques are developed \cite{37}.  However, there are still many obstacles to overcome when studying fatigue-detecting technologies. There is currently no clear-cut solution for measuring fatigue objectively. Various methods have been proposed, including self-report questionnaires, physiological measures, and performance-based tests, but each has limitations and may not provide a comprehensive assessment of fatigue. Additionally, fatigue is a complex and multidimensional phenomenon that can be influenced by various factors, such as physical and mental demands, sleep quality, and overall health, making it difficult to quantify accurately. Therefore, it is essential to consider multiple measures and approaches when assessing fatigue \cite{41}.

Advances in wearable and consumer electronics have significantly improved physiological signal measurements. Wearable devices such as smartwatches, fitness trackers, and bright clothing can now monitor various physiological parameters, including heart rate, respiration, and body temperature. Additionally, many devices now include sensors that can measure other necessary physiological signals, such as electrodermal activity, which can provide insight into stress levels, and EMG, which can provide information about muscle activity. Consumer electronics like smart glasses and head-mounted displays are also being used for physiological signal measurements in research settings like EEG \cite{42}, ECG \cite{43} , \cite{44}, PPG \cite{45}, EOG \cite{46}, EEG \cite{47} and many others. These devices are becoming increasingly sophisticated, with many now including advanced algorithms and machine learning capabilities to provide more accurate and detailed data. This allows for monitoring physiological signals in real-time, providing new opportunities for research, healthcare, and personal use.

On the other hand, the availability of edge and cloud computing has dramatically expanded the capabilities for processing and analyzing data \cite{48}. Edge computing refers to using resources close to the data source, such as on a device or at the network edge, allowing for real-time data processing and analysis \cite{49}. With cloud computing, data can be easily stored, shared, and analyzed by multiple users and devices \cite{50}. This has led to the development of cloud-based platforms and services for data analysis and machine learning, making it easier for organizations and individuals to gain insights from their data \cite{51}. Together, edge and cloud computing provide a powerful combination of capabilities that can be leveraged to process and analyze data in various applications.

AI and wearables for fatigue monitoring have numerous applications, but they are still in their infancy. The ability to continuously monitor physiological signals in real-time using wearable \cite{52} and analyze the data using AI algorithms can revolutionize how we understand and manage fatigue \cite{53}. However, despite the potential, the field of fatigue monitoring using AI and wearable is still in its early stages, and a lot of work needs to be done to develop robust and validated methods. The physiological markers of fatigue are poorly understood, and the algorithms used to analyze the data are still being developed and refined \cite{54}. Additionally, the wearable used for monitoring must be validated for accuracy and reliability. Despite these challenges, the potential benefits of using AI and wearable for fatigue monitoring make it a promising area of research and development.

In light of what has been said before, According to a recent search of Engineering Village©, between 2015 and 2024, 8,121 scholarly articles were published on fatigue measurement and detection based on human medical or physiological data (see Figure \ref{fig:Figure4}).

\subsection{Literature Search Strategy}\label{subsubsec:LiteratureSearchStrategy}
\begin{figure*}[h]
    \centering
    \includegraphics[scale=0.65]{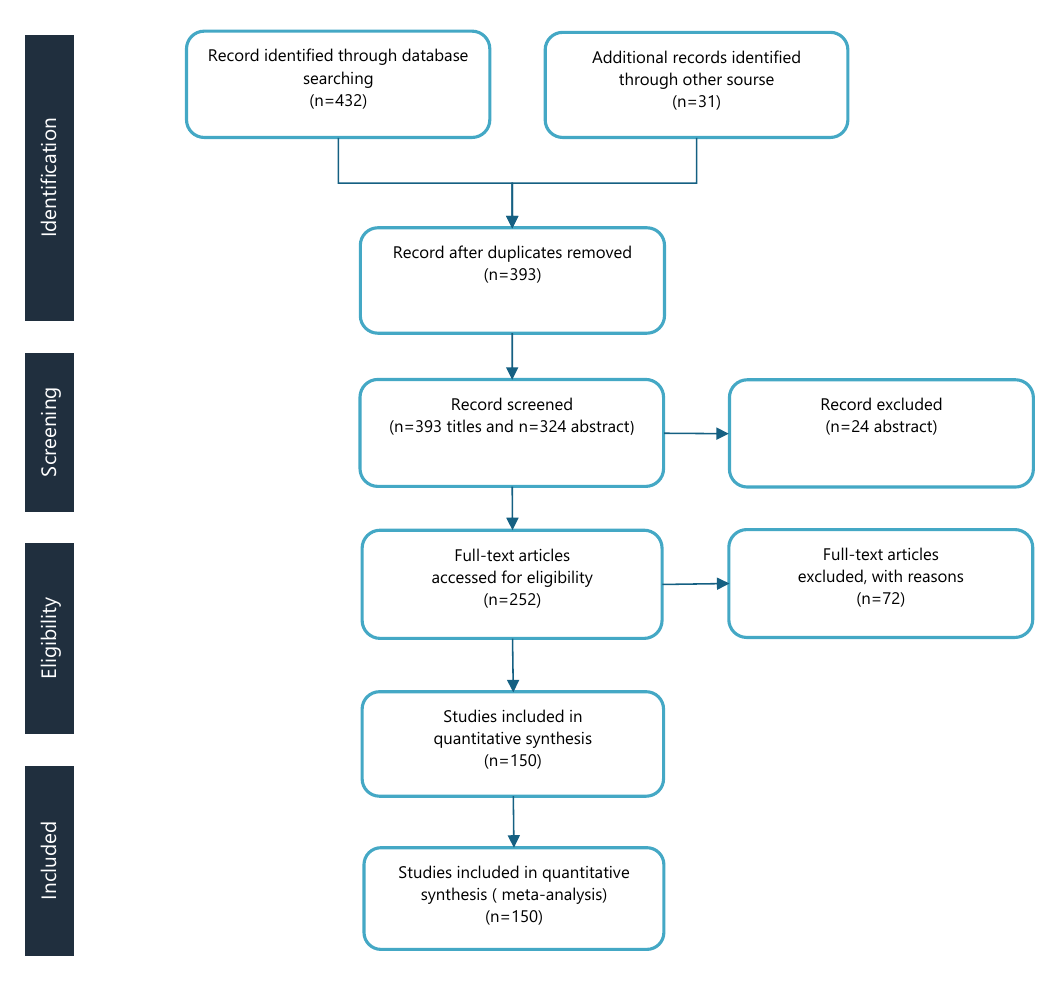}
    \caption{PRISMA flowchart used for our systematic review.}
    \label{fig:Figure5}
\end{figure*}

This section outlines the systematic literature review methodology, including the databases used, search parameters, and inclusion criteria. A four-phase flowchart of the literature search and selection procedure is illustrated in Figure \ref{fig:Figure5} , providing a clear and organized representation of the steps taken in the literature selection process. This framework shows the sequential steps from searching databases, screening, ensuring the relevance of articles, and final inclusion of research papers for a systematic review of articles related to fatigue monitoring using wearable technologies and AI.

\subsubsection{Search Methods}\label{subsubsec:SearchMethods}
For this review, four major databases were utilized to identify and select relevant studies on fatigue monitoring using wearable technologies and AI: Google Scholar, IEEE Xplore Digital Library, Web of Science (WoS), and Scopus. Google Scholar was included due to its broad coverage of academic sources across multiple disciplines. IEEE Xplore Digital Library was selected for its focus on high-quality, peer-reviewed technical and engineering research, particularly in AI and wearable technologies. Scopus was used for its extensive coverage of citations and abstracts from international peer-reviewed literature, while WoS provided access to comprehensive bibliometric analysis and citation metrics.

The following keywords were used in the search process: "Fatigue Monitoring," "Wearable Technology," "Artificial Intelligence," "Machine Learning," "Deep Learning," "Physiological Signals," "ECG," "EEG," "EMG," "PPG," and "EOG." Boolean logical operators were used to combine search terms such as "Fatigue Monitoring" AND "Wearable Technology" to target relevant articles. Additionally, terms like "Artificial Intelligence" OR "Machine Learning" were combined to ensure the inclusion of articles that addressed AI in fatigue monitoring. To broaden the scope, we also included physiological signal modalities like "ECG," "EEG," "EMG," and "PPG."

\subsubsection{Inclusion and Exclusion Criteria}\label{subsubsec:InclusionExclusionCriteria}
We included original research articles written in English that focused on fatigue monitoring through wearable technologies and AI, with an emphasis on fields such as computer science, engineering, and healthcare. No restrictions were placed on publication dates, given the recent advancements in AI and wearable technologies. Secondary studies, reviews, book chapters, editorials, patent documents, and conference papers were excluded to ensure a focus on primary research. This strategy ensured that our review centered on the most relevant and high-quality studies directly addressing our research questions.

\subsubsection{Data Extraction and Structured Meta-analysis}\label{subsubsec:DataExtractionStructuresMetaAnalysis}
During the first screening stage, titles and abstracts were carefully reviewed to identify papers related to fatigue monitoring using wearable technologies and AI. Multiple reviewers independently conducted full-text screening, with disagreements resolved through discussion. Data extraction was managed using Google Sheets, which recorded details of inclusion or exclusion, enabling systematic data management. The extracted data focused on critical issues such as wearables, AI methodologies, and physiological signal types in fatigue monitoring. Studies that lacked transparent methodologies or did not explicitly focus on fatigue detection through AI or wearables were excluded. Tabulation techniques supported the data extraction process, guaranteeing that the study’s outcomes were consistent with the research objectives. Methodological approaches from the selected articles were evaluated, and those with vague definitions or unclear methods were disqualified.

The final data was systematically organized, allowing for a structured meta-analysis and presentation of findings in this review. Each article’s methodological data were thoroughly assessed and organized to ensure relevance to the research focus on fatigue monitoring using AI and wearables.

\section{Datasets Availability for Fatigue Monitoring}\label{sec:DBFatigueMonitoring}
High-quality datasets are essential for advancing the efficacy of fatigue monitoring systems that leverage wearable technologies and AI. They help to improve the accuracy of detecting fatigue through physiological signals like ECG, EMG, EEG, PPG, and EOG. This section summarizes significant publicly accessible datasets pertinent to the study of fatigue detection.

\subsection{Physiological Signal-Based Datasets}\label{subsubsec:PhysioSignalDB}
Datasets based on physiological signals concentrate on the acquisition of vital physiological indicators, which play a significant role in identifying fatigue and examining the body’s reactions to physical and mental stress.

\subsubsection{PhysioNet}\label{subsubsec:PhysioNet}
PhysioNet is an extensive repository that provides access to various physiological signal datasets, encompassing ECG, EEG, and PPG \cite{55}. It is extensively utilized in numerous health-related investigations, particularly fatigue monitoring. PhysioNet provides a wide range of data that can be used to create fatigue detection models by combining different physiological signals \cite{56}. This makes the models more accurate and valuable in healthcare and fitness.

\subsubsection{DREAMER}\label{subsubsec:DREAMER}
The DREAMER dataset includes EEG and ECG signals \cite{57}. Initially created for emotion recognition, it is also a valuable resource for fatigue detection. This tool effectively records physiological responses across various cognitive and emotional tasks, an essential resource for exploring mental fatigue and cognitive workload \cite{58}. The dataset uses machine learning and deep learning models to evaluate mental states like fatigue.

\subsubsection{SEED}\label{subsubsec:SEED}
The SJTU Emotion EEG Dataset (SEED) comprises EEG data collected during cognitive tasks and is commonly utilized to detect mental fatigue \cite{59}. The dataset comprises recordings from participants engaged in emotionally stimulating tasks, facilitating an analysis of the effects of fatigue on cognitive functions. Although the main emphasis is on mental fatigue, SEED is an essential resource for investigating cognitive workload \cite{60}.

\subsubsection{SPhysio}\label{subsubsec:SPhysio}
SPhysio encompasses various physiological signals, such as ECG and PPG, which are instrumental in investigating physical fatigue in exercise and fitness environments \cite{61}. This dataset facilitates the development of models to detect and manage fatigue in environments characterized by physical demands, including sports and labor-intensive occupations.

\subsection{Behavioral and Video-Based Datasets}\label{subsec:BehavioralVideoBasedDatasets}
Datasets that encompass behavioral and video-based elements are designed to capture visual and behavioral markers of fatigue, including facial expressions, eye movements, and yawning. These datasets are frequently utilized in real-time fatigue detection systems, leveraging computer vision and AI methodologies.

\subsubsection{YawDD}\label{subsubsec:YawDD}
The Yawning Detection Dataset (YawDD) \cite{62} comprises a series of video recordings designed to document yawning behaviors, which have been gathered to detect driver fatigue. The dataset comprises labeled yawning events, significant markers of drowsiness within driving scenarios \cite{63}. This dataset is often utilized alongside computer vision methodologies to create systems for real-time fatigue detection by analyzing facial expressions.

\subsubsection{CEW}\label{subsubsec:CEW}
The Closed Eyes in the Wild (CEW) dataset comprises images depicting individuals with both open and closed eyes, serving as a resource for detecting drowsiness in practical environments \cite{64}. The detection of fatigue is particularly significant in contexts where eye closure may serve as an indicator of diminished alertness, notably in driving scenarios. This dataset serves as a valuable resource for the training of deep learning models that utilize facial cues to evaluate levels of drowsiness and fatigue.
\subsection{Multi-Modal Datasets}\label{subsubsec:MultiModalDataset}

\subsubsection{Driver Fatigue}\label{subsubsec:Driver Fatigue}
The Driver Fatigue Dataset comprises EEG, EOG, and behavioral data from drivers engaged in extended driving sessions \cite{65}. The development of fatigue detection systems in automotive safety is of significant importance, particularly due to the necessity for real-time monitoring to mitigate the risk of accidents resulting from drowsiness. This dataset facilitates the investigation of machine learning algorithms aimed at identifying early indicators of fatigue through the analysis of neural and ocular activities.

\subsubsection{DROZY}\label{subsubsec:DROZY}
The DROZY dataset includes different types of multi-modal data, like EEG, ECG, EOG, and video recordings that were collected in certain ways to make people sleepy \cite{66}. This dataset has a lot of potential for studies that look at how to detect fatigue, especially when it comes to situations where combining physiological signals and behavioral data is important for making accurate predictions about fatigue levels \cite{67}. This approach frequently assesses machine learning models for real-time fatigue monitoring within practical applications.

Although various datasets are accessible, there are ongoing challenges related to data quality, the applicability of data in real-time scenarios, and the need for standardization across different studies. Numerous datasets exhibit inconsistencies that hinder the development of generalizable models. Variations in sensor quality, participant characteristics, and experimental conditions may restrict their applicability across different domains. Furthermore, numerous datasets are structured for static or short-term investigations, which poses challenges for ongoing observation in practical settings.

\section{Fatigue Measurement Methods}\label{Sec:FatigueMeasurementMethods}
This section focuses on the diverse methodologies for assessing fatigue, including subjective self-reports, performance-related tests, bio-mathematical models, behavioral observations, and physiological signal-based approaches. Each offers unique insights into the detection and quantification of fatigue.

The term "fatigue measurement" is used to describe the process of determining how tired a person is. Self-report surveys, physiological markers, and performance assessments are just a few ways fatigue may be evaluated. Lack of sleep, strenuous activity, stress, and medical problems, including depression and chronic fatigue syndrome, are all potential contributors to tiredness. Work schedule, shift work, and poor sleep quality are other causes of fatigue. Recognizing and addressing the root causes of tiredness is crucial because of its detrimental effects on people’s safety, productivity, and quality of life.

\begin{figure}[h]
    \centering
    \includegraphics[scale=0.18]{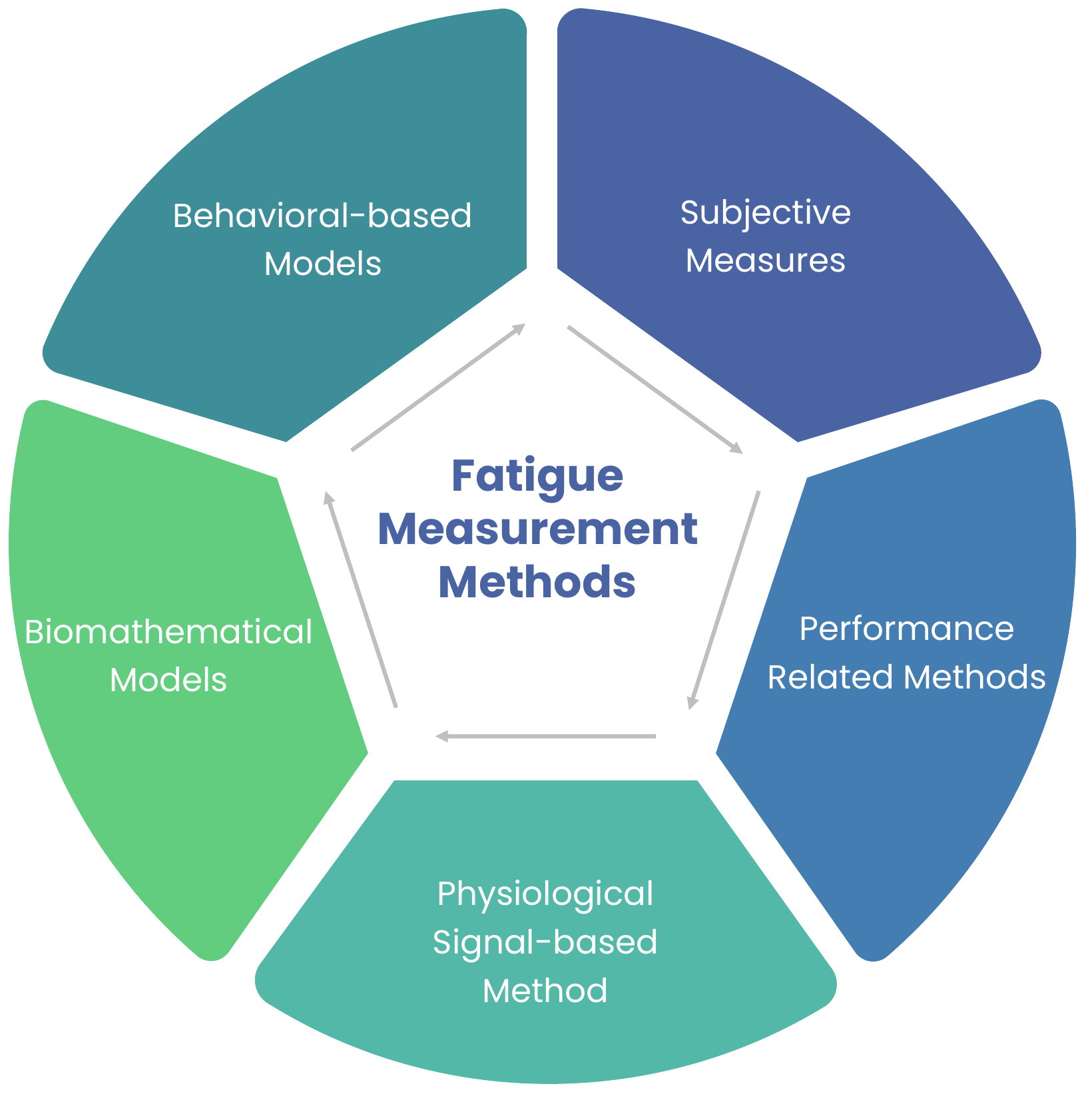}
    \caption{Common techniques and frameworks used in literature for fatigue measurement and monitoring.}
    \label{fig:Figure6}
\end{figure}

Although several methods have been presented for fatigue diagnosis and monitoring, there is currently no agreed-upon method for quantifying fatigue \cite{68}. Figure \ref{fig:Figure6} demonstrates the most common methods used for fatigue monitoring. In this section, we briefly discuss these methods.

\subsection{Subjective Measure}\label{subsec:MultiModalDataset}
Subjective fatigue measurements analyse an individual’s fatigue perception \cite{69}. Self-report questionnaires or surveys ask people to assess their exhaustion on various scales. Visual analog scales, numerical rating scales, and multi-item fatigue inventories assess subjective exhaustion. These scales may ask people to score their fatigue level, frequency, duration, or associated symptoms like drowsiness or energy. Subjective fatigue measurements may reveal an individual’s reported exhaustion and monitor changes over time. However, they are biased and may not appropriately reflect fatigue levels. Therefore, subjective fatigue assessments should be combined with objective performance measures or physiological indicators to provide a complete picture of an individual’s fatigue levels \cite{70}.

\begin{figure}[h]
    \centering
    \includegraphics[scale=0.55]{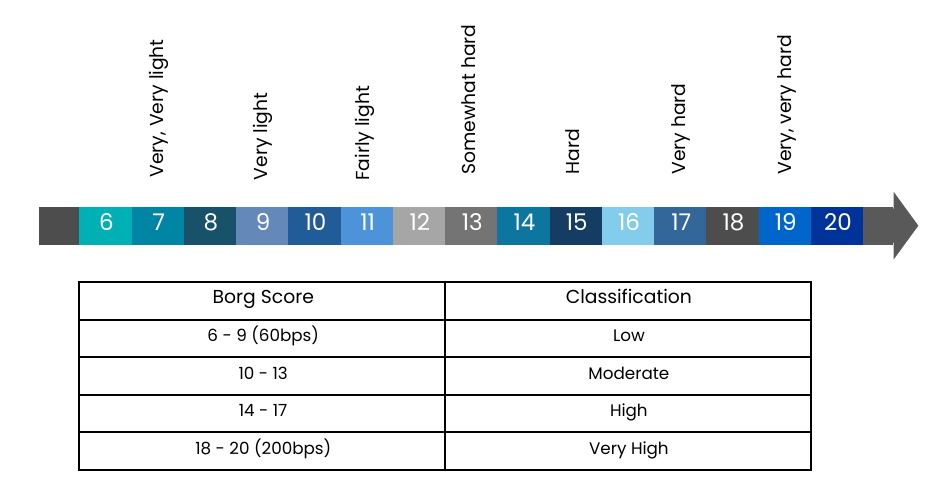}
    \caption{Common techniques and frameworks used in \cite{68} for fatigue measurement and monitoring.}
    \label{fig:Figure7}
\end{figure}
\subsection{Performance Related Methods}\label{subsubsec:PerfRelaMethods}
Methods based on performance depend on the observation that people’s mental and, by extension, physical capacity to carry out a set of activities reflects their degree of exhaustion. Neuro-behavioral tasks are used in these approaches to evaluate a subject’s cognitive abilities (such as alertness, hand-eye coordination, sustained attention, and response time). Although performance-related procedures may be standardized with relative ease, they cannot be utilized to identify the onset of fatigue in time to avoid an accident.

\subsection{Bio-mathematical Models}\label{subsubsec:BioModels}
In line with \cite{68}, subjects’ fatigue levels are predicted by bio-mathematical models using data on their sleep-wake schedules, work-rest patterns, and circadian rhythms. The aviation industries of both the military and the civilian sectors are frequently employed to evaluate the danger of fatigue. These models are the product of research on the effects of both partial and complete sleep deprivation. Wrist-mounted actigraphy has replaced subjective reporting to improve the accuracy of sleep and waking time estimations.

Several bio-mathematical models have been developed to measure fatigue, each with advantages and limitations. Some of the most commonly used models include:

\begin{itemize}
    \item The Fatigue Severity Scale (FSS): This is a self-report questionnaire that measures the impact of fatigue on various
aspects of daily life, such as work, social activities, and mental health. It is a widely used tool for assessing fatigue in various medical conditions, including chronic fatigue syndrome and cancer-related fatigue.

    \item The Chalder Fatigue Scale (CFQ): This is another self-report questionnaire that measures fatigue in 11 items. It is widely used in clinical research and in primary care to diagnose chronic fatigue syndrome.

    \item The Piper Fatigue Scale (PFS): This self-report questionnaire measures the frequency and severity
of different types of fatigue. It is widely used in cancer-related fatigue research and in primary care.

    \item The Fatigue Impact Scale (FIS): This is a self-report questionnaire that measures the impact of fatigue on various aspects of daily life, such as physical, emotional and social
functioning.

    \item The Multidimensional Assessment of Fatigue (MAF): This is a self-report questionnaire that measures the frequency, severity and impact of fatigue on various aspects of daily life, such as physical, emotional, and cognitive functioning.

\end{itemize}

These models are widely used in various fields, including research and clinical practice, to evaluate fatigue and its impact on quality of life. They can be used in different conditions as a way of measuring the fatigue levels of patients, but ultimately, the choice of the right tool depends on the condition, population, and goals of the assessment.

\subsection{Behavioural-Based Methods}\label{subsubsec:BehaveMethods}
Behavioral-based methods for fatigue monitoring involve observing and measuring an individual’s behavior to detect signs of fatigue. As per authors in \cite{68} \cite{45} \cite{71} \cite{72}, external symptoms of exhaustion, such as yawning, sighing, eyelids closing, or head nodding, are observed and accounted for using behaviorally-based procedures (see Figure \ref{fig:Figure7}). Therefore, metrics associated with eye movements, head motion, and facial expression are often used as input characteristics in technologies belonging to this class.

\subsection{Physiological Signal-based Methods}\label{subsubsec:PhysioSignalMethods}
Fatigue monitoring using physiological signals involves measuring an individual’s physiological responses to detect signs of fatigue. Some commonly used physiological signals for fatigue monitoring include EEG, ECG, EMG, and EOG. These methods are highly accurate, non-invasive, and can objectively measure fatigue. They are widely used in various fields, such as transportation, aviation, and shift work, to monitor fatigue levels and prevent accidents. Choosing the right physiological signals to monitor fatigue depends on the context, population, and assessment objectives (see Section \ref{sec:FatigueMonitoringUsingWearablesAI} for a comprehensive review of these methods).

Figure \ref{fig:Figure8} shows the overall AI/ML pipeline for fatigue monitoring using physiological signals. Once wearables collect the data,  pre-processed, filtered, and then cleaned. Then an AI/ML model is developed, optimized, and validated to predict the response variable (here fatigue). This model is then fully tested and deployed for real-world applications.

\section{Physiological Signals for Fatigue Monitoring}\label{sec:PhysiologicalSignalsFatigueMonitoring}
Physiological signs are often used as a reliable and noninvasive technique to determine the level of fatigue \cite{73}. unfortunately, most other non-physiological signal-based techniques primarily concentrate on accuracy rather than forecasts for fatigue effectiveness. The resulting procedures are less reliable and accurate, making them inappropriate for fatigue diagnosis and monitoring. To solve this problem, the research world has proposed magnificent, complete frameworks for the early identification of fatigue by modeling data collected from wearables, relying on physiological signal-based approaches. Below are the core benefits of utilizing an approach based on physiological signals.

\subsection{Data-driven (objective vs subjective)}\label{subsec:DataDrivenvsSubj}
As a quantitative record of the human body’s dynamic physiological processes, physiological-driven data provides an accurate and impartial snapshot of the human body in its present setting \cite{38}. Researchers can better identify and predict fatigue using physiological data-driven methods because they can act on the most relevant information at the most relevant moment. Furthermore, data-driven is a cornerstone of predictive analytics. Physiological data as objective-driven data should allow for a neutral and evidence-based evaluation of fatigue. Important judgments should not be made solely on subjective data for recognizing fatigue since it requires making assumptions and interpretations based on non-data observations without verified facts.

\subsection{Real-Time Monitoring}\label{subsec:WearableRTData}
Collection of physiological signals allows the end user to monitor fatigue in real time \cite{42} \cite{46} \cite{74} \cite{75} \cite{76}. Real-time fatigue monitoring using physiological signals can provide several benefits, including improved safety, increased efficiency, better understanding of fatigue, better treatment, and greater performance.

\subsection{Edge Processing}\label{subsec:EdgeCompDP}
Edge computing improves the speed and accuracy of fatigue detection using physiological data. Think about the possibility of using the analysis of sensor physiological data collected at the distant wearable’s application cloud to keep tabs on the wearable’s health. Such wearables have the potential to generate dozens, if not millions, of data points each second. Raw data transmission to a faraway data center may be excessively time-consuming and resource-intensive without pre-processing at the edge to remove irrelevant information. This is particularly true if the network link is unreliable.
\begin{figure*}[t]
    \centering
    \includegraphics[scale=0.44]{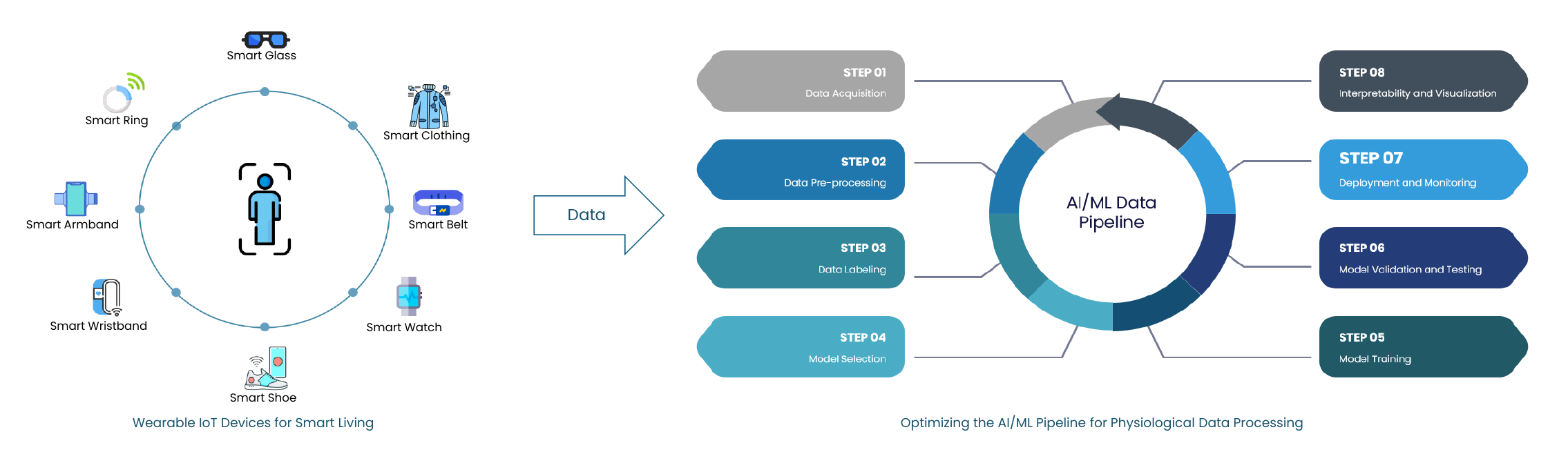}
    \caption{The overall AI/ML pipeline for processing physiological data measured by wearables.}
    \label{fig:Figure8}
\end{figure*}

\subsection{Using Advanced AI and DL}\label{subsec:UseAIDL}
The AI revolution for processing physiological signals for human performance monitoring has been driven by the emergence of wearables, big data, and the ability to perform real-time analysis. The availability of advanced AI/ML techniques for data processing has greatly expanded in recent years, enabling organizations to gain insights from a wide range of data sources, including time series data and multi-source data. These techniques include machine learning, deep learning, natural language processing (NLP), and computer vision. These techniques are widely used in various industries, such as healthcare, finance, transportation, and manufacturing, and are available through various commercial and open-source software platforms and libraries. Furthermore, with the ongoing research and development in the field, new techniques and improvements are continually being developed, making them even more powerful and accessible, especially with the fast data processing capabilities that are now possible. This allows individuals to quickly gain insights from large and complex datasets (here, mainly time-series of physiological signals), which can be used to make data-driven decisions and improve operations (see Fig. \ref{fig:Figure8} for the pipeline).

In addition to commercial software platforms and libraries, there are also a number of open-source Python packages, such as sci-kit-learn and Pytorch, that are widely used for processing physiological data. These packages provide a wide range of tools for data analysis, machine learning, and deep learning and are widely used by researchers and practitioners in the field. They are also very flexible and can be easily integrated with other tools and packages, making them a popular choice for physiological data analysis.

\section{Fatigue Monitoring using Wearables and AI}\label{sec:FatigueMonitoringUsingWearablesAI}
The goal of fatigue monitoring using wearables and AI is to employ cutting-edge technology to reliably and continually evaluate an individual’s state of fatigue \cite{68}. Refer to Figure \ref{fig:Figure8}, Smart-watches, fitness trackers, and sleep monitors are all examples of wearable technology that may give continuous, objective data on physiological and behavioural indicators of fatigue, such as heart rate, activity levels, and sleep patterns \cite{77}. Next, artificial intelligence systems may examine these records to glean useful insights about a person’s fatigue.

One of the key benefits of fatigue monitoring using wearables and AI is the provision of continuous real-time data on an individual’s degrees of fatigue \cite{78}. This paves the way for early diagnosis of fatigue-related concerns, including lower performance or increased accident risk, and may guide actions to enhance health and happiness. Also, those at a greater risk for fatigue-related problems, such as shift workers, truck drivers, and athletes, may have their muscle fatigue tracked using wearables and AI.

Although wearables and AI have many potential benefits for fatigue monitoring, they are not without their drawbacks. There are several variables, such as an individual’s degree of physical activity, might distort the data obtained by wearables, making it possible that they do not always provide an accurate reflection of a person’s genuine fatigue levels. The quality of the data, the complexity of the algorithm, and the existence of outliers or noise in the data may all impact the accuracy of AI algorithms while evaluating the data \cite{79}. Therefore, before deploying wearables and AI-based fatigue monitoring systems in the wild, testing and assessing their efficacy is crucial.

\begin{figure*}[t]
    \centering
    \includegraphics[scale=0.44]{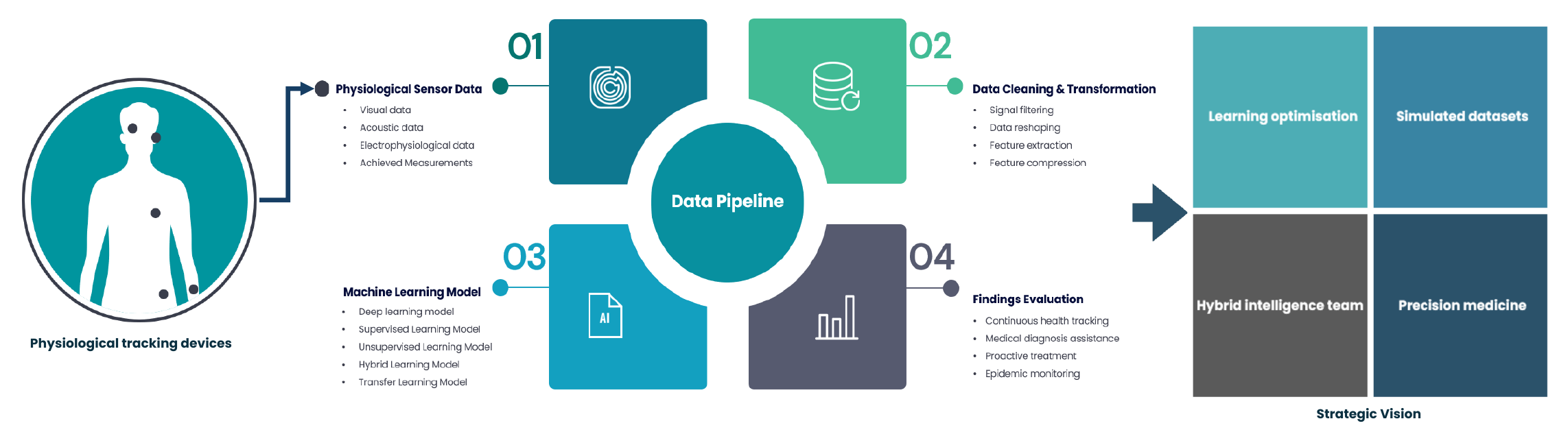 }
    \caption{The overall pipeline of fatigue monitoring using wearable and AI.}
    \label{fig:Figure9}
\end{figure*}

Figure \ref{fig:Figure9}  shows the overall process of using wearables and AI to monitor the performance of an individual. An individual’s wearables passively measure physiological data such as heart rate, skin temperature, and movement patterns. This data is then processed and cleaned to remove noise or irrelevant information. After the data is cleaned, AI algorithms such as machine learning and deep learning are used to model the data. These AI models are then used for prediction, providing insights to the end-users on fatigue level, sleep quality, and stress level. This process allows for non-invasive and continuous monitoring of an individual’s physiological state, providing valuable insights for healthcare professionals, researchers, and individuals to improve their overall health and well-being.

Wearable sensor fatigue monitoring is a method in which tiny, portable devices are worn on the body to gather information on several physiological and behavioral indicators of exhaustion, such as heart rate, activity levels, and sleep patterns \cite{80}. Smartwatches, fitness trackers, and sleep monitors are all examples of wearable sensors that may collect data via EEG, ECG, EMG, and PPG.

Figure \ref{fig:Figure10} shows a detailed workflow of the entire process of physiological signal management, starting from patient preparation, where the patient is readied for signal data acquisition. The raw physiological signals are then captured and undergo initial signal processing to extract and label key features indicative of cardiac activity. The processed signals are further transformed through feature extraction techniques, preparing them for advanced analysis. These features are subsequently mapped into a high-dimensional space using deep learning methods, which facilitate the detailed modeling of the signals. The model is fine-tuned to maximize the similarity between the extracted features and known patterns of various conditions. The final step involves using this trained model to predict specific heart conditions from new signals, categorizing them into Low, Moderate, High, or Very High fatigue conditions. This streamlined depiction emphasizes integrating clinical practices with advanced computational techniques to enhance diagnostic accuracy in fatigue.

\subsection{ECG-based Methods}\label{sec:ECG-basedMethods}
\begin{figure*}[t]
    \centering
    \includegraphics[scale=0.44]{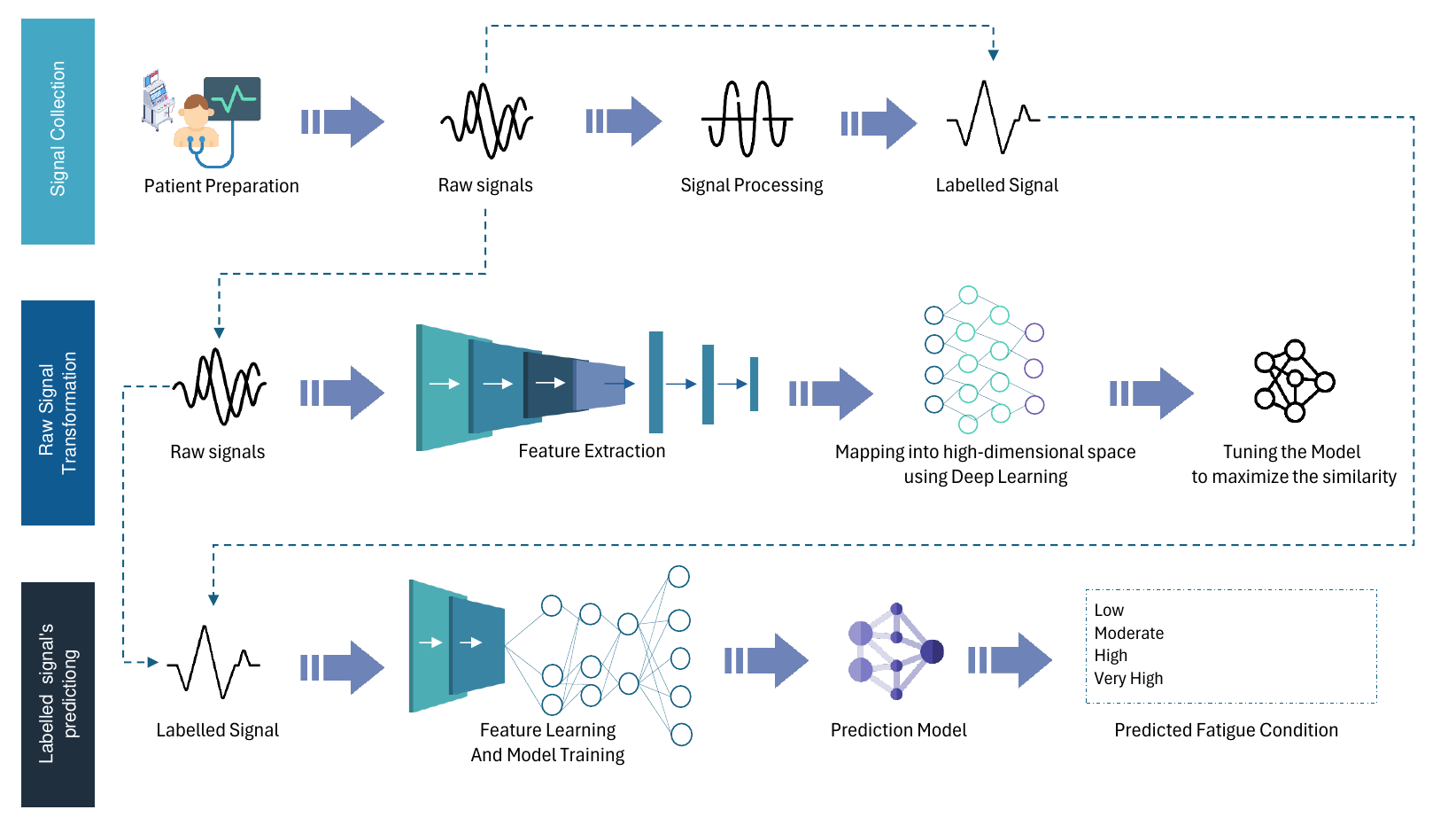 }
    \caption{Process of collecting and transforming physiological signals to predict fatigue using AI.}
    \label{fig:Figure10}
\end{figure*}

Fatigue detection utilizing ECG and AI technologies incorporates complex machine learning algorithms to conduct a detailed analysis of an individual’s HRV. This method involves examining the temporal variations in heart rate, capturing and analyzing the minutiae of its fluctuation patterns. Decreased HRV, indicated by fewer variations between consecutive heartbeats, serves as a critical biomarker for fatigue. It reflects diminished activity and adaptability of the autonomic nervous system, particularly the sympathetic and parasympathetic branches, under stress or fatigue. By mapping these variations and linking them to physiological responses, this approach provides an insightful, quantifiable indicator of an individual’s fatigue levels over time (See Figure \ref{fig:Figure10}).

\begin{figure*}[t]
    \centering
    \includegraphics[scale=0.2]{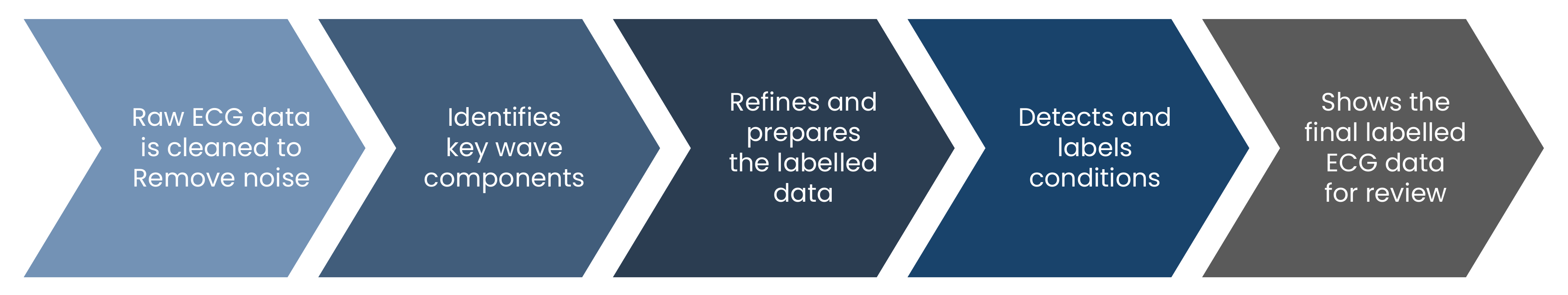 }
    \caption{AI-based ECG analysis: cleans raw data, identifies wave components, refines and labels conditions, and displays results for review.}
    \label{fig:Figure11}
\end{figure*}

The AI-based ECG analysis process begins with patient preparation and collection of raw ECG signals, which are then preprocessed to remove noise and artifacts. After pre-processing, important heart wave parts like atrial depolarization waves, ventricular depolarization complexes, and ventricular re-polarization waves are found.  This wave information is then refined and prepared to produce labeled ECG data. The next step involves feature extraction, where deep learning techniques map significant features from the ECG signals into a high-dimensional space. The model is subsequently tuned to enhance similarity with known cardiac conditions, optimizing its predictive accuracy. The labeled data is analyzed in the detection and labeling phase to identify cardiac abnormalities, such as arrhythmias and ischemic events. Finally, healthcare professionals display the labeled ECG data for further analysis and review. This comprehensive process ensures accurate and efficient ECG analysis, leveraging advanced AI techniques to develop robust prediction models that can accurately identify conditions such as normal heart function (NORM), myocardial infarction (MI), congestive heart failure (CHF), and hypertrophy (HYP), showcasing the efficacy of AI in improving cardiac diagnostics and patient outcomes (See Figure \ref{fig:Figure11}).

Current research in fatigue assessment highlights the critical role of ECG in advancing automated and objective monitoring methods. Wearable ECG devices have been shown to continuously and non-invasively monitor HRV indicators, which are essential for figuring out how tired someone is in real time without getting in the way of daily life \cite{81}. Researchers have shown that ECG signals can accurately pick up on different levels of driver hyper-vigilance, such as drowsiness and cognitive inattention, using advanced classifiers \cite{43}. The integration of ECG with video signals has shown significant promise in educational settings, achieving detection accuracies as high as 99.6\%, making it a powerful tool for identifying fatigue \cite{82}. Using R-R intervals from ECG data in a new way has helped find tired drivers, and advanced statistical models like AR-GARCH are good at spotting changes from alert to tired states \cite{83}. Real-time fatigue detection using short-time period ECG signals has also been validated, demonstrating practical application potential \cite{35}. Using machine learning models to find mental fatigue using ECG has been successful over 94.5\% of the time, showing that ECG features could be used for accurate fatigue monitoring \cite{84}. Techniques such as wavelet transform have further improved the accuracy of driver drowsiness detection by minimizing noise interference \cite{85}. High accuracy in driver drowsiness detection using wrist-worn ECG devices underscores the potential for integrating ECG technology into consumer devices and vehicles \cite{86}. Smart bracelets with ECG sensors have effectively assessed driving fatigue levels, demonstrating their practical application \cite{76}. Predicting exhaustion thresholds using ECG features and artificial neural networks has shown high accuracy, benefiting athletes through precise monitoring \cite{87}.  The integration of ECG in FDA-approved wearable devices for heart health monitoring further underscores its potential to revolutionize healthcare through continuous and personalized data analysis \cite{75}. Additionally, research has identified specific ECG parameters correlating with fatigue, providing valuable insights for clinical diagnosis and highlighting gender differences in fatigue characteristics \cite{81}. These advancements collectively underscore the indispensable role of ECG in developing reliable, objective, and interpretable methods for fatigue assessment.

In addition to being non-invasive and ideal for continuous monitoring in various settings, this cost-effective approach reduces consumables and medical personnel expenses. It enhances patient comfort and compliance, particularly in long-term monitoring, and is easier to use, requires less training, and facilitates broader adoption. The absence of invasive procedures also minimizes the risk of infections and other complications, enhancing safety. Moreover, non-invasive devices tend to be portable, allowing for monitoring in diverse environments, including homes and remote areas, and they can be seamlessly integrated with other health technologies. Furthermore, the rich data generated from continuous, real-time monitoring is well-suited for integration with future AI advancements, enhancing the potential for predictive analytics and adaptive learning algorithms to improve health outcomes over time. This combination of features makes non-invasive ECG monitoring a valuable tool for comprehensive health management and positions it favorably for future technological integrations.

While non-invasive ECG monitoring offers numerous benefits, it also has significant disadvantages, including its inability to distinguish between fatigue and other conditions that affect HRV, such as sleep apnea, potentially leading to false positives and confusion or misdiagnosis. Additionally, elements like the person’s general health, the caliber of the ECG signal, and the particular machine learning algorithm used can impair the detection accuracy. Other challenges include sensitivity to external noise and interference, which can compromise reading accuracy. The variability in signal quality due to individual anatomical differences, such as skin thickness and body fat percentage, may lead to inconsistent results. Non-invasive ECG may not provide the detailed data necessary for diagnosing more complex cardiac conditions and relies heavily on the accuracy of underlying algorithms for signal interpretation, which can result in inaccuracies or missed diagnoses. Continuous wear-ability can also be inconvenient or uncomfortable over long periods, affecting patient compliance and data reliability. Moreover, the continuous transmission and storage of health data raise significant privacy and security concerns, necessitating robust protections to ensure patient confidentiality. These limitations highlight the need for continuous improvement and careful integration with other diagnostic approaches in comprehensive cardiac care.

Regarding practicality, fatigue detection using ECG and AI has wide applications across several industries, notably enhancing safety and efficiency. In the transportation industry, this technology is crucial for monitoring the fatigue levels of drivers, pilots, and train operators, significantly improving safety by potentially preventing accidents caused by drowsiness. Beyond transportation, its utility extends to industrial settings, where continuous monitoring of employees can greatly enhance operational safety. In healthcare, the technology can monitor patients with chronic illnesses, while in athletics, it assesses athletes’ fatigue levels, potentially preventing over-training and aiding in optimized performance.

Other potential application areas include the technology and entertainment industries, where managing fatigue can increase productivity and mitigate health-related liabilities during extensive computer use or production activities. Despite these promising applications, it is important to note that further research is needed to enhance the accuracy and reliability of the detection methods before they can be widely adopted. This includes developing robust and accurate algorithms, studying the physiological and psychological changes that occur with fatigue, and establishing normative data for different population groups. These steps are critical to ensure the efficacy and practicality of ECG and AI in fatigue detection across diverse industries.

Authors in \cite{75} \cite{88} \cite{84} \cite{82} \cite{81} \cite{86} \cite{85} \cite{43} \cite{87} \cite{35} \cite{76} \cite{83} \cite{89} have used EEG signals and AI methods for identifying fatigue in subjects in the lab and real-world settings. Further information about the performance of AI methods for fatigue monitoring using EEG signals can be found in Table \ref{tab:FatigueMonitoring1}.

\subsection{EMG-based Methods}\label{sec:EMG-basedMethods}
Fatigue detection using electromyography (EMG) is a method for identifying signs of fatigue in individuals by analyzing the electrical activity of their muscles. This is typically done by measuring the electrical activity of the muscles via surface electrodes and analyzing the resulting EMG signals using machine learning algorithms or other analytical methods. The use of EMG for fatigue detection offers a non-invasive and practical approach to monitoring muscle function in real-time, making it valuable in various applications such as sports performance, rehabilitation, occupational health, and driver safety. Recent advancements have significantly enhanced the accuracy and reliability of fatigue detection by integrating advanced signal-processing techniques and machine-learning models. We used support vector machines (SVM), convolutional neural networks (CNN), and extended short-term memory networks (LSTM) along with the time-domain and frequency-domain properties of EMG signals to find the essential parts. This helped researchers make strong models that accurately tell the difference between different levels of muscle fatigue.

One of the main advantages of this approach is that it can provide a direct measure of muscle activity related to muscle fatigue. Additionally, EMG can be used with other physiological measures, such as HRV or electroencephalography (EEG), to provide a more complete picture of fatigue. These innovations contribute to a better understanding of muscle physiology and offer practical solutions to preventing fatigue-related injuries and optimizing performance in various settings.

The paper \cite{40} developed an automated muscle fatigue detection system by analyzing cyclostationary properties of sEMG signals under dynamic muscle fatiguing contractions, achieving high accuracy using geometric features and machine learning models. Similarly, \cite{90} suggested a better wavelet threshold method to clean up sEMG signals, improving the accuracy of feature extraction and classification using a CNN-SVM algorithm. The study \cite{90} combined time-frequency analysis, ICA, and neural networks to classify muscle fatigue from EMG signals, demonstrating effective handling of non-stationary EMG signals and achieving over 90\% accuracy. In \cite{91}, an LSTM-based model, combined with an improved wavelet packet threshold function, was used to classify muscle fatigue with high accuracy, showcasing the potential of advanced signal processing and deep learning models. The paper \cite{92} utilized EMG and GSR signals to detect driver fatigue, with an SVM classifier achieving high accuracy, indicating the practicality of mobile-based fatigue detection solutions. The study \cite{93} introduced a novel F-GRU model for sEMG-based muscle fatigue detection, achieving over 98\% classification accuracy by leveraging feature extraction and deep learning techniques. The review \cite{94} discussed how sEMG can measure exercise-induced fatigue. It focused on signal processing, feature extraction, and how sEMG can be combined with other physiological signals to get a complete picture of fatigue.

In the study \cite{95}, machine learning and high-resolution time-frequency methods were used to sort sEMG signals into groups based on muscle fatigue. This shows that combining these methods works well. The paper \cite{96} integrated EMG data from wearable sensors with subjective user reports, using machine learning to develop a model for physical fatigue detection, achieving high accuracy with a Gradient-boosting classifier. The study \cite{97} examined how to tell if someone is tired during rehabilitation after a stroke using sEMG signals and advanced machine learning algorithms. HMM and ANN classifiers were used to get very good results. Also, \cite{98} used sEMG signals to figure out how the elbow moved when it was fatigued in different ways. They trained a machine learning model to  guess joint angles even when the fatigue level changed correctly. This shows that sEMG could be used for real-time monitoring and evaluation in sports performance and rehabilitation. These studies collectively emphasize the efficacy of integrating advanced signal processing and machine learning techniques for reliable and accurate muscle fatigue detection using EMG and sEMG signals.

There are also several challenges associated with using EMG for fatigue monitoring.  Movement of the muscle or noise from outside the body can change EMG signals. This lowers the quality of the signals and makes it harder to accurately measure muscle activity, which requires cleaning and filtering the data in a way that takes a lot of time and effort. Muscle fatigue levels vary significantly between individuals, complicating the establishment of normative data across different population groups, and different muscles may fatigue at different rates, complicating the assessment of overall muscle fatigue. Additionally, EMG is typically used to measure activity in specific muscle groups and is not well-suited for monitoring overall body fatigue levels. The electrodes required for EMG can be uncomfortable for subjects, particularly if worn for extended periods. The accuracy of EMG in measuring muscle fatigue is limited, and the correlation between EMG signals and fatigue levels is not always clear. Furthermore, EMG equipment can be bulky and requires a power source, making it difficult to use in portable or mobile settings. To address these challenges, researchers have developed alternative methods, such as wireless EMG systems, which enhance comfort and allow for greater freedom of movement.

It is important to note that in many cases, EMG is used to complement other techniques, such as HRV or electroencephalography (EEG), as it can provide a more complete picture of the subject’s fatigue. For example, the combination of EEG and EMG can be used for fatigue monitoring in drivers, where the EEG measures the cognitive load and the EMG measures muscle fatigue.

EMG signals and AI approaches have been employed by the authors of \cite{40} \cite{90} \cite{91} \cite{92} \cite{93} \cite{94} \cite{95} \cite{96} \cite{97} \cite{98}

\subsection{EEG-based Methods}\label{sec:EEG-basedMethods}
Fatigue detection using electroencephalography (EEG) combined with artificial intelligence (AI) represents a sophisticated method for identifying fatigue in individuals through the analysis of brain activity. This technique involves placing electrodes on the scalp to measure the brain’s electrical activity, producing EEG signals. These signals are then subjected to advanced machine learning algorithms, which process and analyze the data to detect patterns indicative of fatigue. The integration of AI allows for the development of complex models capable of interpreting the nuanced and often subtle variations in EEG signals that correspond to different fatigue levels. This method provides a reliable and efficient approach to real-time fatigue assessment, which is critical for applications in high-stakes environments such as transportation, aviation, and occupational health. This technology uses AI-driven analysis to make fatigue detection more accurate and make it easier to start early intervention plans that lower the risks of impairments caused by fatigue.

Studies highlight significant advancements in EEG-based fatigue detection and drowsiness monitoring. In the paper \cite{99}, an innovative iterative cross-subject negative-unlabeled (NU) learning algorithm is described. It achieves an average accuracy of 93.77\% using a dry wearable EEG headband and shows negative correlations with beta and alpha bands, improving driver safety by preventing accidents related to fatigue. In the same way, \cite{100} uses several different feature extraction methods and deep learning models, such as AlexNet and VGGNet, along with the tunable Q-factor wavelet transform (TQWT) for sub-band decomposition to get a 94.31\% success rate in detecting driver sleepiness. The study \cite{101} explores a highly wearable in-ear EEG sensor, achieving feasible automatic light sleep classification with accuracies between 80.0\% and 82.9\%, beneficial for out-of-clinic drowsiness monitoring in various operators. In \cite{38}, a simulated driving experiment using EEG signals and advanced classifiers like PSO-H-ELM achieved the highest accuracy, underscoring the potential of EEG-based real-time fatigue detection systems to enhance road safety. The paper \cite{102} combines kernel principal component analysis (KPCA) and Hidden Markov Model (HMM) with complexity parameters to estimate mental fatigue, achieving an accuracy of 84\%, highlighting the efficacy of complexity parameters and KPCA-HMM. In \cite{103}, TQWT-based features effectively distinguish between alertness and drowsiness states, with the extreme learning machine (ELM) classifier achieving an accuracy of 91.842\%. The study \cite{104} tackles noise interference in EEG signals due to vehicle vibrations, providing a reliable measure of driving workload by analyzing frequency components, contributing to traffic safety. Finally, \cite{105} provides a comprehensive review of EEG-based MF detection systems, emphasizing the importance of preprocessing, feature extraction, and various classification algorithms, recommending the Kernel Partial Least Squares - Discrete Linear Regression (KPLS-DLR) model for its outstanding performance and minimal intrusiveness, suggesting opportunities for deep learning models and online implementations for broader applications. Collectively, these studies demonstrate the robust and efficient use of EEG in detecting and monitoring fatigue and drowsiness, paving the way for enhanced safety in various fields.

One of the primary advantages of using EEG is its ability to provide a more direct measure of brain activity compared to other methods, such as self-reported fatigue or physiological measures like HRV. Moreover, EEG-based monitoring can offer real-time assessment, enabling the identification of fatigue at its early stages before it escalates into a significant issue.

However, the presence of artifacts in the EEG signal due to muscle activity and other movements presents a significant challenge when using EEG for fatigue detection in mobile subjects. These artifacts can compromise the accuracy of brain activity measurements, potentially leading to false positives or false negatives in fatigue detection.

Another significant issue is the sensitivity of EEG electrodes to movement. If the subject moves excessively, these electrodes can become dislodged, degrading the quality of the EEG signal. Additionally, movement can cause discomfort for the subject, reducing their compliance with the measurement process.

Furthermore, subjects’ need to remain seated or still during EEG measurement can be restrictive and impractical, particularly in applications involving physical activity such as transportation, sports, and industrial work. This limitation underscores the need for advancements in EEG technology that can accommodate movement without compromising data integrity.

Researchers have developed alternative methods to overcome these issues, such as wireless EEG systems, which can be more comfortable for the subject and allow for greater freedom of movement. Additionally, advanced signal processing techniques such as independent component analysis (ICA) can remove artifacts caused by movement from the EEG signal.

Regarding the EEG, the authors in \cite{100} \cite{106} \cite{107} \cite{99} \cite{102} \cite{104} \cite{105} \cite{103} \cite{38} \cite{101} have employed EEG data and AI algorithms to detect weariness in participants in both laboratory and real-world scenarios. Table \ref{tab:FatigueMonitoring1} and Table \ref{tab:FatigueMonitoring2} provide more details on the efficacy of AI approaches for fatigue monitoring using EEG data.

\subsection{PPG-based Methods}\label{sec:PPG-basedMethods}
Photoplethysmography (PPG) is an invaluable physiological signal for monitoring and detecting fatigue, leveraging its technical prowess to provide detailed cardiovascular insights. PPG operates by emitting light onto the skin and measuring the variations in light absorption and reflection, which correspond to changes in blood volume. This method is particularly effective for real-time heart rate monitoring and calculating other cardiovascular parameters such as HRV.

PPG sensors, often embedded in wearable devices, capture these fluctuations in blood volume with high precision, offering a non-invasive means to track physiological changes continuously. Unlike electroencephalography (EEG), which monitors electrical activity in the brain, PPG focuses on peripheral cardiovascular metrics. When used together, PPG and EEG provide complementary data that delivers a holistic view of an individual’s physiological state, enhancing the accuracy of fatigue detection systems. The integration of PPG sensors in wearable technology facilitates continuous, real-time monitoring, making it possible to detect early signs of fatigue, manage stress, and improve overall cardiovascular health. This synergy of PPG with other sensors underscores its critical role in advancing health monitoring technologies, particularly in applications requiring accurate, non-intrusive, and continuous physiological assessments.

One of the primary advantages of utilizing photoplethysmography (PPG) to measure fatigue is its noninvasive nature, ease of use, and integration capability with various devices such as smartwatches and smartphones. PPG facilitates continuous monitoring of cardiovascular activity, making it particularly valuable in contexts where fatigue poses significant safety risks, such as for drivers, pilots, and other professionals whose performance may be compromised by fatigue.

PPG is highly effective in measuring HRV, a crucial indicator of autonomic nervous system activity. HRV provides insights into the balance between the sympathetic and parasympathetic nervous systems, with variations in HRV reflecting changes in fatigue levels. Specifically, low HRV is associated with increased fatigue and stress, making it a reliable metric for assessing fatigue.

Additionally, PPG can measure blood oxygenation levels, offering further information about an individual’s physiological state. Monitoring blood oxygenation helps detect changes due to fatigue, providing a comprehensive view of the individual’s health status. The integration of PPG in wearable technology thus enables real-time, continuous assessment of cardiovascular health and fatigue, enhancing safety and performance in various high-stakes environments.

The following studies provide robust support for the integration and efficacy of photoplethysmography (PPG) sensors in various applications related to fatigue research, highlighting significant advancements and practical implementations in wearable technology. The paper \cite{108} presents a driver condition recognition system that utilizes Support Vector Machine (SVM) classifiers to detect driver drowsiness by analyzing heart rate and RR interval data obtained from a PPG sensor embedded in a wearable device. The system achieved an impressive classification accuracy of 96.3\%, with data preprocessing steps including segmentation and normalization to remove noise and standardize data. Similarly, \cite{109} extends this application by integrating PPG sensors with additional physiological and motion-related data such as galvanic skin response (GSR), temperature (TEM), acceleration (ACCEL), and gyroscope (GYRO) data. The study implemented preprocessing steps to segment data into 10-second intervals and filter out noise, achieving an accuracy of 68.31\% for distinguishing between normal, stressed, fatigued, and drowsy states, which improved to 84.46\% when fatigue and drowsiness were combined into a single class.

The Cardiopulmonary Care Platform (C2P) described in \cite{110} utilizes a smartwatch equipped with PPG sensors to capture HRV and respiratory cycles, addressing the challenge of noise and motion artifacts by incorporating data from the smartwatch’s accelerometer and applying zero-phase forward-reverse filtering with a Butterworth IIR filter. The system achieved a high correlation with Pearson’s correlation coefficient, averaging 0.987 across sessions for breathing cycles and demonstrated accuracies of 84.2\% and 94\% for detecting stair climbing and brisk walking, respectively. In \cite{111}, a PPG-based system for real-time heart rate monitoring during running exercises utilizes an easy pulse plug-in PPG sensor placed on the user’s fingertips, with heart rate data processed by an Arduino microcontroller and transmitted to an Android smartphone via Bluetooth. The system calculates heart rate by detecting intervals between successive peaks in the PPG signal, providing alerts when the heart rate exceeds predefined thresholds. The experimental setup demonstrated a high accuracy of 95.923\% in heart rate detection, ensuring user safety through timely alerts.

Additionally, the innovative Activity-Aware Recurrent Neural Network (AcRoNN) framework in \cite{112} integrates PPG sensors with accelerometers and electrodermal activity (EDA) sensors to enhance cognitive fatigue assessment. The system processes signals using stationary wavelet transform (SWT) and convex optimization (cvxEDA) techniques to remove noise and motion artifacts. The AcRoNN model, designed in two stages, involves feature extraction, activity recognition, and initial cognitive fatigue scoring based on gestural and postural activities, followed by refinement using an LSTM with Consistency Self-Attention (LSTM-CSA) model. This approach achieved up to a 19\% improvement in accuracy over baseline models, demonstrating the effectiveness of integrating PPG data with other sensor signals. Collectively, these studies underscore the potential of PPG sensors to provide accurate, real-time monitoring of fatigue, stress, and other physiological states, significantly contributing to safety and productivity in various domains. However, it is worth noting that PPG has some limitations for fatigue monitoring. PPG signals can be affected by various factors such as motion, skin pigmentation, and hydration level, which can introduce noise and bias into the measurements. Additionally, PPG is not able to provide information about the neural or cognitive aspects of fatigue, which may be important in certain situations.

In lab and real-world situations, authors in \cite{108} \cite{109} \cite{110} \cite{111} \cite{112} have employed PPG data and AI to detect exhaustion in people. Table \ref{tab:FatigueMonitoring1} and Table \ref{tab:FatigueMonitoring2} show the performance of AI techniques for fatigue monitoring using PPG signals.

\subsection{EDA-based Methods}\label{sec:EDA-basedMethods}
Fatigue detection through electrodermal activity (EDA) is an advanced method for identifying signs of fatigue by analyzing the electrical conductance changes in the skin. This approach involves the use of electrodes to measure the skin’s electrical activity, capturing fluctuations caused by the sympathetic nervous system’s activation of sweat glands. These EDA signals are then subjected to sophisticated signal processing techniques and machine learning algorithms to detect and quantify fatigue correctly. By leveraging these advanced analytical methods, EDA-based fatigue detection provides a non-intrusive, real-time assessment of physiological states.  This capability is crucial for early identification and management of fatigue, enhancing safety and performance in high-stakes environments such as driving, aviation, and demanding workplace settings. The continuous monitoring enabled by EDA offers critical insights into the autonomic nervous system’s functioning, making it an invaluable tool for comprehensive fatigue management and overall health monitoring.

One of the primary advantages of using electrodermal activity (EDA) for measuring fatigue is its ability to provide a non-invasive measure of changes in the sympathetic nervous system, which are directly related to fatigue. EDA sensors capture the electrical conductance of the skin, reflecting the activity of sweat glands controlled by the sympathetic nervous system. This physiological measure can be particularly valuable for monitoring fatigue because it directly correlates with the body’s stress and arousal responses. Moreover, EDA can be effectively combined with other physiological measures, such as HRV or electroencephalography (EEG), to provide a comprehensive assessment of fatigue. This multimodal approach enhances the accuracy and reliability of fatigue detection, offering a more complete picture of an individual’s physiological state and improving the potential for timely interventions.

However, the use of EDA in measuring fatigue also presents several disadvantages. One significant challenge is the susceptibility of EDA signals to artifacts such as sweat, motion, or other external noise sources, which can degrade the quality of the signal. This necessitates rigorous data cleaning and filtering processes to remove these artifacts, which can be time-consuming and complex. Furthermore, EDA is influenced by various factors, including emotional state, stress, and other physiological conditions, which can complicate the interpretation of results. These factors can introduce variability and potential confounders into the EDA data, making it difficult to isolate fatigue-specific signals. Despite these challenges, advancements in signal processing and machine learning techniques continue to enhance the robustness and reliability of EDA-based fatigue assessment, making it a valuable tool in the field of physiological monitoring.

The advancements in electrodermal activity (EDA) signal processing and its applications in fatigue assessment are comprehensively reviewed in several studies, underscoring the critical role of EDA in real-time physiological monitoring. The paper titled \cite{113} highlights innovations in EDA devices and signal processing techniques that enhance measurement accuracy, even amidst artifacts and noise, making it possible to derive sophisticated measures from EDA signals. These advancements enable the assessment of cognitive and physical states by quantifying sudomotor activity linked to the sympathetic nervous system’s response to fatigue, thus facilitating real-time fatigue monitoring through changes in skin conductance responses (SCRs) and skin conductance levels (SCL). Complementing this, the study \cite{114} demonstrates the high accuracy (approximately 89\%) of a wearable EDA device in classifying emotional states such as calm and distress, highlighting its utility in non-intrusive monitoring of arousal, stress, and fatigue. Addressing the challenge of motion artifacts, \cite{115} develops an advanced framework for detecting and mitigating these artifacts in EDA signals, achieving a remarkable mean accuracy of 94.82\% with the GradBoost classifier, thereby enhancing the reliability of EDA-based fatigue measurements in real-world settings.  Additionally, \cite{116} leverages EDA alongside other physiological sensors in a multimodal approach to induce and detect cognitive and physical fatigue, with the integration of EDA data significantly contributing to high classification accuracies using Random Forest and LSTM models. Lastly, \cite{117} integrates EDA with functional near-infrared spectroscopy (fNIRS), ECG, respiratory inductance plethysmography (RIP), and accelerometers to refine cognitive fatigue detection, achieving an average classification accuracy of 70.91\%, with some models exceeding 80\%. Collectively, these studies underscore the pivotal role of EDA in delivering a comprehensive, non-intrusive approach for real-time monitoring and assessment of fatigue, demonstrating its efficacy and potential for practical applications in various high-stakes environments.

In terms of practicality, fatigue detection using EDA has potential applications in transportation, industrial, and sports settings, as well as  healthcare. However, it is important to note that further research is needed to improve the accuracy and reliability of the detection method before it can be widely adopted in practical settings. This includes developing robust and accurate algorithms, studying the physiological and psychological changes that occur with fatigue, and establishing normative data for different population groups.

It is important to note that in many cases, EDA is used to complement other techniques, such as HRV or electroencephalography (EEG), as it can provide a more complete picture of the subject’s fatigue.  For example, the combination of EEG and EDA can be used for fatigue monitoring in drivers, where the EEG measures the cognitive load and the EDA measures the physiological changes caused by fatigue.

Regarding the EDA-Based methods, the authors in \cite{113} \cite{114} \cite{115} \cite{116} \cite{117} have employed EDA to detect weariness in participants in both laboratory and real-world scenarios. Table 1 provides more details on the efficacy of AI approaches for fatigue monitoring using EDA sensors.

\subsection{IMU-based Methods}\label{sec:IMU-based Methods}
Fatigue detection using inertial measurement units (IMUs) is a sophisticated method for identifying signs of fatigue in individuals by analyzing biomechanical changes in body movements and postures. This method typically involves measuring angular and linear acceleration, angular velocity, and magnetic field strength via devices embedded with one or more IMUs, such as smartwatches or accelerometers attached to the body. The high-resolution data captured from these IMUs are then subjected to advanced machine learning algorithms or other analytical techniques to identify and quantify signs of fatigue accurately. Recent research highlights the critical role of IMUs in enabling real-time, continuous monitoring of physical fatigue by capturing detailed motion data and identifying subtle variations indicative of fatigue onset. These advancements have facilitated the development of robust predictive models that integrate IMU data with sophisticated analytical frameworks, achieving high accuracy in fatigue detection. This methodological approach offers significant practical applications across various high-stakes environments, including sports, industrial safety, and occupational health, enhancing managing fatigue proactively and improving overall safety, performance, and productivity.

Recent research highlights the significant progress in employing Inertial Measurement Unit (IMU) sensors for fatigue assessment, emphasizing their pivotal role in real-time physiological monitoring. The academic paper \cite{118} systematically reviews the application of IMUs in sports performance, emphasizing their role in capturing and analyzing detailed motion data, such as acceleration, velocity, and jerk, to detect subtle changes in movement patterns indicative of fatigue. Complementing this, the study \cite{119} demonstrates a robust data-driven methodology to develop a model capable of detecting physical fatigue by analyzing IMU data from the lower back, identifying key features like gait maximum acceleration and mean back rotational position, with a classification accuracy of 88\%. In an industrial context, the paper \cite{120} explores the use of a Smart Safety Helmet (SSH) equipped with IMU and EEG sensors to detect fatigue-related head motions and assess accident risks, providing precise measurements of acceleration and orientation to identify fatigue-induced movements. Additionally, \cite{121} introduces an innovative system that uses forearm IMU sensors to continuously monitor construction workers’ fatigue by predicting the aerobic fatigue threshold (AFT) through the analysis of muscle activity and motion data, achieving superior accuracy in fatigue evaluation compared to traditional metrics. Finally, the research \cite{122} presents an AI model that integrates IMU-generated data to predict fatigue and stamina in athletes, effectively capturing triaxial acceleration, angular velocity, and magnetic orientation to train machine learning models like Random Forest, Gradient Boosting Machines, and LSTM networks, showing high predictive accuracy and enabling timely interventions. Collectively, these studies highlight the practicality, reliability, and effectiveness of IMU technology for continuous fatigue monitoring across various fields, enhancing safety, performance, and productivity through precise and real-time fatigue assessment.

One of the main advantages of using inertial measurement units (IMUs) for measuring fatigue is that they provide a non-invasive means of capturing changes in body movements and postures related to fatigue. IMUs can continuously monitor and record angular and linear acceleration, angular velocity, and magnetic fields, offering high-resolution data that reflect the physical state of an individual. Furthermore, IMUs can be integrated with other physiological measures, such as HRV or electroencephalography (EEG), to provide a comprehensive assessment of fatigue. This multimodal approach enhances the robustness and reliability of fatigue detection by combining biomechanical data with physiological indicators.

However, this approach also has several disadvantages. One significant challenge is the susceptibility of IMU data to artifacts, such as muscle movements or external noise sources, which can degrade the signal quality and complicate the accurate measurement of body movements and postures. Addressing these artifacts requires extensive data cleaning and filtering, which can be time-consuming and complex. Additionally, the accuracy of fatigue detection using IMUs may be influenced by factors such as the placement of the sensors, the specific machine learning algorithms employed, and individual variations in biomechanics. These variables can introduce variability and potential inaccuracies in the fatigue assessment, necessitating careful calibration and validation to ensure reliable results.

Researchers in \cite{118} \cite{119} \cite{120} \cite{121} [\cite{122} have applied IMU signals and AI techniques to detect fatigue in humans in both laboratory and real-world scenarios.

\subsection{EOG-based Methods}\label{sec:EOG-based Methods}
EOG stands for Electrooculography, a technique used to measure the electrical potential difference between the front and back of the human eye. It is commonly used in research and clinical settings for monitoring eye movements and detecting changes in the electrical activity of the eye, including fatigue.

The technique works by placing electrodes around the eyes and measuring the electrical activity generated by the eye muscles as they move. Changes in this electrical activity can indicate changes in the state of the eye, including fatigue. EOG can be used to monitor fatigue in  various settings, including in drivers, pilots, and other individuals whose performance may be affected by fatigue.

Recent research papers have showcased significant progress in fatigue detection through the use of Electrooculography (EOG) sensors. The paper \cite{123} introduces a novel wearable system that combines EOG sensors and a three-axis gyroscope to monitor fatigue and provide timely wake-up interventions through physical stimuli like music. The system utilizes a single-lead electrode for EOG signal collection and gyroscope data to analyze blink and head posture characteristics using a Backpropagation (BP) neural network, effectively identifying fatigue-related features and ensuring real-time monitoring suitable for high-vigilance environments.

In \cite{124}, the study employs EOG sensors to assess driver hypo-vigilance, including fatigue, drowsiness, visual inattention, and cognitive inattention. The research uses a rigorous experimental protocol to collect EOG data from ten participants during simulated driving sessions. Seventeen significant features are extracted and analyzed using advanced machine learning algorithms such as Support Vector Machine (SVM), k-nearest neighbor (KNN), and ensemble classifiers, achieving a maximum accuracy of 98.7\% for binary classification and 90.9\% for five-class detection. Principal component analysis (PCA) is used for feature reduction, enhancing classification performance.

Similarly, \cite{125} provides a comprehensive review of EOG-based driver fatigue detection, highlighting the integration of EOG with other physiological signals to improve detection accuracy.  The review discussed the development of a multi-feature fusion technique and advanced pre-processing methods like wavelet transforms and singular spectrum analysis to address challenges such as baseline drift and signal noise. Machine learning classifiers like SVM, Artificial Neural Networks (ANN), and Extreme Learning Machines (ELM) are emphasized for their high classification accuracy in detecting fatigue states.

The paper \cite{126} focuses on office environments, developing an algorithm to detect fatigue through EOG signal analysis during a 60-minute N-back task with 24 participants. The study captures and analyzes blink duration, amplitude, and the time between blinks, achieving an average classification accuracy of 93\% in user-dependent mode and 89\% in user-independent mode. Sophisticated preprocessing techniques ensure the accuracy and reliability of the detected signals, providing a practical solution for real-time fatigue management in repetitive task environments.

In \cite{127}, EOG glasses are used to continuously monitor fatigue in real-world settings over a two-week period with
16 participants. The study collects over 2,860 hours of EOG recordings and 1,047 ground truth assessments using Psychomotor Vigilance Tasks (PVT). The results demonstrate a statistically significant positive correlation between increased blink frequency and longer reaction times, validating  using EOG glasses as a practical and unobtrusive tool for real-time fatigue monitoring. The development of a robust predictive model based on continuous EOG data allows for accurate detection of fatigue-related changes in alertness, enhancing user safety and performance.

Finally, \cite{128} proposes a multimodal signal fatigue detection method employing deep learning techniques to  identify driver fatigue effectively states using both EOG and EEG sensors. Convolutional autoencoders (CAE) are used to fuse EEG and EOG signal features, while convolutional neural networks (CNN) maintain spatial locality. The fused features are input into a recurrent neural network (RNN) for fatigue recognition, achieving superior performance with an RMSE/COR of 0.08/0.96 compared to single modality features. This multimodal approach significantly enhances the accuracy and robustness of fatigue detection systems, demonstrating practical applications in improving road safety through timely and accurate fatigue monitoring.

Collectively, these studies underscore the versatility, reliability, and practical applications of EOG-based fatigue monitoring systems across various environments, offering comprehensive solutions for enhancing safety and performance. One of the main advantages of EOG is that it is non-invasive and relatively easy to use. It can be used to monitor fatigue in real-time, which can be useful in situations where fatigue is a potential safety concern. Additionally, EOG can be used with other techniques, such as EEG, to provide a more complete picture of an individual’s level of fatigue.

However, there are also some limitations to using EOG for fatigue monitoring. The technique is sensitive to head movements, making it difficult to use in specific settings. Additionally, EOG may not be able to detect all types of fatigue, particularly those not caused by changes in the state of the eye. Regarding practicality, EOG is still mainly used in clinical or laboratory settings and is not yet a widely available technology in practical applications, such as cars or airplanes. The cost and complexity of the equipment and lack of standardization in the techniques used to analyze the data obtained by EOG could be some of the reasons behind this. However, as technology is advancing, it is possible that EOG will become more widely
used in the future.

Authors in \cite{123} \cite{124} \cite{125} \cite{126} \cite{127} \cite{128} utilised EOG signals and AI to detect fatigue. Tables 1 and 2 provide information on AI fatigue monitoring approaches utilising EOG signals.

\subsection{Hybrid Models}\label{sec:HybridModels}
Hybrid methods for fatigue monitoring combine multiple physiological signals and camera images to provide a more comprehensive assessment of an individual’s level of fatigue. These methods aim to overcome the limitations of using a single physiological signal or camera image by combining the strengths of multiple techniques.

One example of a hybrid method for fatigue monitoring is the combination of electroencephalography (EEG) and electrooculography (EOG) to measure brain activity and eye movements, respectively. EEG can provide information about brain activity and fatigue, while EOG can provide information about eye movements and changes in the state of the eye. By combining these two techniques, researchers can get a more complete picture of an individual’s level of fatigue.

Another example is using physiological signals such as electrocardiography (ECG) and photoplethysmography (PPG)  combined with camera images to monitor fatigue. ECG and PPG can provide information about heart rate and blood flow, respectively, while camera images can provide information about facial expressions, head movements and blink rate. The combination of these signals can provide a more accurate assessment of an individual’s level of fatigue, and other factors that may be related to fatigue, such as stress.

Additionally, using a combination of physiological signals and camera images can allow for the monitoring of fatigue in real-time and in naturalistic settings without the need for the individual to wear a lot of equipment. However, it is worth noting that the combination of multiple signals and images also increases the complexity of the system and the need for sophisticated algorithms for data analysis. Additionally, it may also increase the cost of the system and the need for trained personnel to operate and interpret the results.

The recent body of research demonstrates remarkable progress in the field of fatigue detection through the integration of multimodal sensors and sophisticated machine learning algorithms. The study \cite{28} leverages Electroencephalography (EEG) and Electrooculography (EOG) signals, employing a fast Support Vector Machine (FSVM) algorithm to process features such as power spectral density (PSD) and differential entropy from EEG and blink characteristics and independent component analysis (ICA) derived features from EOG, achieving superior recognition accuracy in real-time driver fatigue detection. Similarly, \cite{129} develops a real-time fatigue monitoring system using lightweight Convolutional Neural Network (CNN) architectures on EEG and EOG signals, extracting features such as PSD and saccades and achieving 94\% classification accuracy on the Jetson TX2 platform while utilizing IoT technology for enhanced road safety.

The study \cite{130} integrates multiple sensors, including Electrocardiography (ECG), Electromyography (EMG), Electrodermal Activity (EDA), and EEG, to monitor cognitive and physical fatigue. The system employs machine learning models such as Random Forest (RF) and Long Short-Term Memory (LSTM) networks, achieving 80.5\% accuracy for physical fatigue and 84.1\% for cognitive fatigue, demonstrating the efficacy of multimodal sensor fusion. In \cite{72}, driver sleepiness detection is enhanced using Continuous Wavelet Transform (CWT) to extract time-frequency features from EEG and EOG signals and augmented with Generative Adversarial Networks (GAN) to overcome data limitations, resulting in a mean accuracy of 98\% with LSTM networks.

The study \cite{131} focuses on miner fatigue detection by integrating ECG and EMG signals, employing Principal Component Analysis (PCA) and Grey Relational Analysis (GRA) for feature optimization, and utilizing machine learning models such as XG-Boost, SVM, and Random Forest (RF), with XG-Boost achieving the highest accuracy of 89.47\%. Similarly, \cite{132} uses a multi-feature information fusion approach, combining ECG, EMG, pulse, blood pressure, reaction time (RT), and vital capacity (VC) signals, and employs SVM and RF classifiers, achieving 90\% accuracy with RF for fatigue detection in extreme conditions.

The study \cite{36} explores fatigue detection in Pilates rehabilitation by fusing ECG and surface EMG (sEMG) signals, using an Improved Particle Swarm Optimization-SVM (IPSO-SVM) classifier to achieve an average recognition rate of 93.58\% across relaxed, transition, and tired states. This approach demonstrates the complementary nature of ECG and sEMG in monitoring muscle fatigue accurately. Lastly, \cite{71} introduces an IoT-based multimodal fatigue monitoring system combining EEG, EOG, and facial data, using deep learning models such as Convolutional Neural Networks (CNNs) and Long Short-Term Memory (LSTM) networks to classify eye and mouth states with accuracies of 99.01\% and 99.5\%, respectively, significantly enhancing real-time driver fatigue detection and road safety.

These studies show that multimodal sensor systems and advanced machine learning techniques can provide complete, real-time solutions for monitoring fatigue in many areas, such as healthcare, transportation, workplace safety, and rehabilitation. Therefore, the hybrid methods for fatigue monitoring that combine physiological signals with camera images can provide a more comprehensive assessment of an individual’s level of fatigue. However, when implementing these methods, it is important to consider practical issues such as cost and complexity.

Authors in \cite{28} \cite{36} \cite{71} \cite{72} \cite{129} \cite{131} \cite{132} \cite{130} have used a hybrid model for identifying fatigue. Further information about the performance of AI methods for fatigue monitoring using hybrid models can be found in Table \ref{tab:FatigueMonitoring1} and Table \ref{tab:FatigueMonitoring2}.

\subsection{Discussion}\label{sec:Discussion}
\begin{figure}[h]
    \centering
    \includegraphics[scale=0.5]{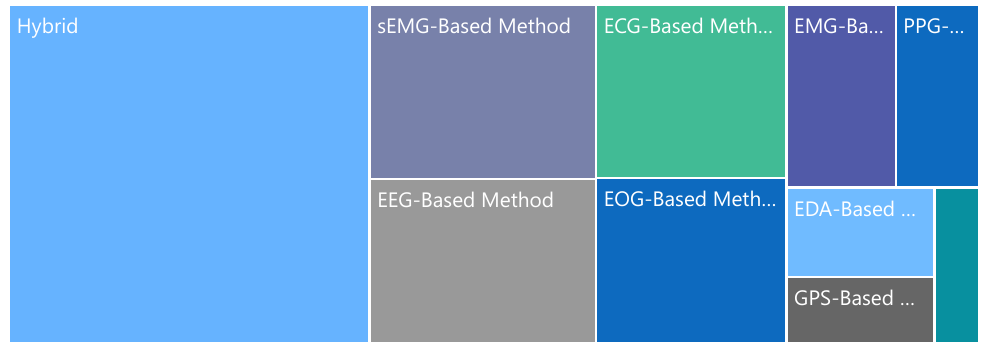 }
    \caption{The popularity of different physiological signals used for fatigue monitoring.}
    \label{fig:Figure12}
\end{figure}

Figure \ref{fig:Figure12} illustrates the popularity of various physiological signals and data sources used for fatigue monitoring in different studies. EEG and ECG are the most commonly used signals, with a significant proportion of studies utilizing them. On the other hand, EDA/GSR is the least used source for fatigue monitoring among the options presented. This highlights the trend towards using EEG and ECG in fatigue monitoring research while using EDA/GSR remains limited.

\begin{figure}[h]
    \centering
    \includegraphics[scale=0.6]{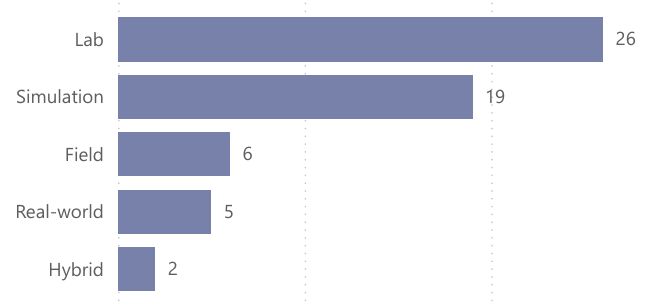 }
    \caption{The count of environment studies conducted for fatigue monitoring using wearables and physiological signals.}
    \label{fig:Figure13}
\end{figure}

Figure \ref{fig:Figure13} shows the count of field and lab tests for fatigue monitoring using physiological signals. Most studies utilize lab tests, rather than field tests, for monitoring fatigue. This is likely due to the convenience and practicality of conducting lab tests and the ability to collect data in a controlled environment. However, field tests are still used in a few studies and may provide more practical solutions for measuring fatigue. This representation demonstrates that field and lab tests are used to monitor fatigue using physiological signals, but lab tests are more commonly used.

\begin{figure}[h]
    \centering
    \includegraphics[scale=0.6]{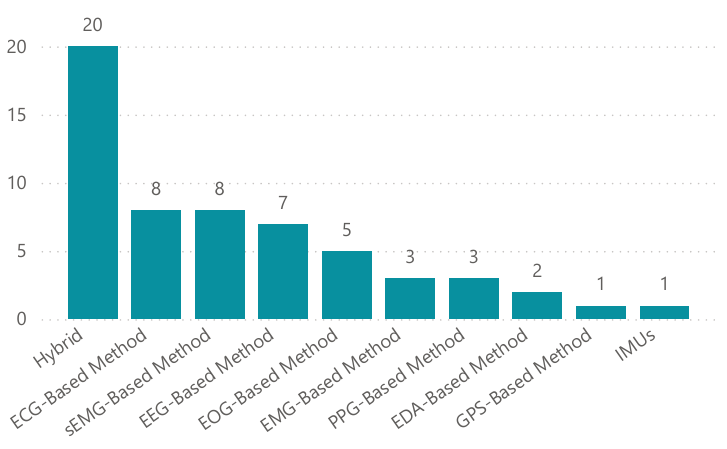 }
    \caption{Popular AI methods for fatigue monitoring using wearables and physiological signals.}
    \label{fig:Figure14}
\end{figure}

According to Figure \ref{fig:Figure14}, the hybrid approach is more prevalent among other AI methods for measuring fatigue using wearables. A hybrid model, which combines multiple approaches, may provide a more robust and comprehensive measurement of fatigue. ECG-based methods are in the second rank. ECG is a suitable method for monitoring fatigue because it can provide information about the heart rate and rhythm, which fatigue can affect. One important measure that can be obtained from ECG is HRV, which is the variation in time between heartbeats. HRV can indicate the body’s ability to adapt to stressors; lower HRV has been associated with increased fatigue.

\begin{figure}[h]
    \centering
    \includegraphics[scale=0.6]{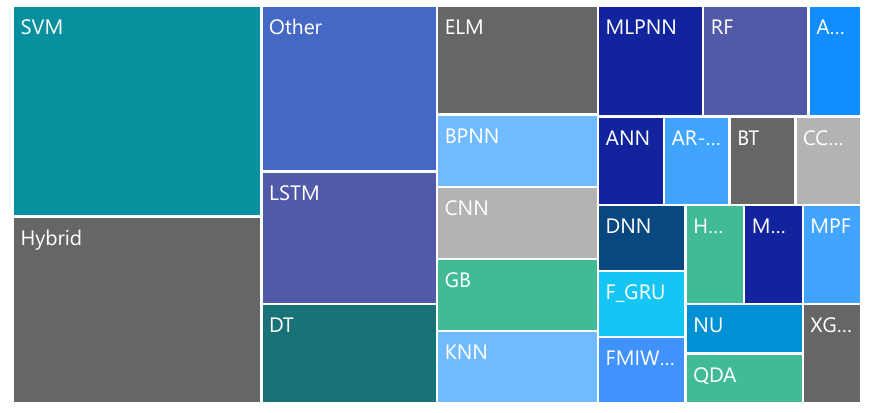 }
    \caption{Popular AI models for fatigue monitoring.}
    \label{fig:Figure15}
\end{figure}

Figure \ref{fig:Figure15} shows that Support Vector Machine (SVM) has emerged as the most popular model for modeling and identifying fatigue. This is due to the excellent capabilities of SVM for handling large amounts of data, versatility (being able to process both linear and nonlinear samples), and excellent generalization power (robust to noise). Other popular methods for fatigue monitoring are kNN, RF, and DL models.

Ultimately, the analysis highlights significant trends and methodologies in fatigue monitoring research. EEG and ECG signals emerge as the most widely utilized modalities, offering crucial insights into physiological parameters like heart rate variability and brain activity. In contrast, methods leveraging EDA/GSR remain less explored, suggesting opportunities for further investigation in this area.

Lab-based studies dominate the field, highlighting the preference for controlled environments that enable precise data collection. Nonetheless, the inclusion of simulation, field, and hybrid approaches reflects a growing interest in extending research to more realistic and diverse settings, striking a balance between control and ecological validity.

The increasing prevalence of hybrid models, which integrate multiple modalities and machine learning methods, demonstrates the push toward comprehensive and robust fatigue assessment systems. Support Vector Machines (SVM) are the leading AI method due to their adaptability and robustness, while deep learning approaches such as CNNs and LSTMs, along with hybrid frameworks, are gaining traction for their effectiveness in processing complex, multimodal datasets.

These findings underscore the evolution of fatigue monitoring research, emphasizing the role of advanced machine learning techniques, diverse experimental frameworks, and multimodal physiological data in developing reliable and effective solutions.

\section{Research Challenges}\label{sec:ResearchChallenges}
\subsection{Access to Real- Time Data}\label{sec:AccessRealTimeData}
The data transmission needed for fatigue assessment is a multi-step procedure that might take a lengthy time. The first step is to sync the wearable gadget and upload the data. Suppose this device is on a mobile network. In that case, the data must go via that network provider’s infrastructure and out onto the internet before ending up in the service provider’s network, potentially through the cloud. The data might arrive at its destination later than expected if access to the Cloud \cite{143} is disrupted or if an internet outage happens. Mobile networks
\cite{144} are not guaranteed to be constantly accessible. If mobile users can’t connect to the internet for routine tasks, that’s a problem; but when someone’s life is on the line, as it may be in a dangerous situation, information must be sent quickly and accurately \cite{145} \cite{129}.

\onecolumn
\begin{landscape}
\small 
\begin{longtable}{@{}ccccccc@{}}
\caption{Summary of fatigue monitoring techniques using ECG, EDA, EEG, EMG, EOG, GPS, PPG, IMUs, sEMG, and hybrid methods.}
\label{tab:FatigueMonitoring1}\\
\hline
\textbf{Reference} & \textbf{Modality} & \textbf{Machine Learning Methods} & \textbf{Sample Size} & \textbf{Feature Count} & \textbf{Experiment} & \textbf{Performance Result} \\ \hline
\endfirsthead

\hline
\textbf{Reference} & \textbf{Modality} & \textbf{Machine Learning Methods} & \textbf{Sample Size} & \textbf{Feature Count} & \textbf{Experiment} & \textbf{Performance Result} \\ \hline
\endhead

\hline
\hline
\endfoot

\hline
\endlastfoot

\cite{35}  & ECG-Based Method   & SVM         & 12  & 4    & Simulation & Accuracy = 83.9\% \\ \hline
\cite{37}  & sEMG-Based Method  & CNN         & 64  & 2    & Lab        & \begin{tabular}[c]{@{}c@{}}Accuracy = 91.38\% (60°/s)\\ Accuracy = 89.87\% (180°/s)\end{tabular} \\ \hline
\cite{40}  & sEMG-Based Method  & ELM         & 58  & 7    & Lab        & \begin{tabular}[c]{@{}c@{}}Accuracy = 94.09\%\\ F-Score = 93.75\%\\ Recall = 93.33\%\end{tabular} \\ \hline
\cite{42}  & EMG-Based Method   & GB          & 10  & 26   & Lab        & \begin{tabular}[c]{@{}c@{}}Precision = 69.40\%\\ F-Score = 76.60\%\\ Recall = 74.10\%\end{tabular} \\ \hline
\cite{43}  & ECG-Based Method   & Hybrid      & 10  & 20   & Simulation & Accuracy = 100\% \\ \hline
\cite{76}  & ECG-Based Method   & BPNN        & 9   & 8    & Simulation & N/A \\ \hline
\cite{83}  & ECG-Based Method   & AR-GARCH    & 10  & 1    & Field      & N/A \\ \hline
\cite{85}  & ECG-Based Method   & KNN         & 25  & 32   & Lab        & \begin{tabular}[c]{@{}c@{}}Accuracy = 88.3\% (For males)\\ Accuracy = 85.7\% (For females)\end{tabular} \\ \hline
\cite{87}  & ECG-Based Method   & ANN         & 5   & 56   & Lab        & Accuracy = 89.3\% \\ \hline
\cite{89}  & Hybrid             & MERF        & 21  & 58   & Real-world & N/A \\ \hline
\cite{131} & Hybrid             & XG-Boost    & 55  & 23   & Field      & \begin{tabular}[c]{@{}c@{}}Accuracy = 89.47\%\\ Precision = 80\%\\ F-Score = 88.89\%\\ Recall = 100\%\\ AUD = 0.9\end{tabular} \\ \hline
\cite{119} & Hybrid             & RF          & 24  & 43   & Lab        & \begin{tabular}[c]{@{}c@{}}Accuracy = 96\%\\ Precision = 93\%\\ F-Score = 93\%\\ Recall = 92\%\end{tabular} \\ \hline
\cite{135} & EMG-Based Method   & MLPNN       & 14  & N/A  & Lab        & \begin{tabular}[c]{@{}c@{}}Accuracy = 91\%\\ Recall = 93\%\end{tabular} \\ \hline
\cite{136} & Hybrid             & SVM         & 6   & 10   & Lab        & \begin{tabular}[c]{@{}c@{}}Accuracy = 92\%\\ Precision = 90\%\end{tabular} \\ \hline
\cite{137} & sEMG-Based Method  & F\_GRU      & 8   & 9    & Simulation & Accuracy = 98\% \\ \hline
\cite{38}  & EEG-Based Method   & ELM         & 6   & 96   & Simulation & Accuracy = 91.78\% \\ \hline
\cite{45}  & PPG-Based Method   & SVM         & 10  & 2    & Lab        & \begin{tabular}[c]{@{}c@{}}Accuracy = 96.3\%\\ Precision = 97.80\%\\ Recall = 94.70\%\end{tabular} \\ \hline
\cite{46}  & EOG-Based Method   & CCNF        & 30  & 36   & Hybrid     & N/A \\ \hline
\cite{72}  & Hybrid             & LSTM        & 12  & N/A  & Simulation & \begin{tabular}[c]{@{}c@{}}Accuracy = 98\%\\ F-Score = 95\%\end{tabular} \\ \hline
\cite{109} & Hybrid             & SVM         & 28  & 38   & Simulation & Accuracy = 98.43\% \\ \hline
\cite{99}  & EEG-Based Method   & NU          & 29  & 96   & Simulation & Accuracy = 93.77\% \\ \hline
\cite{100} & EEG-Based Method   & LSTM        & 16  & 4100 & Simulation & Accuracy = 94.31\% \\ \hline
\cite{101} & EEG-Based Method   & SVM         & 23  & 60   & Lab        & \begin{tabular}[c]{@{}c@{}}Accuracy = 82.9\%\\ Precision = 84.40\%\\ Recall = 66.40\%\end{tabular} \\ \hline
\cite{103} & EEG-Based Method   & ELM         & N/A & 5    & Lab        & \begin{tabular}[c]{@{}c@{}}Accuracy = 91.84\%\\ F-Score = 93.3\%\\ Recall = 96.50\%\end{tabular} \\ \hline
\cite{140} & Hybrid             & CNN         & 6   & N/A  & Simulation & \begin{tabular}[c]{@{}c@{}}Accuracy = 70\% for ECG\\ Accuracy = 64\% for PPG\end{tabular} \\ \hline
\cite{141} & Hybrid             & FMIWPT      & 31  & N/A  & Simulation & Accuracy = 95\% - 97\% \\ \hline
\cite{82}  & Hybrid             & Hybrid      & \begin{tabular}[c]{@{}c@{}}35 (DROZY)\\ 92 (Self-Collected Dataset)\end{tabular} & N/A & Lab & \begin{tabular}[c]{@{}c@{}}Accuracy = 99.6\% (DROZY)\\ Accuracy = 91.8\% (Self-Collected Dataset)\end{tabular} \\ \hline
\cite{81}  & ECG-Based Method   & KNN         & 35  & 6    & Lab        & \begin{tabular}[c]{@{}c@{}}Accuracy = 75.5\%\\ AUC = 0.74\end{tabular} \\ \hline
\cite{84}  & ECG-Based Method   & RF          & 60  & 9    & Lab        & \begin{tabular}[c]{@{}c@{}}Accuracy = 94.5\%\\ F-Score = 95\%\end{tabular} \\ \hline
\cite{170} & EMG-Based Method   & MPF         & 10  & N/A  & Simulation & N/A \\ \hline
\cite{102} & EEG-Based Method   & Hybrid      & 50  & 10   & Lab        & \begin{tabular}[c]{@{}c@{}}Accuracy = 84\%\\ Precision = 83\%\\ Recall = 84\%\end{tabular} \\ \hline
\cite{107} & Hybrid             & BT          & 12  & 21   & Simulation & Accuracy = 82.6\% \\ \hline
\cite{91}  & sEMG-Based Method  & LSTM        & 20  & 4    & Lab        & \begin{tabular}[c]{@{}c@{}}Accuracy = 95.18\%\\ Precision = 96.65\%\\ Recall = 94.08\%\end{tabular} \\ \hline
\cite{98}  & sEMG-Based Method  & MLPNN       & 30  & 2    & Lab        & Accuracy = 60.12\% \\ \hline
\cite{97}  & sEMG-Based Method  & HMM         & 6   & 8    & Lab        & Accuracy = 95.3\% \\ \hline
\cite{127} & EOG-Based Method   & Other       & 16  & 1    & Real-world & N/A \\ \hline
\cite{128} & Hybrid             & Hybrid      & 23  & N/A  & Simulation & N/A \\ \hline
\cite{126} & EOG-Based Method   & QDA         & 24  & 3    & Real-world & \begin{tabular}[c]{@{}c@{}}Accuracy = 93\% (user-dependent)\\ Accuracy = 89\% (user-independent)\\ Precision = 86\%\\ Recall = 92\%\\ AUC = 0.94\end{tabular} \\ \hline
\cite{12}  & Hybrid             & AcRoNN      & 5   & 53   & Real-world & N/A \\ \hline
\cite{111} & PPG-Based Method   & Other       & 7   & 1    & Lab        & Accuracy = 95.92\% \\ \hline
\cite{110} & PPG-Based Method   & Other       & 19  & 3    & Hybrid     & N/A \\ \hline
\cite{130} & Hybrid             & Hybrid      & 32  & 269  & Lab        & \begin{tabular}[c]{@{}c@{}}Accuracy = RF: 80.5\%\\ Accuracy = LSTM: 84.1\%\\ Recall = 90\% (LSTM)\\ Recall = 88\% (RF)\end{tabular} \\ \hline
\cite{115} & EDA-Based Method   & GB          & 20  & 40   & Lab        & \begin{tabular}[c]{@{}c@{}}Accuracy = 94.7\%\\ F-Score = 91.61\%\\ Recall = 92.90\%\end{tabular} \\ \hline
\cite{114} & EDA-Based Method   & DT          & 50  & 36   & Lab        & \begin{tabular}[c]{@{}c@{}}Accuracy = 89.18\%\\ Recall = 93.9\% (Distress)\\ Recall = 85.36\% (Calm)\end{tabular} \\ \hline
\cite{142} & Hybrid             & DT          & 19  & 3    & Simulation & \begin{tabular}[c]{@{}c@{}}Accuracy = 89.5\%\\ Recall = 100\%\\ AUC = 0.833 (Nose temperature)\\ AUC = 0.975 (Wrist temperature)\end{tabular} \\ \hline
\cite{74}  & Hybrid             & SVM         & 28  & 190  & Simulation & \begin{tabular}[c]{@{}c@{}}Accuracy = 68.31\% (Four states)\\ Accuracy = 84.46\% (Three states)\end{tabular} \\ \hline
\cite{171} & GPS-Based Method   & Other       & 30  & 6    & Field      & N/A \\ \hline
\cite{132} & Hybrid             & Hybrid      & 45  & 90   & Field      & \begin{tabular}[c]{@{}c@{}}Accuracy = 85-90\%\\ Precision = 90\%\\ F-Score = 85.71\%\\ Recall = 88.89\%\\ AUC = 0.854\end{tabular} \\ \hline
\cite{71}  & Hybrid             & DNN         & N/A & N/A  & Simulation & \begin{tabular}[c]{@{}c@{}}Accuracy = 99.5\% (Mouth)\\ Accuracy = 99.01\% (Eye)\\ Precision = 98.51\% (Eye)\\ Recall = 99.81\% (Eye)\\ AUC = 0.99\end{tabular} \\ \hline
\cite{119} & Hybrid             & RF          & 24  & 43   & Lab        & \begin{tabular}[c]{@{}c@{}}Accuracy = 96\%\\ Precision = 93\%\\ F-Score = 93\%\\ Recall = 92\%\end{tabular} \\ \hline
\cite{122} & IMUs               & LSTM        & 19  & N/A  & Field      & \begin{tabular}[c]{@{}c@{}}Accuracy = 96\%\\ Precision = 93\%\\ F-Score = 92\%\\ Recall = 93\%\end{tabular} \\ \hline
\cite{121} & Hybrid             & DT          & 10  & 12   & Field      & \begin{tabular}[c]{@{}c@{}}Accuracy = 92.31\%\\ F-Score = 92.40\%\\ Recall = 100\%\end{tabular} \\ \hline
\cite{104} & EEG-Based Method   & Other       & 30  & 2    & Real-world & N/A \\ \hline
\cite{90}  & sEMG-Based Method  & SVM         & 20  & 5    & Lab        & Accuracy = 86.69\% \\ \hline
\cite{95}  & sEMG-Based Method  & Hybrid      & 52  & 4    & Lab        & Accuracy = 91.39\% \\ \hline
\cite{124} & EOG-Based Method   & Hybrid      & 10  & 16   & Simulation & \begin{tabular}[c]{@{}c@{}}Accuracy = 98.7\% (Two-class detection)\\ Accuracy = 90.9\% (Five-class detection)\\ Precision = 81.60\%\\ F-Score = 80.60\%\\ Recall = 79.80\%\end{tabular} \\ \hline
\cite{123} & EOG-Based Method   & BPNN        & 24  & 8    & Lab        & N/A \\ \hline
\cite{28} & Hybrid             & SVM         & \begin{tabular}[c]{@{}c@{}}SEED-VIG dataset\\ Simulated driving data\end{tabular} & N/A & Simulation & Accuracy = 95.17\% \\ \hline
\cite{36}  & Hybrid             & SVM         & 20  & 7    & Lab        & \begin{tabular}[c]{@{}c@{}}Accuracy = 93.58\%\\ Precision = 94.25\% (Relaxed State)\\ Precision = 92.25\% (Transition State)\\ Precision = 94.25\% (Tired State)\\ F-Score = 94.25\% (Relaxed State)\\ F-Score = 92.25\% (Transition State)\\ F-Score = 94.25\% (Tired State)\\ Recall = 94.25\% (Relaxed State)\\ Recall = 92.25\% (Transition State)\\ Recall = 94.25\% (Tired State)\end{tabular} \\ \hline
\end{longtable}
\end{landscape}


\begin{landscape}
\small 
\begin{longtable}[c]{@{}cccccccc@{}}
\caption{Summary of Proposed Method and Overall Performance of Various Modalities for Fatigue Monitoring.}
\label{tab:FatigueMonitoring2}\\
\hline
\textbf{Reference} & \textbf{Paper Topic} & \textbf{Proposed Method} & \textbf{Overall Performance} \\ \hline
\endfirsthead

\hline
\textbf{Reference} & \textbf{Paper Topic} & \textbf{Proposed Method} & \textbf{Overall Performance} \\ \hline
\endhead

\hline
\hline
\endfoot

\hline
\endlastfoot

\cite{35} & \begin{tabular}[c]{@{}c@{}}Study on Driving Fatigue Evaluation \\ System Based on Short Time Period \\ ECG Signal\end{tabular} & \begin{tabular}[c]{@{}c@{}}Support Vector Machine (SVM) model \\ with RBF kernel, using\\ HRV time-frequency domain features\end{tabular} & \begin{tabular}[c]{@{}c@{}}Effective fatigue detection in real-time with 5-second \\ ECG intervals, achieving high accuracy compared to\\ long-period   ECG methods\end{tabular} \\\hline
\cite{37} & \begin{tabular}[c]{@{}c@{}}Research on the Recognition of \\ Various Muscle Fatigue States in \\ Resistance Strength   Training\end{tabular} & \begin{tabular}[c]{@{}c@{}}Convolutional Neural Network (CNN) model, \\ Multi-SVM, Multi-LDA\end{tabular} & \begin{tabular}[c]{@{}c@{}}CNN achieved the highest accuracy and AUC, \\ outperforming Multi-SVM and Multi-LDA in fatigue \\ recognition across four fatigue levels.\end{tabular} \\\hline
\cite{40} & \begin{tabular}[c]{@{}c@{}}Automated Detection of Muscle \\ Fatigue Conditions Using \\ Cyclostationary-based Geometric \\ Features of sEMG\end{tabular} & \begin{tabular}[c]{@{}c@{}}Extreme Learning Machine (ELM), \\ Multilayer Perceptron (MLP)\end{tabular} & \begin{tabular}[c]{@{}c@{}}ELM model with geometric features demonstrated \\ high accuracy and F-score, proving effective for \\ muscle fatigue   detection.\end{tabular}\\\hline
\cite{42} & \begin{tabular}[c]{@{}c@{}}Physical Fatigue Detection Using \\ EMG Wearables and User Reports - \\ A Machine Learning Approach \\ Towards Adaptive Rehabilitation\end{tabular} & \begin{tabular}[c]{@{}c@{}}Linear   SVM, SVM with RBF kernel, \\ Gradient Boosting (GB), \\ Extra-Trees (ET), Random Forests (RF), \\ with post-processing using\\ median filtering and temporal grouping\end{tabular} & \begin{tabular}[c]{@{}c@{}}Best performance achieved with Gradient Boosting (GB), \\ supported by post-processing for temporal accuracy \\ improvement. F1 Score improved from   initial \\ 67.2\% to 76.6\% with post-processing\end{tabular}\\\hline
\cite{43} & \begin{tabular}[c]{@{}c@{}}Detection and Analysis of Driver \\ State with Electrocardiogram (ECG)\end{tabular} & \begin{tabular}[c]{@{}c@{}}Support Vector Machine (SVM), \\ K-Nearest Neighbors (KNN), \\ Ensemble Classifier\end{tabular} & \begin{tabular}[c]{@{}c@{}}Ensemble classifier achieved the highest accuracy for \\ two-class states; reduced accuracy (58.3\%) in five-class   \\ detection (normal, drowsy, fatigue, visual inattention, \\ cognitive   inattention)\end{tabular}\\\hline
\cite{76} & \begin{tabular}[c]{@{}c@{}}Research   on Driving Fatigue \\ Level Using ECG Signal from \\ Smart Bracelet\end{tabular} & \begin{tabular}[c]{@{}c@{}}BP Neural Network for data compensation, \\ multi-index fusion theory\\ with principal component analysis (PCA)\end{tabular} & \begin{tabular}[c]{@{}c@{}}BP Neural Network and PCA successfully distinguished \\ four fatigue levels (wide awake, mild, moderate, severe)\end{tabular}\\\hline
\cite{83} & \begin{tabular}[c]{@{}c@{}}Modeling and Recognition of \\ Driving Fatigue   State Based \\ on R-R Intervals of ECG Data\end{tabular} & \begin{tabular}[c]{@{}c@{}}AR(1)-GARCH(1,1) model for time series \\ analysis\end{tabular} & \begin{tabular}[c]{@{}c@{}}Effective detection of fatigue with a recognition delay \\ of less than 5 minutes for all drivers\end{tabular}\\\hline
\cite{85} & \begin{tabular}[c]{@{}c@{}}Driver   Drowsiness Detection \\ Algorithms Using Electrocardiogram \\ Data Analysis\end{tabular} & \begin{tabular}[c]{@{}c@{}}Support   Vector Machine (SVM) and \\ K-Nearest Neighbors (KNN) \\ with Wavelet Transform (WT) and \\ Short Fourier Transform (STFT) \\ for feature extraction\end{tabular} & \begin{tabular}[c]{@{}c@{}}KNN with Wavelet Transform achieved the highest \\ accuracy, demonstrating improved   detection performance \\ over other methods.\end{tabular}\\\hline
\cite{87} & \begin{tabular}[c]{@{}c@{}}Prediction of Exhaustion Threshold \\ Based on   ECG Features Using \\ Artificial Neural Network (ANN)\end{tabular} & \begin{tabular}[c]{@{}c@{}}Artificial Neural Network (ANN) \\ with Levenberg-Marquardt \\ (trainlm) training algorithm\end{tabular} & \begin{tabular}[c]{@{}c@{}}ANN model effectively predicts exhaustion threshold \\ with high correlation to Borg scale and \\ time to exhaustion (TTE) indicators.\end{tabular}\\\hline
\cite{89} & \begin{tabular}[c]{@{}c@{}}Towards   Automated Fatigue \\ Assessment Using Wearable \\ Sensing and Mixed-Effects \\ Models\end{tabular} & \begin{tabular}[c]{@{}c@{}}Random Forest Mixed-Effects Model (MERF) \\ using age and BMI clustering,\\ Random Forest, and \\ Linear Regression for comparison\end{tabular} & \begin{tabular}[c]{@{}c@{}}The MERF model demonstrated improved performance \\ by incorporating demographic factors,   \\ achieving the lowest error and highest correlation \\ among the methods tested.\end{tabular}\\\hline
\cite{131} & \begin{tabular}[c]{@{}c@{}}Information fusion and multi-classifier   \\ system for miner fatigue recognition in \\ plateau environments based on   \\ electrocardiography and electromyography \\ signals\end{tabular} & \begin{tabular}[c]{@{}c@{}}Support Vector Machine (SVM), \\ Random Forest (RF), \\ and Extreme Gradient Boosting (XG-Boost), using \\ Principal Component Analysis (PCA) and \\ Grey Relational Analysis (GRA) for \\ feature selection\end{tabular} & \begin{tabular}[c]{@{}c@{}}This research highlights that the XG-Boost model with \\ PCA-based feature fusion achieved the best results, \\ especially in   a high-altitude mining context.\end{tabular}\\\hline
\cite{119} & \begin{tabular}[c]{@{}c@{}}A   data-driven approach to Physical \\ Fatigue Management Using Wearable \\ Sensors   for Four Fatigue Levels \\ Classification\end{tabular} & \begin{tabular}[c]{@{}c@{}}Random Forest (RF), \\ Artificial Neural Network (ANN), \\ SVM, Decision Tree (DT),   \\ K-Nearest Neighbors (KNN), \\ Logistic Regression (LR)\end{tabular} & \begin{tabular}[c]{@{}c@{}}This   research achieved reliable fatigue classification \\ in walking tasks with the   RF model using reduced \\ feature sets, which helped optimize model performance.\end{tabular}\\\hline
\cite{135} & \begin{tabular}[c]{@{}c@{}}Muscle Fatigue Detection in EMG \\ Using   Time–Frequency Methods, \\ ICA, and Neural Networks\end{tabular} & \begin{tabular}[c]{@{}c@{}}Multi-Layer Perceptron Neural Network (MLPNN) \\ trained with Levenberg–Marquardt (L-M) and \\ Gradient Descent (GDA) algorithm   \\  using time-frequency analysis and \\ Independent Component Analysis (ICA) \\ for feature extraction\end{tabular} & \begin{tabular}[c]{@{}c@{}}The research demonstrated the effectiveness of \\ time-frequency methods combined with \\ neural networks for detecting muscle fatigue, \\ particularly with CWT preprocessing and \\ L-M algorithm.\end{tabular}\\\hline
\cite{136} & \begin{tabular}[c]{@{}c@{}}Mobile-Based   Wearable-Type \\ Driver Fatigue Detection by GSR \\ and EMG\end{tabular} & \begin{tabular}[c]{@{}c@{}}Support Vector Machine (SVM) classifier with \\ frequency-domain feature analysis for   \\ EMG and GSR\end{tabular} & \begin{tabular}[c]{@{}c@{}}The   research demonstrates that EMG and GSR \\ data processed through a mobile-based   SVM \\ classifier can effectively detect fatigue, achieving \\ high accuracy.\end{tabular}\\\hline
\cite{137} & \begin{tabular}[c]{@{}c@{}}Muscle Fatigue Detection Using \\ Feature   Extraction and Deep \\ Learning\end{tabular} & \begin{tabular}[c]{@{}c@{}}F\_GRU (Gated Recurrent Unit) model for   \\ fatigue classification, with feature extraction \\ from time and frequency   domains\end{tabular} & \begin{tabular}[c]{@{}c@{}}This research highlights the effectiveness of the \\ F\_GRU model for high-accuracy, real-time \\ muscle fatigue detection.\end{tabular}\\\hline
\cite{38} & \begin{tabular}[c]{@{}c@{}}Classifying   Driving Fatigue \\ Using EEG Signals\end{tabular} & \begin{tabular}[c]{@{}c@{}}Particle Swarm Optimization Hybrid Extreme \\ Learning Machine (PSO-H-ELM), \\ Support Vector Machine (SVM), \\ K-Nearest Neighbors (KNN)\end{tabular} & \begin{tabular}[c]{@{}c@{}}The   research indicates that the PSO-H-ELM \\ classifier performed better than   traditional \\ methods for driving fatigue detection using EEG.\end{tabular}\\\hline
\cite{45} & \begin{tabular}[c]{@{}c@{}}Sensor-based driver condition \\ recognition using support vector \\ machine for the detection of driver \\ drowsiness\end{tabular} & \begin{tabular}[c]{@{}c@{}}Support Vector Machine (SVM) classifier   \\ with data normalization and segmentation \\ for enhanced classification\end{tabular} & \begin{tabular}[c]{@{}c@{}}This research demonstrated the high effectiveness \\ of using SVM with PPG data from a wearable \\ device to recognize   drowsiness in a controlled \\ driving simulation environment.\end{tabular}\\\hline
\cite{46} & \begin{tabular}[c]{@{}c@{}}Vigilance Estimation Using a \\ Wearable EOG Device in Real \\ Driving Environments\end{tabular} & \begin{tabular}[c]{@{}c@{}}Continuous Conditional Neural Field (CCNF) \\ and Continuous Conditional Random Field (CCRF) \\ for temporal modeling of vigilance states\end{tabular} & \begin{tabular}[c]{@{}c@{}}This   research highlights the effectiveness of a \\ wearable EOG device for continuous vigilance \\ monitoring in real-world and simulated environments, \\ with high correlation values indicating accurate \\ vigilance estimation.\end{tabular}\\\hline
\cite{72} & \begin{tabular}[c]{@{}c@{}}Driver Sleepiness Detection from \\ EEG and   EOG Signals Using \\ GAN and LSTM Networks\end{tabular} & \begin{tabular}[c]{@{}c@{}}Long Short-Term Memory (LSTM) networks,   \\ augmented with Conditional Wasserstein GAN (CWGAN)\end{tabular} & \begin{tabular}[c]{@{}c@{}}This Research used CWGAN to enhance training data \\ for LSTM, achieving high accuracy in drowsiness \\ detection by   monitoring alpha waves in EEG and EOG.\end{tabular}\\\hline
\cite{109} & \begin{tabular}[c]{@{}c@{}}Wearable   Device-Based System to \\ Monitor Driver’s Stress, Fatigue, and \\ Drowsiness\end{tabular} & \begin{tabular}[c]{@{}c@{}}Support   Vector Machine (SVM) classifier with \\ feature selection using ANOVA and   \\ Sequential Floating Forward Selection (SFFS)\end{tabular} & \begin{tabular}[c]{@{}c@{}}This   research demonstrates the effectiveness of using \\ wearable sensors with SVM   classification for \\ monitoring driver states in a simulated environment.\end{tabular}\\\hline
\cite{99} & \begin{tabular}[c]{@{}c@{}}Iterative Cross-Subject Negative-Unlabeled   \\ Learning Algorithm for Quantifying Passive \\ Fatigue\end{tabular} & \begin{tabular}[c]{@{}c@{}}Negative-Unlabeled (NU) Learning Algorithm   \\ using Support Vector Machine (SVM) with an \\ RBF kernel\end{tabular} & \begin{tabular}[c]{@{}c@{}}This research highlights the effectiveness   \\ of the NU learning approach for cross-subject \\ passive fatigue detection.\end{tabular}\\\hline
\cite{100} & \begin{tabular}[c]{@{}c@{}}An   Effective Hybrid Model for EEG-Based \\ Drowsiness Detection\end{tabular} & \begin{tabular}[c]{@{}c@{}}Hybrid   model combining signal-processing-based \\ features and deep learning-based   features, using \\ Long Short-Term Memory (LSTM) and \\ deep convolutional networks (AlexNet and VGGNet)\end{tabular} & \begin{tabular}[c]{@{}c@{}}This   research combines traditional signal \\ processing and deep learning features,   \\ showing high accuracy for EEG-based \\ drowsiness detection.\end{tabular}\\\hline
\cite{101} & \begin{tabular}[c]{@{}c@{}}Automatic Detection of Drowsiness Using   \\ In-Ear EEG\end{tabular} & Support Vector Machine (SVM) with RBF kernel & \begin{tabular}[c]{@{}c@{}}This research confirms the feasibility of using \\ in-ear EEG for drowsiness detection, achieving \\ substantial accuracy in   a controlled setting.\end{tabular}\\\hline
\cite{103} & \begin{tabular}[c]{@{}c@{}}Feature   Extraction Method for Classification \\ of Alertness and Drowsiness States in   \\ EEG Signals\end{tabular} & \begin{tabular}[c]{@{}c@{}}Tunable Q-Factor Wavelet Transform (TQWT) \\ for feature extraction and various classifiers \\ (Decision Tree, Logistic Regression, \\ Fine Gaussian SVM, \\ Weighted   KNN, Ensemble Boosted Trees, \\ Extreme Learning Machine)\end{tabular} & \begin{tabular}[c]{@{}c@{}}This Research confirms the effectiveness of \\ TQWT-based features in distinguishing between \\ alertness and drowsiness, achieving high \\ classification performance with ELM.\end{tabular}\\\hline
\cite{140} & \begin{tabular}[c]{@{}c@{}}Using Wearable ECG/PPG Sensors \\ for Driver   Drowsiness Detection \\ Based on Recurrence Plots\end{tabular} & \begin{tabular}[c]{@{}c@{}}Convolutional Neural Network (CNN) with three  \\ types of Recurrence Plots (Bin-RP, Cont-RP, ReLU-RP) \\ for feature   extraction and drowsiness classification\end{tabular} & \begin{tabular}[c]{@{}c@{}}This research highlights the effectiveness of \\ using recurrence plots with CNN for drowsiness \\ detection in a virtual driving setup using ECG \\ and PPG data.\end{tabular}\\\hline
\cite{141} & \begin{tabular}[c]{@{}c@{}}Driver   Drowsiness Classification \\ Using Fuzzy Wavelet-Packet-Based \\ Feature-Extraction Algorithm\end{tabular} & \begin{tabular}[c]{@{}c@{}}Fuzzy Mutual Information Wavelet Packet Transform (FMIWPT) \\ for feature extraction, combined with various classifiers (LDA, \\ kNN, LIBLINEAR, LIBSVM)\end{tabular} & \begin{tabular}[c]{@{}c@{}}This research developed a fuzzy-entropy-based \\ wavelet packets transform for driver drowsiness \\ classification, achieving high accuracy in \\ simulated conditions.\end{tabular}\\\hline
\cite{82} & \begin{tabular}[c]{@{}c@{}}Data-driven learning fatigue detection   \\ system: A multimodal fusion approach \\ of ECG (electrocardiogram) and video   \\ signals\end{tabular} & \begin{tabular}[c]{@{}c@{}}XGBoost classifier with a hybrid of handcrafted and\\ deep learning features from ECG and video data\end{tabular} & \begin{tabular}[c]{@{}c@{}}This research combines ECG and video data to achieve\\ high detection accuracy in different fatigue conditions,   \\ validating the approach across datasets.\end{tabular}\\\hline
\cite{81}& \begin{tabular}[c]{@{}c@{}}Detection   of Mental Fatigue State \\ with Wearable ECG Devices\end{tabular} & \begin{tabular}[c]{@{}c@{}}HRV (Heart Rate Variability) indicators used with \\ classifiers (SVM, KNN, Naïve Bayes, Logistic Regression)\end{tabular} & \begin{tabular}[c]{@{}c@{}}This   research validates the use of HRV-based features \\ from wearable ECG devices for mental fatigue detection, \\ achieving notable accuracy with the KNN classifier.\end{tabular}\\\hline
\cite{84} & \begin{tabular}[c]{@{}c@{}}ECG Signal Features Classification for   \\ Mental Fatigue Recognition\end{tabular} & \begin{tabular}[c]{@{}c@{}}Principal Component Analysis (PCA) for feature selection, \\ Random Forest (RF), K-Nearest Neighbor (KNN), \\ Linear Discriminant Analysis (LDA), and \\ Decision Tree (DT) classifiers\end{tabular} & \begin{tabular}[c]{@{}c@{}}This research demonstrates effective mental fatigue \\ detection using ECG features and a machine learning \\ classification   model, with high accuracy achieved \\ through the RF classifier.\end{tabular}\\\hline
\cite{170} & \begin{tabular}[c]{@{}c@{}}Electromyographical   Manifestations of \\ Muscle Fatigue During Different Levels \\ of Simulated Light   Manual Assembly \\ Work\end{tabular} & \begin{tabular}[c]{@{}c@{}}EMG frequency and amplitude analysis with \\ regression analysis to measure fatigue   \\ over two low-intensity conditions \\ (8\% and 12\% MVC)\end{tabular} & \begin{tabular}[c]{@{}c@{}}This research explores EMG-based fatigue indicators \\ in low-intensity manual tasks, highlighting that muscle \\ fatigue can be detected through MPF decreases across tasks.\end{tabular}\\\hline
\cite{102} & \begin{tabular}[c]{@{}c@{}}EEG-Based Estimation of Mental Fatigue   \\ Using KPCA–HMM and Complexity \\ Parameters\end{tabular} & \begin{tabular}[c]{@{}c@{}}Kernel Principal Component Analysis (KPCA)   \\ combined with Hidden Markov Model (HMM) \\ for mental fatigue classification\end{tabular} & \begin{tabular}[c]{@{}c@{}}This research effectively combines KPCA and HMM \\ for detecting mental fatigue, leveraging complexity \\ measures to classify fatigue states with high accuracy.\end{tabular}\\\hline
\cite{107} & \begin{tabular}[c]{@{}c@{}}Monitoring   Fatigue in Construction \\ Workers Using Physiological Measurements\end{tabular} & \begin{tabular}[c]{@{}c@{}}Boosted tree classifier using physiological \\ features from heart rate, skin   temperature, \\ and EEG measurements to classify fatigue levels\end{tabular} & \begin{tabular}[c]{@{}c@{}}This   research found that monitoring skin temperature \\ from the temple region was   particularly effective, \\ achieving higher accuracy than heart rate alone.\end{tabular}\\\hline
\cite{91} & \begin{tabular}[c]{@{}c@{}}A Muscle Fatigue Classification Model Based   \\ on LSTM and Improved Wavelet Packet Threshold\end{tabular} & \begin{tabular}[c]{@{}c@{}}LSTM (Long Short-Term Memory) network,   \\ Improved Wavelet Packet Threshold function \\ for denoising sEMG signals\end{tabular} & \begin{tabular}[c]{@{}c@{}}This research showcases a high-accuracy muscle \\ fatigue detection model with LSTM, leveraging \\ enhanced signal denoising through wavelet techniques.\end{tabular}\\\hline
\cite{98} & \begin{tabular}[c]{@{}c@{}}Estimation   of Elbow Kinematics Under \\ Fatigue and Non-Fatigue Conditions \\ Using MLPN   Network Based on \\ sEMG Signal\end{tabular} & \begin{tabular}[c]{@{}c@{}}Multi-layered   Perceptron Neural Network (MLPNN), \\ focusing on IEMG and Zero Crossing (ZC) as   \\ key features for elbow kinematics and \\ fatigue classification\end{tabular} & \begin{tabular}[c]{@{}c@{}}This research attempts to estimate elbow movement \\ and classify fatigue conditions, achieving moderate \\ accuracy with MLPNN based on specific time-domain \\ sEMG features.\end{tabular}\\\hline
\cite{97} & \begin{tabular}[c]{@{}c@{}}Fatigue Status Recognition in a Post-Stroke   \\ Rehabilitation Exercise with sEMG Signal\end{tabular} & \begin{tabular}[c]{@{}c@{}}Hidden Markov Model (HMM) and \\ Artificial Neural Network (ANN) \\ for fatigue status classification\end{tabular} & \begin{tabular}[c]{@{}c@{}}This study demonstrates the effectiveness of HMM \\ for accurate fatigue detection in post-stroke rehabilitation   \\ exercises.\end{tabular}\\\hline
\cite{127} & \begin{tabular}[c]{@{}c@{}}Continuous   Alertness Assessments Using \\ EOG Glasses to Unobtrusively Monitor Fatigue   \\ Levels In-The-Wild\end{tabular} & \begin{tabular}[c]{@{}c@{}}Eye blink frequency analysis to assess \\ fatigue levels; measurements correlated with   \\ Psychomotor Vigilance Task (PVT) reaction \\ times as a ground truth for   alertness levels\end{tabular} & \begin{tabular}[c]{@{}c@{}}This research utilizes EOG-based eye blink frequency in a \\ wearable device to   continuously monitor alertness levels \\ in real-world settings, showing a   correlation between \\ increased blink frequency and decreased alertness.\end{tabular}\\\hline
\cite{128} & \begin{tabular}[c]{@{}c@{}}Fatigue Driving Detection Method Based on   \\ Time-Space-Frequency Features of Multimodal \\ Signals\end{tabular} & \begin{tabular}[c]{@{}c@{}}Convolutional Autoencoder (CAE) for feature fusion and \\ Convolutional Neural Network (CNN) followed \\ by Long Short-Term   Memory (LSTM) for classification\end{tabular} & \begin{tabular}[c]{@{}c@{}}This research demonstrates the effectiveness of multimodal \\ EEG and EOG feature fusion in detecting fatigue,   \\ with strong performance metrics using CAE and Bi-LSTM.\end{tabular}\\\hline
\cite{126} & \begin{tabular}[c]{@{}c@{}}Fatigue   Detection Caused by Office Work \\ With the Use of EOG Signal\end{tabular} & \begin{tabular}[c]{@{}c@{}}Quadratic Discriminant Analysis (QDA) classifier, \\ with features based on blink   analysis, \\ including blink duration, blink amplitude, \\ and time between blinks\end{tabular} & \begin{tabular}[c]{@{}c@{}}This research demonstrates the potential for using EOG signals \\ to monitor fatigue   in an office environment with high \\ classification accuracy.\end{tabular}\\\hline
\cite{12} & \begin{tabular}[c]{@{}c@{}}Activity-Aware Deep Cognitive Fatigue   \\ Assessment Using Wearables\end{tabular} & \begin{tabular}[c]{@{}c@{}}Activity-Aware Recurrent Neural Network (AcRoNN) \\ with two-stage LSTM and Consistency Self-Attention\end{tabular} & \begin{tabular}[c]{@{}c@{}}This research develops a novel AcRoNN model incorporating \\ activity-awareness for more accurate cognitive fatigue   \\ estimation across varying contexts.\end{tabular}\\\hline
\cite{111} & \begin{tabular}[c]{@{}c@{}}Heart Rate   Monitoring System During \\ Physical Exercise for Fatigue Warning Using   \\ Non-Invasive Wearable Sensor\end{tabular} & \begin{tabular}[c]{@{}c@{}}Android-based application with an \\ Easy Pulse PPG sensor for real-time monitoring, \\ with alerts provided based on heart rate thresholds\end{tabular} & \begin{tabular}[c]{@{}c@{}}This research introduces a real-time heart rate monitoring \\ system with warning notifications based on threshold limits \\ to help manage fatigue during   exercise.\end{tabular}\\\hline
\cite{110} & \begin{tabular}[c]{@{}c@{}}An Unobtrusive Smartwatch-Based Platform   \\ for Automatic Background Monitoring of Fatigue\end{tabular} & \begin{tabular}[c]{@{}c@{}}Heart rate and breathing pattern extraction   \\ from PPG, activity detection using \\ accelerometer and pressure sensors\end{tabular} & \begin{tabular}[c]{@{}c@{}}This research leverages PPG and other smartwatch sensors \\ to monitor fatigue unobtrusively, validating physiological   \\ markers with high correlation to established tools.\end{tabular}\\\hline
\cite{130} & \begin{tabular}[c]{@{}c@{}}Assessing   Fatigue with Multimodal \\ Wearable Sensors and Machine Learning\end{tabular} & \begin{tabular}[c]{@{}c@{}}Random Forest (RF) for physical fatigue (PF) \\ detection and LSTM for cognitive fatigue (CF)\\ detection\end{tabular} & \begin{tabular}[c]{@{}c@{}}This research leverages a comprehensive multimodal setup \\ to assess physical and cognitive fatigue, demonstrating high \\ accuracy with RF for PF and LSTM for CF.\end{tabular}\\\hline
\cite{115} & \begin{tabular}[c]{@{}c@{}}Automatic Motion Artifact Detection in   \\ Electrodermal Activity Data Using \\ Machine Learning\end{tabular} & \begin{tabular}[c]{@{}c@{}}Gradient Boosting (GradBoost) classifier,   \\ evaluated with Leave-One-Subject-Out (LOSO) \\ cross-validation\end{tabular} & \begin{tabular}[c]{@{}c@{}}The research emphasizes high accuracy and F1 Score in \\ detecting motion artifacts within EDA data, but it does not   \\ provide a specific value for Precision or AUC.\end{tabular}\\\hline
\cite{114} & \begin{tabular}[c]{@{}c@{}}Electrodermal Activity Sensor for \\ Classification of Calm/Distress \\ Condition\end{tabular} & \begin{tabular}[c]{@{}c@{}}Statistical analysis and classification using \\ a tree-based model; Stratified \\ 10-fold cross-validation\end{tabular} & \begin{tabular}[c]{@{}c@{}}This research uses a wearable EDA-based device to classify \\ calm versus distress   states. It demonstrates the device's high \\ accuracy, making it suitable for   real-time, long-term mental \\ health monitoring applications.\end{tabular}\\\hline
\cite{142} & \begin{tabular}[c]{@{}c@{}}Feature Extraction and Evaluation for   \\ Driver Drowsiness Detection Based on \\ Thermoregulation\end{tabular} & \begin{tabular}[c]{@{}c@{}}Decision tree classification based on   \\ changes in nose and wrist temperature \\ and heart rate, using slopes in   \\ temperature and HR trends\end{tabular} & \begin{tabular}[c]{@{}c@{}}This research explores using thermoregulatory changes to detect \\ driver drowsiness in simulations, showing the potential of \\ temperature-based detection for monitoring drowsiness.\end{tabular}\\\hline
\cite{74} & \begin{tabular}[c]{@{}c@{}}Wearable Device-Based System to \\ Monitor a Driver’s Stress, Fatigue, and \\ Drowsiness\end{tabular} & \begin{tabular}[c]{@{}c@{}}Support Vector Machine (SVM) classifier \\ with preprocessing for noise reduction and   \\ optimal feature selection using ANOVA and \\ Sequential Floating Forward Selection (SFFS)\end{tabular} & \begin{tabular}[c]{@{}c@{}}This research proposes a wearable system for real-time monitoring \\ of driver conditions, achieving practical accuracy for distinguishing \\ stress, fatigue, and drowsiness states.\end{tabular}\\\hline
\cite{171} & \begin{tabular}[c]{@{}c@{}}A Deep Learning Approach for Fatigue   \\ Prediction in Sports Using GPS Data \\ and Rate of Perceived Exertion\end{tabular} & \begin{tabular}[c]{@{}c@{}}Convolutional Neural Network (CNN) with   \\ Gated Recurrent Units (GRU), named FatigueNet, \\ for time-series prediction based on GPS data features\end{tabular} & \begin{tabular}[c]{@{}c@{}}This research introduces a CNN-GRU-based deep learning \\ model to predict fatigue through GPS-tracked movement,   \\ providing insights into fatigue monitoring without relying \\ on manual RPE data   collection.\end{tabular}\\\hline
\cite{132} & \begin{tabular}[c]{@{}c@{}}Psychophysiological Data-Driven \\ Multi-Feature Information Fusion and \\ Recognition of Miner Fatigue   \\ in High-Altitude and Cold Areas\end{tabular} & \begin{tabular}[c]{@{}c@{}}Feature extraction with Pearson correlation, \\ t-tests, and ROC analysis.   \\ classification using Support Vector Machine (SVM) \\ and Random Forest (RF) models\end{tabular} & \begin{tabular}[c]{@{}c@{}}This research demonstrates a high-accuracy approach \\ for identifying miner fatigue   by combining physiological \\ factors in challenging environments.\end{tabular}\\\hline
\cite{71} & \begin{tabular}[c]{@{}c@{}}Towards Safer Roads: A Deep Learning-Based   \\ Multimodal Fatigue Monitoring System\end{tabular} & \begin{tabular}[c]{@{}c@{}}Fully Convolutional Recurrent Deep Neural Network \\ (FCR-DNN) with IoT-based architecture for \\ real-time monitoring; parallel CNN and \\ LSTM layers for eye and mouth classification\end{tabular} & \begin{tabular}[c]{@{}c@{}}This research proposes a high-accuracy fatigue monitoring \\ system using IoT and deep learning for enhanced driver safety \\ by detecting fatigue indicators through facial features and   \\ physiological data.\end{tabular}\\\hline
\cite{119} & \begin{tabular}[c]{@{}c@{}}A   Data-Driven Approach to Physical Fatigue \\ Management Using Wearable Sensors to   \\ Classify Four Diagnostic Fatigue States\end{tabular} & \begin{tabular}[c]{@{}c@{}}Feature   extraction from IMU and EMG data; \\ classification using Random Forest (RF) as   \\ the optimal model\end{tabular} & \begin{tabular}[c]{@{}c@{}}This   research highlights a high-accuracy approach for \\ managing physical fatigue   through wearable sensors \\ and machine learning, with a specific focus on   \\ rehabilitation and exercise intensity control.\end{tabular}\\\hline
\cite{122} & \begin{tabular}[c]{@{}c@{}}AI-Assisted Fatigue and Stamina Control for   \\ Performance Sports on IMU-Generated \\ Multivariate Time Series Datasets\end{tabular} & \begin{tabular}[c]{@{}c@{}}AI model integrating Random Forest,   \\ Gradient Boosting Machines, and \\ LSTM networks for fatigue and stamina prediction, \\ with real-time feedback loops for \\ adaptive training adjustments\end{tabular} & \begin{tabular}[c]{@{}c@{}}This research emphasizes AI-driven, real-time fatigue and \\ stamina monitoring in athletes using multivariate time series \\ from IMUs, focusing on preventing overtraining and optimizing   \\ performance.\end{tabular}\\\hline
\cite{121} & \begin{tabular}[c]{@{}c@{}}Automated   and Continuous Fatigue \\ Monitoring in Construction Workers Using \\ Forearm EMG and IMU Wearable Sensors\end{tabular} & \begin{tabular}[c]{@{}c@{}}Aerobic Fatigue Threshold (AFT) prediction \\ using BiLSTM-based recurrent neural network; \\ Decision Tree classifier used for \\ fatigue level classification\end{tabular} & \begin{tabular}[c]{@{}c@{}}This research validates a real-time fatigue monitoring system in \\ construction, effectively using forearm EMG and IMU data \\ for continuous fatigue assessment during labor-intensive tasks.\end{tabular}\\\hline
\cite{104} & \begin{tabular}[c]{@{}c@{}}Highly Reliable Driving Workload Analysis \\ Using Driver Electroencephalogram (EEG) \\ Activities During Driving\end{tabular} & 
\begin{tabular}[c]{@{}c@{}}Noise removal using Half $\beta$-wave correction; \\ statistical analysis of $\alpha$ and $\beta$ waves to calculate \\ EEG activity; paired $t$-tests to analyze driving \\ workload across different road types\end{tabular} & 
\begin{tabular}[c]{@{}c@{}}This research proposes a novel method to quantify driving \\ workload using EEG by isolating engine vibration noise and \\ analyzing $\alpha$ and $\beta$ waves. It demonstrates significant findings \\ regarding increased workload during left-turn driving \\ compared to straight driving.\end{tabular} \\ \hline

\cite{90} & \begin{tabular}[c]{@{}c@{}}Recognition of Muscle Fatigue Status Based \\ on Improved Wavelet Threshold and CNN-SVM\end{tabular} & \begin{tabular}[c]{@{}c@{}}Improved wavelet threshold for sEMG denoising; \\ feature extraction (RMS, IEMG, MF, MPF,   BSE); \\ classification using CNN-SVM, SVM, PSO-SVM, \\ and CNN.\end{tabular} & \begin{tabular}[c]{@{}c@{}}This   research proposes a robust method for sEMG signal \\ denoising and fatigue   classification, integrating wavelet \\ thresholding with CNN-SVM for improved performance \\ in identifying muscle fatigue states.\end{tabular}\\\hline
\cite{95} & \begin{tabular}[c]{@{}c@{}}Surface Electromyography-Based Muscle   \\ Fatigue Detection Using High-Resolution \\ Time-Frequency Methods and \\ Machine Learning Algorithms\end{tabular} & \begin{tabular}[c]{@{}c@{}}High-resolution time-frequency analysis using \\ S-transform, B-distribution (BD), and \\ Extended Modified B-Distribution   (EMBD); \\ feature selection via Genetic Algorithm (GA) and \\ Binary Particle   Swarm Optimization (BPSO); \\ classification using SVM, Random Forest, and   \\ Rotation Forest\end{tabular} & \begin{tabular}[c]{@{}c@{}}This research emphasizes the capability of high-resolution \\ time-frequency methods, particularly EMBD, in detecting   \\ dynamic muscle fatigue with high accuracy, offering robust \\ applications for   sports, ergonomics, and rehabilitation.\end{tabular}\\\hline
\cite{124} & \begin{tabular}[c]{@{}c@{}}An   Electro-Oculogram (EOG) Sensor’s \\ Ability to Detect Driver Hypovigilance \\ Using Machine Learning\end{tabular} & \begin{tabular}[c]{@{}c@{}}Preprocessing using Butterworth filter (0.1–30 Hz); \\ feature extraction (16 features: statistical, higher-order \\ statistics, non-linear); feature selection via   \\ ANOVA and PCA; classifiers used include \\ SVM, KNN, and Ensemble methods\end{tabular} & \begin{tabular}[c]{@{}c@{}}This research demonstrates the efficacy of EOG signals \\ in detecting driver hypovigilance across multiple \\ behavioral states, with Ensemble classifiers offering \\ the best performance for multiclass detection.\end{tabular}\\\hline
\cite{123} & \begin{tabular}[c]{@{}c@{}}Fatigue Monitoring and Awakening System   \\ Based on Eye Electrical and Head Movement \\ Parameters Monitoring\end{tabular} & \begin{tabular}[c]{@{}c@{}}EOG signal acquisition using three electrodes; \\ head posture detection via gyroscope sensor; \\ fatigue level   classification using \\ BP Neural Network; wake-up intervention \\ via music and   physical stimulation\end{tabular} & \begin{tabular}[c]{@{}c@{}}This research introduces a wearable system combining \\ EOG and head movement analysis for fatigue detection \\ and intervention, validated through sleep deprivation \\ experiments and dual-task   performance tests.\end{tabular}\\\hline
\cite{28} & \begin{tabular}[c]{@{}c@{}}A Novel Fatigue Driving State Recognition \\ and Warning Method Based on EEG and EOG \\ Signals\end{tabular} & 
\begin{tabular}[c]{@{}c@{}}Preprocessing: Noise filtering and extraction \\ of Power Spectral Density (PSD) and \\ Differential Entropy (DE) features from \\ EEG signals, and Independent Component Analysis \\ (ICA) from EOG signals. \\ Classification: \\ Fast Support Vector Machine (FSVM) \\ with feature-level fusion.\end{tabular} & 
\begin{tabular}[c]{@{}c@{}}This research presents a real-time fatigue detection \\ and warning system combining EEG and EOG signals, \\ validated in simulated driving conditions and leveraging \\ IoT technology for alert dissemination.\end{tabular} \\ \hline

\cite{36} & \begin{tabular}[c]{@{}c@{}}Research on Exercise Fatigue Estimation \\ Method in Pilates Rehabilitation Based on \\ ECG and sEMG Feature Fusion\end{tabular} & \begin{tabular}[c]{@{}c@{}}Feature extraction from ECG and sEMG using   \\ time-domain, frequency-domain, and advanced \\ decomposition techniques.Classification using \\ Improved Particle Swarm Optimization \\ Support Vector Machine (IPSO-SVM) \\ to detect three states: Relaxed, Transition, and Tired.\end{tabular} & \begin{tabular}[c]{@{}c@{}}This research demonstrates the potential of combining \\ ECG and sEMG data for accurate fatigue detection in \\ rehabilitation, paving the way for safer and more effective \\ man-machine interfaces.\end{tabular}\\\hline

\end{longtable}
\end{landscape}

\twocolumn
\subsection{Ergonomics and Portability}\label{sec:ErgonomicsPortability}
When it comes to wearable devices, ergonomics and comfort are of utmost significance, particularly if they are to be worn for When it comes to wearable devices, ergonomics and comfort are of utmost significance, particularly if they are to be worn for extended periods. The ideal degree of comfort would prevent the user from ever noticing that they have anything extra attached to their body. Industrial design may achieve such a high degree of ergonomics and comfort by experimenting with various forms and shapes until they are just right and then improving the design depending on user input. To prevent issues like irritation or allergies, the device’s materials should be carefully selected \cite{146}.

Most healthcare wearables are either single-use or have replaceable batteries that cannot be recharged. For this reason, mobility and size are being compromised by the battery’s strict overall form-factor requirements. The battery capacity of wearable devices is usually limited. Therefore, making effective use of available battery life is crucial for the devices to function for a fair period \cite{146}.

\subsection{Unobtrusiveness}\label{sec:Unobtrusiveness}
When it comes to fatigue monitoring, there are a variety of scenarios in which prolonged monitoring of a user (for either immediate evaluation or a more in-depth ongoing watch) is both essential and beneficial. Additionally, it is preferable, particularly for long-term fatigue monitoring, that the selected equipment has a minimum effect on the user \cite{147}. To be considered unobtrusive, a wearable gadget must not interfere with the user’s normal activities. This may be accomplished via minimalism in size and style or by disguising the item as commonplace \cite{147}. The bulkiness of the equipment being worn is also a key issue. Users seldom welcome large, bulky, and cumbersome-to-use wearables. The development of chip-scale integrated circuit packaging is notoriously difficult and is often reserved for large corporations or production runs.

Unobtrusive wearable devices like wristbands, phones, and glasses have been the focus of much recent research in wireless physiological sensing and monitoring. Recently, advanced smart apparel (fabric/textile as a sensor) has also received attention as one of the best unobtrusive data collection
\cite{148} \cite{149}.

\subsection{Continuous Monitoring}\label{sec:Continuous Monitoring}
The remote monitoring platform should exist and provide individualized monitoring plans for each kind of wearable to guarantee  obtaining sufficient data while keeping linked to a person, which is necessary for effective fatigue management and monitoring. Patients will be under constant and active supervision with a smart ecosystem that connects the Wearables and the might of Artificial Intelligence. In the event of any unexpected or developing fatigue, this might even save their lives \cite{150}.

\subsection{Data Quality}\label{sec:Data Quality}
Patient-reported exhaustion outcome metrics have become more popular in fatigue monitoring, reflecting the complex interdependence of many variables. Considering the importance of accurately measuring and monitoring fatigue, data acquired by patient-reported fatigue data must be of high quality. The subjective character of the data on fatigue raises concerns about the reliability of the results because of the many confounding variables that might affect patients’ replies. This might happen during the making and using of the instrument, or it could  result from the patient’s reaction patterns. Such considerations are crucial for producing reliable fatigue monitoring results based on data that accurately represents patients’ assessments \cite{151}.

Data accuracy is essential for monitoring and analysing fatigue levels. False information cannot assist in diagnosing weariness and developing a remedy for it since it does not provide a realistic picture. Data inaccuracies may be traced back to several causes, such as missing data, corrupted measurements owing to noise or data Inconsistency and significant drops for mobile users when using wearable devices.

\section{Research Gaps and Future Opportunities}\label{sec:ResearchGapsandFutureOpportunities}
The use of AI and wearables to track fatigue has a lot of unexplored potentials. Despite progress in these areas, numerous obstacles must be overcome before viable fatigue monitoring systems can be developed, validated, and deployed. The absence of standardized ways for assessing fatigue, the limited accuracy of existing wearables, and the need for more robust and adaptive AI algorithms are all mentioned in this section as examples of research gaps. The use of wearables and AI to monitor fatigue raises ethical and privacy problems that must also be addressed.

\subsection{Predictive Accuracy}\label{sec:Predictive Accuracy}
The intricacy of the fatigue concept necessitates that the research community comprehend how the offered AI algorithms could enhance fatigue monitoring. Most AI methods have a long way to go before they can reliably generalise, much alone be clinically applicable, for most forms of fatigue monitoring data \cite{152}. There is concern that blind spots in artificial intelligence would reinforce biases already present in the performance of the current methods, leading to inaccurate predictions about fatigue monitoring. Although there are presently few instances of fatigue monitoring, In the research domain, AI approaches for fatigue monitoring are being proposed with positive results but still, A lot of effort has to go into creating a framework for reliable fatigue monitoring in the clinical domain \cite{152}.

Testing an AI system using sufficiently substantial datasets obtained from institutions other than those that contributed the data for model training is an integral part of external validation and is necessary for the accurate evaluation of real-world clinical performance linked to fatigue monitoring.

Methods, such as independent test sets, need to be developed to allow for direct comparisons of AI systems. It is crucial for those working on AI algorithms to keep an eye out for the danger posed by a lack of fatigue datasets \cite{152}.

\subsection{Multimodal Information Fusion}\label{sec:Multimodal Information Fusion}
In the context of fatigue monitoring, multimodal information fusion is a method of combining data from multiple sources to accurately detect and monitor fatigue. This can include physiological data such as heart rate and eye movement, as well as behavioural data and reaction time. Multimodal information fusion allows for more robust and reliable detection of fatigue than using a single modality alone. The AI algorithms can also continuously learn and adapt to changes in the data, making the system more accurate over time. The usage of this technology can potentially improve fatigue monitoring \cite{153}.

Wearables are being used to gather many modalities for fatigue monitoring since a single modality may not be enough for capturing the whole scope of fatigue. For that, we need to apply more research into the model’s interpretability, exploration of the correlation distribution between features and classification outcomes, break the black box barrier, and demonstrate the model’s decision-making mechanism’s rationality in the task of muscle fatigue detection as well as algorithm optimisation \cite{154}. For that, we need to deepen our fatigue detection framework by fusing environmental and psychophysiological signals with optical data for a comprehensive assessment of the overall fatigue detection framework design.

\subsection{Explainable AI for Fatigue Measurements}\label{sec:Explainable AI for Fatigue Measurements}
Explainable AI (XAI) has become one of the most important topics in the field of AI and its applications in the real world. For instance, AI algorithms applied for medical diagnosis and prediction must be fully explicable and generated by transparent AI models (rather than black boxes) \cite{154}. The simplest AI model cannot properly explain the logic of how they process data/features and reach a decision. The nonconvexity problem of AI algorithms further justifies the requirement for explainability. AI models developed for fatigue measurement and monitoring must be able to simply yet accurately describe how they make detection and prediction. It is especially  required if those AI-empowered devices for fatigue monitoring will be approved by relevant authorities in different countries. For instance, there are a set of guidelines issued by the US Federal Trade Commission related to the explainability of AI systems \cite{154}. To the best of our knowledge, the current literature related to using AI for fatigue monitoring fully ignores this path. Explainability for fatigue monitoring could be partially addressed using recently developed tools such as SHAP could be used for this purpose \cite{155}.

\begin{figure}[h]
    \centering
    \includegraphics[scale=0.6]{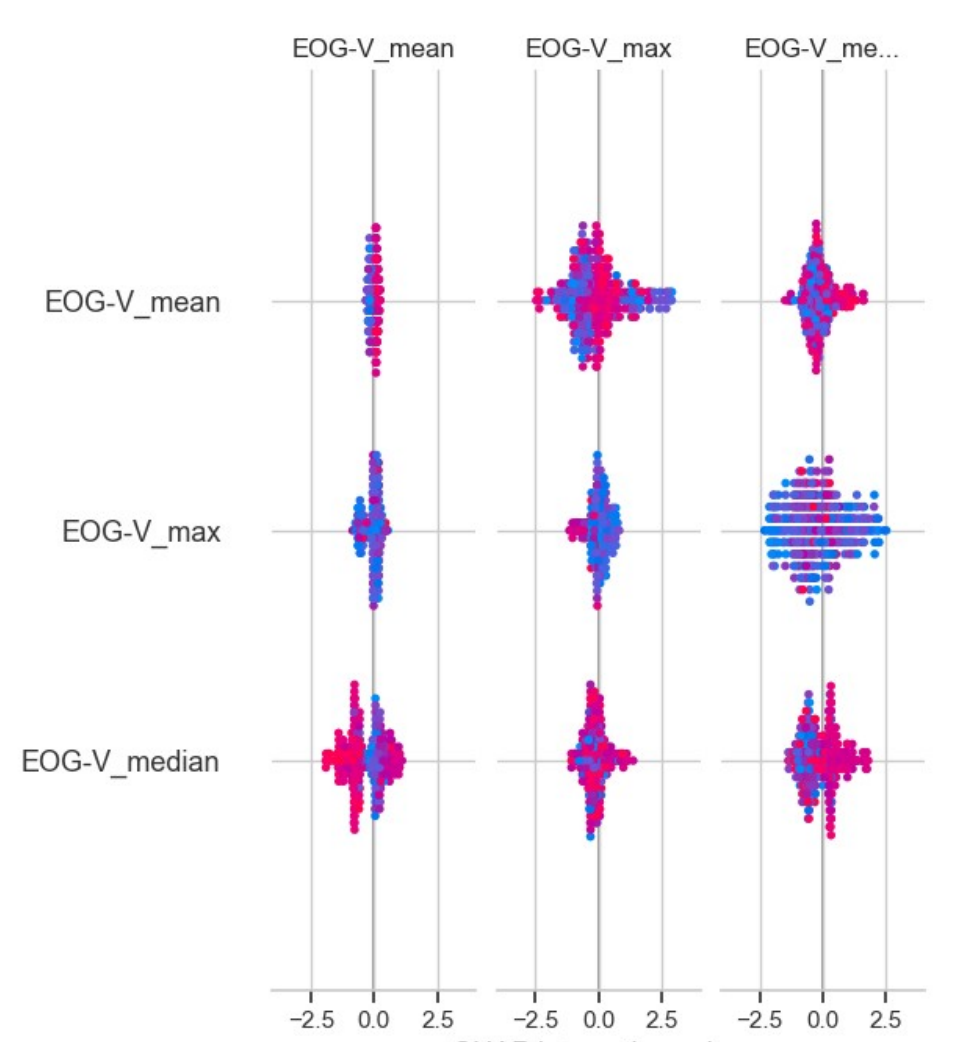 }
    \caption{SHAP values describing the contribution and importance.}
    \label{fig:Figure16}
\end{figure}

\begin{figure}[h]
    \centering
    \includegraphics[scale=0.4]{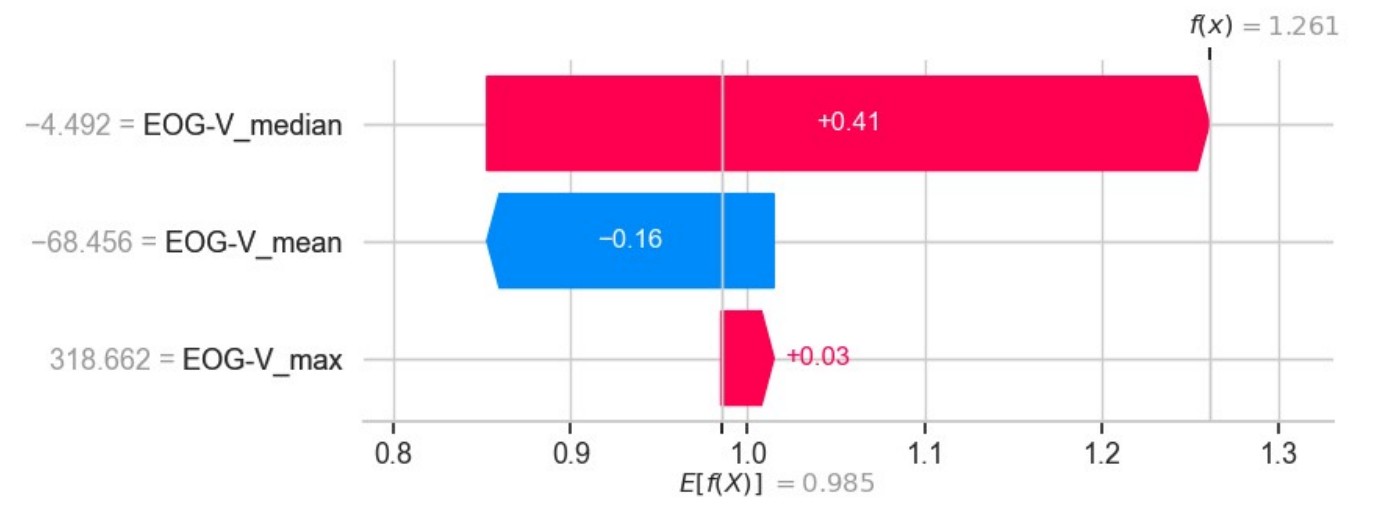 }
    \caption{SHAP Waterfall Plot showing the contribution of features to the model's prediction. }
    \label{fig:Figure17}
\end{figure}

To improve the explainability and readability of AI/ML models, researchers have developed a technique called SHAP values (Shapley Additive exPlanations) \cite{156} that is based on cooperative game theory. SHAP displays the relevance or contribution of each feature to the model’s prediction, but it does not assess the accuracy of the forecast. Authors in \cite{157} used SHAP and gradient boosting methods for predictive fatigue monitoring in drivers (Figure \ref{fig:Figure16} and Figure \ref{fig:Figure17}). The importance of a feature in a model refers to how much it contributes to the model’s output or prediction. The more important a feature is, the more impact it has on the model’s output. The model’s performance can be improved by focusing on the most important features and giving them more weight in the model. Conversely, features with little or no impact on the model’s output can be removed or given less weight. It’s also important to note that feature importance can change as the dataset changes, so it’s important to  monitor and adapt the feature selection process constantly.

\subsection{Uncertainty Quantification and Trust}\label{sec:Uncertainty Quantification and Trust}
Fatigue definition and measurement are intrinsically fuzzy and uncertain. Experts from different fields can’t agree on how AI algorithms and signals from wearables can be used to define and measure fatigue in a precise way. All AI techniques applied in literature for fatigue estimation fully ignore issues related to predictive uncertainties associated with AI models. There are two types of uncertainty associated with predictions generated by AI models for fatigue monitoring. These are aleatoric and epistemic uncertainties \cite{158}. Aleatoric uncertainty is irreducible and caused by noise in recordings or sample mislabelling. The lack of data in different regions of input space causes epistemic (knowledge) uncertainty. It is of practical importance to quantify these uncertainties and flag uncertain outcomes/decisions/predictions generated by AI models for fatigue monitoring. Ignoring uncertainties could lead to misleading results, as AI models could easily make wrong decisions when confronted with unknown or noisy samples \cite{159}.

There are also important questions about how AI models can be trusted, how fair they are, and how reliable they are. These issues are the focus of AI deployment for mission-critical applications including autonomous vehicles, cyber-security, and healthcare \cite{160} \cite{161}. These issues are also required to be theoretically and practically considered when designing fatigue monitoring systems.

\subsection{Edge Computing}\label{sec:Edge Computing}
Processing multimodal datasets at scale on edge using limited computing resources. In the context of future research, answers to how can one distinguish mental exhaustion from physical exhaustion, what factors could lead to exhaustion, and what implications learning fatigue has on mental performance are essential to be explored. A comprehensive review of edge computing technologies for wearables can be found in \cite{162}.

Research gaps related to fatigue monitoring include information perception, sensing, storage, and computing.
The study presented a detailed methodology for detecting driver hypovigilance using ECG data analysis. ECG signals were meticulously recorded and pre-processed to remove noise and artifacts, ensuring accurate HRV analysis—a key factor in monitoring driver states\cite{7}\cite{13}. Twenty distinct features were extracted from the ECG data, including time-domain linear features like mean and median heart rates and frequency-domain nonlinear features such as Approximate Entropy\cite{24}\cite{25}. The relevance of each feature was statistically evaluated using one-way ANOVA, and feature reduction was performed using PCA\cite{26}\cite{27}. The processed features were then classified using advanced machine learning techniques: SVM, KNN, and Ensemble methods\cite{28}\cite{29}.

The results indicated high accuracy rates for the two-class detection model, which distinguished between normal and hypovigilant states, including drowsiness and visual or cognitive inattention, faring better than the five-class detection approach\cite{30}. The success of the two-class model, particularly with the Ensemble classifier that achieved accuracies of up to 100\%, underscores the effectiveness of the selected features and the potential of the proposed methodology in practical, real-world applications\cite{31}.

These findings highlight the viability of ECG data in the development of robust driver monitoring systems and suggest further enhancement of the classification algorithms for improved detection of hypovigilance states\cite{32}.

The study’s constraints were noted, particularly the artificial simulation of fatigue through problem-solving tasks and the need for broader research to encompass diverse populations and physical activity levels \cite{163}. Nonetheless, the findings support the viability of wearable ECG devices in monitoring mental fatigue effectively, emphasizing the promise of such technology in mobile health (m-health) applications \cite{164}. The research concluded that a select number of HRV indicators could be sufficient for effective mental fatigue prediction, indicating the potential for employing low-cost wearable devices for efficient data processing and fatigue state detection in real-time \cite{165}.

The study on wrist-worn wearable sensors for detecting driver drowsiness \cite{86} utilized a detailed methodology within a high-fidelity driving simulator to gather physiological data crucial for assessing drowsiness. A key focus of the methodology was the extraction and analysis of HRV from data collected via wrist-worn devices, using sophisticated algorithms designed to isolate relevant HRV metrics indicative of drowsiness \cite{166}. This included time-domain measures such as mean heart rate and standard deviations of RR intervals, along with frequency-domain measures like low and high-frequency components of the HRV spectrum, which reflect autonomic nervous system activity \cite{167}.

\begin{figure}[h]
    \centering
    \includegraphics[scale=0.18]{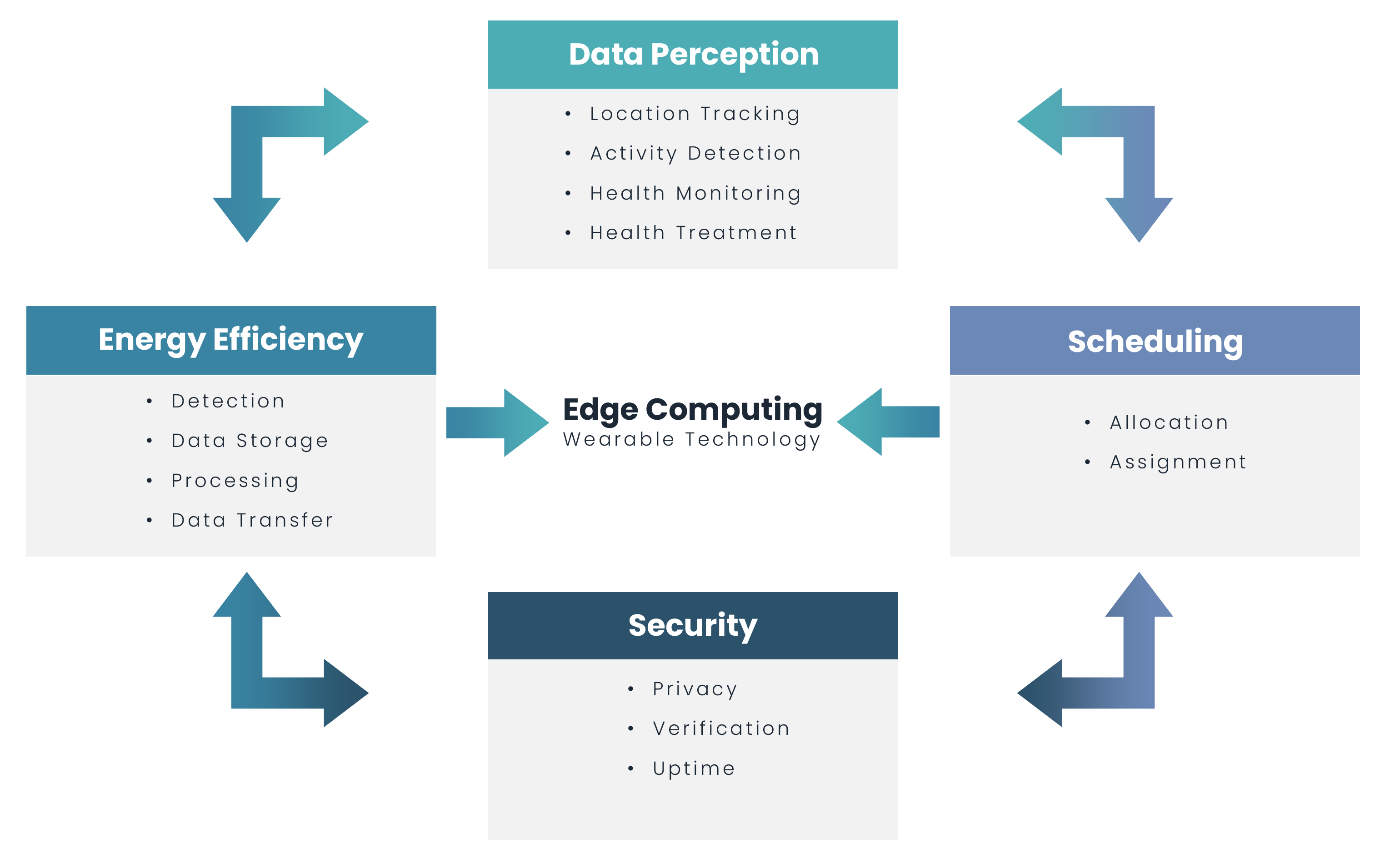 }
    \caption{Existing research efforts on edge computing for wearable technologies.}
    \label{fig:Figure18}
\end{figure}

The efficacy of several machine learning classifiers—Support Vector Machines (SVM), Random Forests, and Decision Trees—was rigorously evaluated in both user-dependent and user-independent frameworks. These classifiers underwent thorough testing for accuracy in predicting drowsiness states based on the extracted features, with training using cross-validation techniques to ensure robustness and prevent over-fitting \cite{168}\cite{169}.

The results revealed that the wearable sensors achieved high accuracy rates, showcasing their potential as viable alternatives to more traditional, invasive methods such as electrocardiography (ECG) \cite{170}\cite{171}. This high level of precision underscores the capability of wearable technologies in the real-time monitoring and detection of driver drowsiness, highlighting significant advancements in driver safety and automated monitoring systems \cite{172}. The operational viability of such wearable devices in practical applications not only confirms their effectiveness but also paves the way for their integration into automotive safety systems to enhance road safety by effectively monitoring driver states \cite{173}.

This paper \cite{174} introduces a proposed model for fatigue detection using multimodal data—ECG and video signals. The model stands out for its integration of various feature extraction techniques and the application of machine learning to precisely classify fatigue states. During the ECG processing phase, the study extracts both traditional handcrafted features and advanced deep learning (DL) features from ECG signals \cite{81}\cite{175}. The handcrafted features encompass time, frequency, and nonlinear domain metrics, providing a comprehensive view of the physiological indicators of fatigue \cite{86}\cite{176}. The advanced DL features are derived from a cascaded network that includes a convolutional neural network (CNN) and a bidirectional long short-term memory (BiLSTM) with an attention mechanism \cite{177}, which together enhance the temporal analysis of ECG data. For video processing, the research utilizes a hybrid attention cascade network to extract facial expression features \cite{178}, incorporating spatial features with ResNet-50 and applying hybrid attention and a gated recurrent unit (GRU) for temporal feature extraction. The model’s classifier module employs an XGBoost classifier that synergizes the features from both ECG and video data \cite{179}. After a detailed feature fusion and selection process, the classifier proficiently categorizes the subject’s state into alert, normal, or tired. Validated on the DROZY public dataset and a self-collected dataset, the model demonstrates substantial superiority over existing methods, achieving 99.6\% accuracy on the DROZY dataset \cite{179} and 91.8\% accuracy on the self-collected dataset \cite{180}, underpinning its potential for reliable application in real-world scenarios (see Figure \ref{fig:Figure18}).

\subsection{Lessons Learned and Novel Insights}\label{sec:Lessons Learned and Novel Insights}
This article offers valuable insights into the current technological advancements and methodologies employed for fatigue monitoring, emphasizing the incorporation of wearable technologies and AI. One of the significant insights gained is the efficacy of integrating various physiological signals, including ECG, EEG, EMG, and PPG, with machine learning and deep learning frameworks. Integrating multi-modal data fusion significantly improves the precision of fatigue detection systems by providing a more holistic perspective on mental and physical fatigue. This combination has demonstrated significant utility in high-demand environments, including aviation and manufacturing, where real-time monitoring is essential for ensuring safety and optimizing performance. The findings indicate that this trend will likely persist as advancements in AI and wearable technology progress, facilitating the development of increasingly sophisticated and precise fatigue detection systems.

Another important discovery is the potential for unobtrusive and real-time fatigue monitoring made possible by wearable technology, such as smartwatches, fitness trackers, and specialized sensors.  Research highlights the significance of ongoing observation conducted in a non-intrusive way, enabling fatigue evaluation while preserving the user’s workflow and daily routines. The transition towards seamlessly incorporating monitoring technologies into daily life presents novel opportunities for proactive fatigue management across various sectors, including sports, healthcare, and transportation. Nonetheless, a significant challenge persists regarding the variability in the quality and availability of real-time data. This situation necessitates the development of more standardized protocols and dependable sensing technologies to achieve consistent and actionable outcomes.

Even though the results of current methods are promising, there is still a lot of room for improvement in making AI models that can be explained and that give more accurate and understandable information about fatigue detection. Furthermore, the progression of edge computing is essential for facilitating real-time data processing on wearable devices, thereby minimizing reliance on cloud-based systems and enhancing response times. Addressing these gaps is essential to developing advanced fatigue monitoring solutions that exhibit accuracy and real-time capabilities while being portable, scalable, and tailored to meet individual requirements.

\section{Conclusion}\label{sec:Conclusion}
This review thoroughly examines the landscape of fatigue monitoring using wearables and AI techniques, showcasing the wide range of methodologies and technologies created to evaluate and address fatigue in different fields. Various approaches, ranging from traditional subjective measures to advanced physiological signal-based methods, provide valuable insights for detecting and understanding fatigue in this field of research. It is evident from the summary that the combination of AI and wearable devices has dramatically improved the precision, ability to monitor in real-time, and overall efficiency of fatigue detection systems. Implementing advanced machine learning and deep learning algorithms has made it possible. Although notable advancement has been in this area, several challenges still need to be addressed. These include the requirement for immediate access to data, the development of user-friendly and inconspicuous device designs, and the assurance of data quality and reliability. In addition, there are areas within this field that require further investigation. These include enhancing the accuracy of predictions, making advancements in the fusion of multimodal information, and creating AI models that can be easily explained. These research gaps provide exciting prospects for future exploration. With the emergence of edge computing and new sensor technologies, there is great potential to overcome these challenges and develop more advanced fatigue monitoring systems. These advancements hold the promise of increased reliability and robustness in the field. Given the significant progress achieved, it is crucial to continue researching and fostering innovation to address the current constraints and unlock the complete capabilities of AI and wearable technologies in fatigue monitoring. This review highlights the significance of collaborative efforts in advancing current methodologies, leading to enhanced safety and efficiency in critical areas such as healthcare, aviation, sports, and workplace safety.



\begin{thebibliography}{100}
\expandafter\ifx\csname url\endcsname\relax
  \def\url#1{\texttt{#1}}\fi
\expandafter\ifx\csname urlprefix\endcsname\relax\def\urlprefix{URL }\fi
\expandafter\ifx\csname href\endcsname\relax
  \def\href#1#2{#2} \def\path#1{#1}\fi

\bibitem{1}
A.~Frasie, M.~Bertrand-Charette, M.~Compagnat, L.~J. Bouyer, J.-S. Roy, \href{http://dx.doi.org/10.1016/j.apergo.2023.104200}{Validation of the borg cr10 scale for the evaluation of shoulder perceived fatigue during work-related tasks}, Applied Ergonomics 116, borg CR10 Scale;Generalized estimating equations;Load effects;Maximal voluntary contractions;Repeated measures;Shoulder;Task effects;Upper-extremity;Work-related;Work-related disorder; (2024).
\newline\urlprefix\url{http://dx.doi.org/10.1016/j.apergo.2023.104200}

\bibitem{2}
B.~Corcelle, F.~D. Silva, F.~Monjo, J.~Gioda, J.-P. Giacomo, G.~M. Blain, S.~S. Colson, E.~Piponnier, Immediate but not prolonged effects of submaximal eccentric vs concentric fatiguing protocols on the etiology of hamstrings’ motor performance fatigue, European Journal of Applied Physiology (2024) 1--12.

\bibitem{3}
A.~Lambay, Y.~Liu, P.~Morgan, Z.~Ji, \href{http://dx.doi.org/10.1109/HORA52670.2021.9461377}{A data-driven fatigue prediction using recurrent neural networks}, Ankara, Turkey, 2021, industrial revolutions;Inertial measurement unit;Manual material handling;Manufacturing industries;Physical fatigues;Recurrent neural network (RNN);Time series prediction;Work-related musculoskeletal disorders;.
\newline\urlprefix\url{http://dx.doi.org/10.1109/HORA52670.2021.9461377}

\bibitem{4}
H.~Zong, W.~Yi, M.~F. Antwi-Afari, Y.~Yu, \href{http://dx.doi.org/10.1016/j.ssci.2024.106529}{Fatigue in construction workers: A systematic review of causes, evaluation methods, and interventions}, Safety Science 176, construction health and safety;Construction workers;Evaluation methods;Health and safety;Mental fatigue;Occupational illness;Occupational injury;Physical fatigues;Systematic Review;Workers'; (2024).
\newline\urlprefix\url{http://dx.doi.org/10.1016/j.ssci.2024.106529}

\bibitem{5}
X.~Li, J.~Zeng, C.~Chen, H.-l. Chi, G.~Q. Shen, \href{http://dx.doi.org/10.1111/mice.12891}{Smart work package learning for decentralized fatigue monitoring through facial images}, Computer-Aided Civil and Infrastructure Engineering 38~(6) (2023) 799 -- 817, block-chain;Centralised;Decentralised;Equipment operators;Facial images;Fatigue monitoring;Learning approach;Occupational health and safety;Technical efficiency;Work packages;.
\newline\urlprefix\url{http://dx.doi.org/10.1111/mice.12891}

\bibitem{6}
D.~Ziakkas, J.~Keller, S.~Vhaduri, The role of artificial intelligence in the fatigue risk management system, in: Human Interaction and Emerging Technologies (IHIET-AI 2024): Artificial Intelligence and Future Applications, Vol. 120, 2024.

\bibitem{7}
Y.~Li, J.~He, \href{http://dx.doi.org/10.1007/s11831-024-10123-5}{A review of strategies to detect fatigue and sleep problems in aviation: Insights from artificial intelligence}, Archives of Computational Methods in EngineeringCatastrophic accidents;Fatigue detection;Fatigue problems;Multi-modal;Multimodal and aviation;Physiological signals;Recent progress;Sleep deprivation;Sleep problems;Visual recognition; (2024).
\newline\urlprefix\url{http://dx.doi.org/10.1007/s11831-024-10123-5}

\bibitem{8}
L.~Kong, K.~Xie, K.~Niu, J.~He, W.~Zhang, \href{http://dx.doi.org/10.3390/s24020455}{Remote photoplethysmography and motion tracking convolutional neural network with bidirectional long short-term memory: Non-invasive fatigue detection method based on multi-modal fusion}, Sensors 24~(2), bidirectional LSTM;Detection methods;Fatigue detection;Features fusions;Heart-rate;Intelligent traffics;Multi-modal;Multi-modal feature fusion;Multi-modal fusion;RGB cameras; (2024).
\newline\urlprefix\url{http://dx.doi.org/10.3390/s24020455}

\bibitem{9}
W.~Taylor, D.~Hill, R.~Adam, J.~Cooper, Q.~H. Abbasi, M.~Ali~Imran, \href{http://dx.doi.org/10.1109/JSEN.2024.3404637}{Hybrid sensing for fatigue detection using wearables and rf}, IEEE Sensors Journal 24~(14) (2024) 22764 -- 22772, fatigue detection;Frequency sensing;Radio frequency sensing;Radiofrequencies;Receiver;Smart healthcare;Wearable devices;Wearables;Wireless fidelities;.
\newline\urlprefix\url{http://dx.doi.org/10.1109/JSEN.2024.3404637}

\bibitem{10}
A.~Biro, A.~I. Cuesta-Vargas, L.~Szilagyi, \href{http://dx.doi.org/10.3390/s24010132}{Ai-assisted fatigue and stamina control for performance sports on imu-generated multivariate times series datasets}, Sensors 24~(1), assessment;Deep learning;Fatigue control;IMU;LSTM;Machine-learning;Multivariate time series;Performance;Stamen;Time-series data; (2024).
\newline\urlprefix\url{http://dx.doi.org/10.3390/s24010132}

\bibitem{11}
J.~Zhang, M.~Chen, Y.~Peng, S.~Li, D.~Han, S.~Ren, K.~Qin, S.~Li, T.~Han, Y.~Wang, Z.~Gao, \href{http://dx.doi.org/10.1002/btm2.10318}{Wearable biosensors for human fatigue diagnosis: A review}, Bioengineering and Translational Medicine 8~(1), biofluids;Biosensing;Communication modules;Deleterious effects;Fatigue diagnose;Human fatigue;Physical health;Real time monitoring;Sensing module;Wearable biosensor; (2023).
\newline\urlprefix\url{http://dx.doi.org/10.1002/btm2.10318}

\bibitem{12}
M.~A. Ul~Alam, Activity-aware deep cognitive fatigue assessment using wearables, Virtual, Online, Mexico, 2021, pp. 7433 -- 7436.

\bibitem{13}
C.~Goumopoulos, N.~Potha, \href{http://dx.doi.org/10.1007/s12652-021-03674-z}{Mental fatigue detection using a wearable commodity device and machine learning}, Journal of Ambient Intelligence and Humanized Computing 14~(8) (2023) 10103 -- 10121, adverse effect;Experimental study;Fatigue detection;Heart rate variability;High-accuracy;Machine-learning;Mental fatigue;Physical health;Quality of life;Wearable devices;.
\newline\urlprefix\url{http://dx.doi.org/10.1007/s12652-021-03674-z}

\bibitem{14}
A.~Lambert, A.~Soni, A.~Soukane, A.~R. Cherif, A.~Rabat, \href{http://dx.doi.org/10.1016/j.neucom.2023.126999}{Artificial intelligence modelling human mental fatigue: A comprehensive survey}, Neurocomputing 567, accurate modeling;Cognitive ability;Fatigue detection;Fatigue model;Intelligence models;Mental fatigue;Mental fatigue detection;Mental fatigue modeling;Neural activity; (2024).
\newline\urlprefix\url{http://dx.doi.org/10.1016/j.neucom.2023.126999}

\bibitem{15}
H.~Yaacob, F.~Hossain, S.~Shari, S.~K. Khare, C.~P. Ooi, U.~R. Acharya, \href{http://dx.doi.org/10.1109/ACCESS.2023.3296382}{Application of artificial intelligence techniques for brain-computer interface in mental fatigue detection: A systematic review (2011-2022)}, IEEE Access 11 (2023) 74736 -- 74758, brain-computer interface;Electroencephalogram;Fatigue detection;Hardware;Mental disorders;Mental fatigue;Mental fatigue detection;Meta-analysis;Preferred reporting item for systematic review and meta-analyse;Sleep;Systematic;Systematic Review;.
\newline\urlprefix\url{http://dx.doi.org/10.1109/ACCESS.2023.3296382}

\bibitem{16}
S.~Mu, S.~Liao, K.~Tao, Y.~Shen, \href{http://dx.doi.org/10.1016/j.bspc.2024.106127}{Intelligent fatigue detection based on hierarchical multi-scale ecg representations and hrv measures}, Biomedical Signal Processing and Control 92, deep learning;Electrocardiogram;Fatigue detection;Heart rate variability;Multi-scales;Multiscale representations;Times series;Variability measures;Wearable devices; (2024).
\newline\urlprefix\url{http://dx.doi.org/10.1016/j.bspc.2024.106127}

\bibitem{17}
B.~Russell, A.~McDaid, W.~Toscano, P.~Hume, \href{http://dx.doi.org/10.3390/s21165442}{Predicting fatigue in long duration mountain events with a single sensor and deep learning model}, Sensors 21~(16), contextual factors;Learning models;Mean absolute error;Multi-attributes;Neural network model;Physical fatigues;Prediction model;Rolling average; (2021).
\newline\urlprefix\url{http://dx.doi.org/10.3390/s21165442}

\bibitem{18}
Q.~Liu, Y.~Liu, C.~Zhang, Z.~Ruan, W.~Meng, Y.~Cai, Q.~Ai, \href{http://dx.doi.org/10.1109/JIOT.2021.3056126}{semg-based dynamic muscle fatigue classification using svm with improved whale optimization algorithm}, IEEE Internet of Things Journal 8~(23) (2021) 16835 -- 16844, classification models;Differential evolution algorithms;Initial population;Optimal parameter;Optimization algorithms;Optimization capabilities;Robot-assisted rehabilitation;Surface electromyography;.
\newline\urlprefix\url{http://dx.doi.org/10.1109/JIOT.2021.3056126}

\bibitem{19}
S.~Narteni, V.~Orani, E.~Cambiaso, M.~Rucco, M.~Mongelli, \href{http://dx.doi.org/10.1109/ACCESS.2022.3191907}{On the intersection of explainable and reliable ai for physical fatigue prediction}, IEEE Access 10 (2022) 76243 -- 76260, explainable artificial intelligence;Fatigue detection;Human safety;Learning machines;Logic learning machine;Physical fatigue detection;Physical fatigues;Reliable artificial intelligence;Service robots;Support vectors machine;.
\newline\urlprefix\url{http://dx.doi.org/10.1109/ACCESS.2022.3191907}

\bibitem{20}
A.~Giorgi, V.~Ronca, A.~Vozzi, P.~Arico, G.~Borghini, R.~Capotorto, L.~Tamborra, I.~Simonetti, S.~Sportiello, M.~Petrelli, C.~Polidori, R.~Varga, M.~van Gasteren, A.~Barua, M.~U. Ahmed, F.~Babiloni, G.~Di~Flumeri, \href{http://dx.doi.org/10.3389/fnbot.2023.1240933}{Neurophysiological mental fatigue assessment for developing user-centered artificial intelligence as a solution for autonomous driving}, Frontiers in Neurorobotics 17, autonomous driving;EEG index;Fatigue assessments;Mental fatigue;Multi-modal;Multimodal assessment;Psychophysical;Road safety;Simulated driving;User-centred; (2023).
\newline\urlprefix\url{http://dx.doi.org/10.3389/fnbot.2023.1240933}

\bibitem{21}
Y.~Jiao, C.~Zhang, X.~Chen, L.~Fu, C.~Jiang, C.~Wen, \href{http://dx.doi.org/10.1109/TITS.2023.3333252}{Driver fatigue detection using measures of heart rate variability and electrodermal activity}, IEEE Transactions on Intelligent Transportation Systems 25~(6) (2024) 5510 -- 5524, biomedical monitoring;Driver fatigue;Electrodermal activity;Fatigue detection;Fatigue level;Features extraction;Features fusions;Heart rate variability;Physiological response;.
\newline\urlprefix\url{http://dx.doi.org/10.1109/TITS.2023.3333252}

\bibitem{22}
S.~A. El-Nabi, W.~El-Shafai, E.-S.~M. El-Rabaie, K.~F. Ramadan, F.~E. Abd El-Samie, S.~Mohsen, \href{http://dx.doi.org/10.1007/s11042-023-15054-0}{Machine learning and deep learning techniques for driver fatigue and drowsiness detection: a review}, Multimedia Tools and Applications 83~(3) (2024) 9441 -- 9477, deep learning;Driver drowsiness;Drowsiness;Drowsiness detection;Eye closure;Facial Expressions;Head movements;Machine-learning;Vehicle accidents;Yawning;.
\newline\urlprefix\url{http://dx.doi.org/10.1007/s11042-023-15054-0}

\bibitem{23}
L.~Zhao, X.~Zhang, X.~Niu, J.~Sun, R.~Geng, Q.~Li, X.~Zhu, Z.~Dai, Remote photoplethysmography (rppg) based learning fatigue detection, Applied Intelligence 53~(23) (2023) 27951--27965.

\bibitem{24}
W.-J. Chang, L.-B. Chen, Y.-Z. Chiou, \href{http://dx.doi.org/10.1109/TCE.2018.2872162}{Design and implementation of a drowsiness-fatigue-detection system based on wearable smart glasses to increase road safety}, IEEE Transactions on Consumer Electronics 64~(4) (2018) 461 -- 469, design and implementations;Fatigue detection;Internet of Things (IOT);On board diagnostics;Road safety;Telematics platforms;Vehicle infotainment;Wearable devices;.
\newline\urlprefix\url{http://dx.doi.org/10.1109/TCE.2018.2872162}

\bibitem{25}
G.~Liu, C.~Dobbins, M.~DSouza, N.~Phuong, \href{http://dx.doi.org/10.1007/s00779-023-01718-z}{A machine learning approach for detecting fatigue during repetitive physical tasks}, Personal and Ubiquitous Computing 27~(6) (2023) 2103 -- 2120, adverse effect;Human bodies;Machine learning approaches;Machine-learning;Muscle strength;Musculoskeletal disorders;Physical movements;Polynomial kernels;Work-related;Workers';.
\newline\urlprefix\url{http://dx.doi.org/10.1007/s00779-023-01718-z}

\bibitem{26}
Z.~Sedighi~Maman, Y.-J. Chen, A.~Baghdadi, S.~Lombardo, L.~A. Cavuoto, F.~M. Megahed, \href{http://dx.doi.org/10.1016/j.eswa.2020.113405}{A data analytic framework for physical fatigue management using wearable sensors}, Expert Systems with Applications 155, analytic approach;Fatigue detection;Fatigue management;Physical fatigues;Random forest modeling;Sensor technologies;Task assignment models;Workplace conditions; (2020).
\newline\urlprefix\url{http://dx.doi.org/10.1016/j.eswa.2020.113405}

\bibitem{27}
M.~S.~R. Hooda, V.~Joshi, A comprehensive review of approaches to detect fatigue using machine learning techniques, 2021 (2021).

\bibitem{28}
L.~Liu, Y.~Ji, Y.~Gao, Z.~Ping, L.~Kuang, T.~Li, W.~Xu, \href{http://dx.doi.org/10.1155/2021/7799793}{A novel fatigue driving state recognition and warning method based on eeg and eog signals}, Journal of Healthcare Engineering 2021, driving state;Driving-state recognition;Early warning;Electro-oculogram;Electroencephalogram signals;Facial Expressions;Fast support vector machines;Internet of things technologies;Recognition accuracy;Recognition time; (2021).
\newline\urlprefix\url{http://dx.doi.org/10.1155/2021/7799793}

\bibitem{29}
\href{https://www.roadsafety.gov.au/action-plan/2018-2020/fatigue#:~:text=According%20to%20the%20Transport%20Accident,severe%20injuries%20on%20the%20road.}{(2020). [link]}.
\newline\urlprefix\url{https://www.roadsafety.gov.au/action-plan/2018-2020/fatigue#:~:text=According%20to%20the%20Transport%20Accident,severe%20injuries%20on%20the%20road.}

\bibitem{30}
Z.~Y.~K. Sadeghniiat-Haghighi, Fatigue management in the workplace, 2015\href {https://doi.org/10.4103/0972-6748.160915} {\path{doi:10.4103/0972-6748.160915}}.

\bibitem{31}
\href{https://www.alaskasleep.com/~alaskasl/the-relationship-between-sleep-and-industrial-accidents/}{(2022). [link]}.
\newline\urlprefix\url{https://www.alaskasleep.com/~alaskasl/the-relationship-between-sleep-and-industrial-accidents/}

\bibitem{32}
A.~Jaiswal, M.~Z. Zadeh, A.~Hebri, F.~Makedon, \href{http://dx.doi.org/10.48550/arXiv.2205.00287}{Assessing fatigue with multimodal wearable sensors and machine learning}, cognitive fatigue;Cognitive performance;Condition;Machine-learning;Multi-modal;Multimodal sensor;Physical fatigues;Physical performance;Physiological factors;Work hours; (2022).
\newline\urlprefix\url{http://dx.doi.org/10.48550/arXiv.2205.00287}

\bibitem{33}
F.~A. D.~E. Talebi, W.~P. Rogers, Environmental and work factors that drive fatigue of individual haul truck drivers, MPDI 2~(3) (2022).
\newblock \href {https://doi.org/10.3390/mining2030029} {\path{doi:10.3390/mining2030029}}.

\bibitem{34}
Y.~Hu, Z.~Liu, A.~Hou, C.~Wu, W.~Wei, Y.~Wang, M.~Liu, \href{http://dx.doi.org/10.1155/2022/4911005}{On fatigue detection for air traffic controllers based on fuzzy fusion of multiple features}, Computational and Mathematical Methods in Medicine (2022).
\newblock \href {https://doi.org/10.1155/2022/4911005} {\path{doi:10.1155/2022/4911005}}.
\newline\urlprefix\url{http://dx.doi.org/10.1155/2022/4911005}

\bibitem{35}
J.~Lei, F.~Liu, Q.~Han, Y.~Tang, L.~Zeng, M.~Chen, L.~Ye, L.~Jin, \href{http://dx.doi.org/10.1109/ITSC.2018.8569409}{Study on driving fatigue evaluation system based on short time period ecg signal}, Vol. 2018-November, Maui, HI, United states, 2018, pp. 2466 -- 2470, driving fatigue;Early warning;ECG signals;Fatigue detection;Frequency resolutions;Pre-processing;Real time;Time-periods;.
\newline\urlprefix\url{http://dx.doi.org/10.1109/ITSC.2018.8569409}

\bibitem{36}
C.~C.~D. Li, Research on exercise fatigue estimation method of pilates rehabilitation based on ecg and semg feature fusion, BMC Medical Informatics and Decision Making (2022).
\newblock \href {https://doi.org/10.1186/s12911-022-01808-7} {\path{doi:10.1186/s12911-022-01808-7}}.

\bibitem{37}
M.~Z. J. W. Z. T.~Y. Wang, C.~Lu, Research on the recognition of various muscle fatigue states in resistance strength training, Healthcare 10 (2022).
\newblock \href {https://doi.org/10.3390/healthcare10112292} {\path{doi:10.3390/healthcare10112292}}.

\bibitem{38}
C.~Zeng, Z.~Mu, Q.~Wang, \href{http://dx.doi.org/10.1155/2022/1885677}{Classifying driving fatigue by using eeg signals}, Computational Intelligence and Neuroscience 2022, brain signals;Driver fatigue;Driving fatigue;Driving platform;Driving simulation;EEG signals;Fatigue level;Interaction methods;Neuromuscular systems;Simulation equipments; (2022).
\newline\urlprefix\url{http://dx.doi.org/10.1155/2022/1885677}

\bibitem{39}
M.~T. K. Y. O. K. H. K. Y. W.~K. Mizuno, Mental fatigue caused by prolonged cognitive load associated with sympathetic hyperactivity, 2011.

\bibitem{40}
D.~B. K, K.~P.~A, R.~S, \href{http://dx.doi.org/10.1080/10255842.2021.1955104}{Automated detection of muscle fatigue conditions from cyclostationary based geometric features of surface electromyography signals}, Computer Methods in Biomechanics and Biomedical Engineering 25~(3) (2022) 320 -- 332, automated detection;Extreme learning machine;Machine learning models;Multi layer perceptron;Neuromuscular condition;Spectral correlation density;Surface electromyography;Surface electromyography signals;.
\newline\urlprefix\url{http://dx.doi.org/10.1080/10255842.2021.1955104}

\bibitem{41}
W.~Gomberg, Measuring the fatigue factor, 2016.

\bibitem{42}
M.~Papakostas, V.~Kanal, M.~Abujelala, K.~Tsiakas, F.~Makedon, \href{http://dx.doi.org/10.1145/3316782.3322772}{Physical fatigue detection through emg wearables and subjective user reports - a machine learning approach towards adaptive rehabilitation}, Rhodes, Greece, 2019, pp. 475 -- 481, daily lives;Dataset;EMG measurement;Machine learning approaches;Multiple sclerosis;Physical fatigues;Physiological monitoring;User Modeling;.
\newline\urlprefix\url{http://dx.doi.org/10.1145/3316782.3322772}

\bibitem{43}
S.~Murugan, J.~Selvaraj, A.~Sahayadhas, \href{http://dx.doi.org/10.1007/s13246-020-00853-8}{Detection and analysis: driver state with electrocardiogram (ecg)}, Physical and Engineering Sciences in Medicine 43~(2) (2020) 525--537.
\newblock \href {https://doi.org/10.1007/s13246-020-00853-8} {\path{doi:10.1007/s13246-020-00853-8}}.
\newline\urlprefix\url{http://dx.doi.org/10.1007/s13246-020-00853-8}

\bibitem{44}
\href{https://www.engineeringvillage.com/}{(2022). [link]}.
\newline\urlprefix\url{https://www.engineeringvillage.com/}

\bibitem{45}
S.-P. Cheon, S.-J. Kang, \href{http://dx.doi.org/10.1109/IVS.2017.7995924}{Sensor-based driver condition recognition using support vector machine for the detection of driver drowsiness}, Vol.~0, Redondo Beach, CA, United states, 2017, pp. 1517 -- 1522, classification accuracy;Condition recognition;Driver drowsiness;Fatal accidents;Feature vectors;Photoplethysmography (PPG);Precision and recall;Wearable devices;.
\newline\urlprefix\url{http://dx.doi.org/10.1109/IVS.2017.7995924}

\bibitem{46}
W.-L. Zheng, K.~Gao, G.~Li, W.~Liu, C.~Liu, J.-Q. Liu, G.~Wang, B.-L. Lu, \href{http://dx.doi.org/10.1109/TITS.2018.2889962}{Vigilance estimation using a wearable eog device in real driving environment}, IEEE Transactions on Intelligent Transportation Systems 21~(1) (2020) 170 -- 184, drowsy driving;Dry electrode;forehead EOG;Real-world drivings;Vigilance estimation;Wearable devices;.
\newline\urlprefix\url{http://dx.doi.org/10.1109/TITS.2018.2889962}

\bibitem{47}
Q.~Massoz, T.~Langohr, C.~Francois, J.~G. Verly, \href{http://dx.doi.org/10.1109/WACV.2016.7477715}{The ulg multimodality drowsiness database (called drozy) and examples of use}, Lake Placid, NY, United states, 2016, facial Expressions;Monitoring system;Multi-modality;Multiple modalities;Near Infrared;Passenger compartment;Range images;Road transportation;.
\newline\urlprefix\url{http://dx.doi.org/10.1109/WACV.2016.7477715}

\bibitem{48}
P.~Maciel, J.~Dantas, C.~Melo, P.~Pereira, F.~Oliveira, J.~Araujo, R.~Matos, \href{http://dx.doi.org/10.1007/s40860-021-00154-1}{A survey on reliability and availability modeling of edge, fog, and cloud computing}, Journal of Reliable Intelligent Environments 8~(3) (2022) 227 -- 245, cloud computing architectures;Cloud environments;Computing paradigm;Context aware services;Internet of thing (IOT);Potential researches;Reliability and availability;Technology landscapes;.
\newline\urlprefix\url{http://dx.doi.org/10.1007/s40860-021-00154-1}

\bibitem{49}
M.~Gusev, \href{http://dx.doi.org/10.1109/COMPSAC51774.2021.00269}{What makes dew computing more than edge computing for internet of things}, Virtual, Online, Spain, 2021, pp. 1795 -- 1800, architectural approach;Cloud architectures;Computing solutions;IOT applications;Research and development;.
\newline\urlprefix\url{http://dx.doi.org/10.1109/COMPSAC51774.2021.00269}

\bibitem{50}
F.~Wulf, M.~Westner, S.~Strahringer, \href{http://dx.doi.org/10.17705/1CAIS.04843}{Cloud computing adoption: A literature review on what is new and what still needs to be addressed}, Communications of the Association for Information Systems 48 (2021) 523 -- 561, decision makers;Determinant factors;Environmental characteristic;Individual characteristics;Literature reviews;Multi-clouds;.
\newline\urlprefix\url{http://dx.doi.org/10.17705/1CAIS.04843}

\bibitem{51}
M.~Kulkarni, P.~Deshpande, S.~Nalbalwar, A.~Nandgaonkar, \href{http://dx.doi.org/10.1007/978-981-19-2719-5_56}{Cloud computing based workload prediction using cluster machine learning approach}, Vol. 303 SIST, Virtual, Online, 2022, pp. 591 -- 601, 'current;Cloud-computing;Cluster machines;Clusterings;Dynamic resources;Machine learning approaches;ML;Specific problems;Workload;Workload predictions;.
\newline\urlprefix\url{http://dx.doi.org/10.1007/978-981-19-2719-5_56}

\bibitem{52}
J.-W. Jeong, W.~Lee, Y.-J. Kim, \href{http://dx.doi.org/10.3390/s22010104}{A realtime wearable physiological monitoring system for home-based healthcare applications}, Sensors 22~(1) (2022).
\newblock \href {https://doi.org/10.3390/s22010104} {\path{doi:10.3390/s22010104}}.
\newline\urlprefix\url{http://dx.doi.org/10.3390/s22010104}

\bibitem{53}
C.~Goumopoulos, N.~Potha, \href{http://dx.doi.org/10.1007/s12652-021-03674-z}{Mental fatigue detection using a wearable commodity device and machine learning}, Journal of Ambient Intelligence and Humanized Computing 14~(8) (2023) 10103 -- 10121, adverse effect;Experimental study;Fatigue detection;Heart rate variability;High-accuracy;Machine-learning;Mental fatigue;Physical health;Quality of life;Wearable devices;.
\newline\urlprefix\url{http://dx.doi.org/10.1007/s12652-021-03674-z}

\bibitem{54}
M.~Kalanadhabhatta, C.~Min, A.~Montanari, F.~Kawsar, \href{http://dx.doi.org/10.1007/978-3-030-99194-4_14}{Fatigueset: A multi-modal dataset for modeling mental fatigue and fatigability}, Vol. 431 LNICST, Virtual, Online, 2022, pp. 204 -- 217, cognitive performance;Fatigue management;Management systems;Management technologies;Mental fatigue;Multi-modal dataset;Multimodal sensing;Performance;Physiological markers;Real-world;.
\newline\urlprefix\url{http://dx.doi.org/10.1007/978-3-030-99194-4_14}

\bibitem{55}
B.~Zhu, Z.~Zhou, S.~Yu, X.~Liang, Y.~Xie, Q.~Sun, Review of phonocardiogram signal analysis: Insights from the physionet/cinc challenge 2016 database, Electronics 13~(16) (2024) 3222.

\bibitem{56}
S.~Hong, W.~Zhang, C.~Sun, Y.~Zhou, H.~Li, Practical lessons on 12-lead ecg classification: Meta-analysis of methods from physionet/computing in cardiology challenge 2020, Frontiers in Physiology 12 (2022) 811661.

\bibitem{57}
S.~Katsigiannis, N.~Ramzan, \href{http://dx.doi.org/10.1109/JBHI.2017.2688239}{Dreamer: A database for emotion recognition through eeg and ecg signals from wireless low-cost off-the-shelf devices}, IEEE Journal of Biomedical and Health Informatics 22~(1) (2018) 98 -- 107, affect;Affect recognition;emotion;Physiological signals;Wireless devices;.
\newline\urlprefix\url{http://dx.doi.org/10.1109/JBHI.2017.2688239}

\bibitem{58}
T.~Song, W.~Zheng, P.~Song, Z.~Cui, \href{http://dx.doi.org/10.1109/TAFFC.2018.2817622}{Eeg emotion recognition using dynamical graph convolutional neural networks}, IEEE Transactions on Affective Computing 11~(3) (2020) 532 -- 541, adjacency matrices;Biological neural networks;dynamical convolutional neural networks (DGCNN);Emotion recognition;graph convolutional neural networks (GCNN);.
\newline\urlprefix\url{http://dx.doi.org/10.1109/TAFFC.2018.2817622}

\bibitem{59}
H.~Zeng, Q.~Wu, Y.~Jin, H.~Zheng, M.~Li, Y.~Zhao, H.~Hu, W.~Kong, \href{http://dx.doi.org/10.1109/TIM.2022.3216829}{Siam-gcan: A siamese graph convolutional attention network for eeg emotion recognition}, IEEE Transactions on Instrumentation and Measurement 71, attention mechanisms;Brain modeling;Convolutional networks;Effective performance;Emotion recognition;Features extraction;Frequency-domain analysis;Graph convolutional network;Siamese network; (2022).
\newline\urlprefix\url{http://dx.doi.org/10.1109/TIM.2022.3216829}

\bibitem{60}
D.~Yuan, J.~Yue, X.~Xiong, Y.~Jiang, P.~Zan, C.~Li, A regression method for eeg-based cross-dataset fatigue detection, Frontiers in Physiology 14 (2023) 1196919.

\bibitem{61}
A.~Moniri, D.~Terracina, J.~Rodriguez-Manzano, P.~H. Strutton, P.~Georgiou, \href{http://dx.doi.org/10.1109/TBME.2020.3012783}{Real-time forecasting of semg features for trunk muscle fatigue using machine learning}, IEEE Transactions on Biomedical Engineering 68~(2) (2021) 718 -- 727, accuracy and precision;Mean absolute percentage error;Muscle activities;Muscle activity and fatigues;Real-time forecasting;Standard deviation;Surface electromyography;Trunk muscle fatigues;.
\newline\urlprefix\url{http://dx.doi.org/10.1109/TBME.2020.3012783}

\bibitem{62}
S.~Abtahi, M.~Omidyeganeh, S.~Shirmohammadi, B.~Hariri, \href{http://dx.doi.org/10.1145/2557642.2563678}{Yawdd: A yawning detection dataset}, Singapore, Singapore, 2014, pp. 24 -- 28, detection methods;Illumination conditions;Testing algorithm;Video datasets;Yawning dataset;.
\newline\urlprefix\url{http://dx.doi.org/10.1145/2557642.2563678}

\bibitem{63}
B.~K. Sava, Y.~Becerikli, \href{http://dx.doi.org/10.1109/ACCESS.2020.2963960}{Real time driver fatigue detection system based on multi-task connn}, IEEE Access 8 (2020) 12491 -- 12498, driver drowsiness;Driver fatigue;Driver's behavior;Fatigue detection;Intelligent vehicle systems;PERCLOS;Single models;Vehicle accidents;.
\newline\urlprefix\url{http://dx.doi.org/10.1109/ACCESS.2020.2963960}

\bibitem{64}
Z.~Sun, Y.~Miao, J.~Y. Jeon, Y.~Kong, G.~Park, \href{http://dx.doi.org/10.1016/j.engappai.2023.106981}{Facial feature fusion convolutional neural network for driver fatigue detection}, Engineering Applications of Artificial Intelligence 126, deep learning;Driver fatigue;Fatigue detection;Features fusions;Interaction modules;Low qualities;Multi-stream;Multi-stream network;Performance;Stream networks; (2023).
\newline\urlprefix\url{http://dx.doi.org/10.1016/j.engappai.2023.106981}

\bibitem{65}
B.~Kir~Sava, Y.~Becerikli, \href{http://dx.doi.org/10.1007/978-3-031-26852-6_42}{Driver fatigue detection via eye state analyses based on deep learning approach}, Vol. 629 LNNS, Castelo Branco, Portugal, 2023, pp. 452 -- 462, analysis system;Convolutional neural network;Detection system;Driver fatigue;Eye state analysis;Eye state classification;Fatigue detection;Learning approach;Learning methods;State classification;.
\newline\urlprefix\url{http://dx.doi.org/10.1007/978-3-031-26852-6_42}

\bibitem{66}
Q.~Massoz, T.~Langohr, C.~Francois, J.~G. Verly, \href{http://dx.doi.org/10.1109/WACV.2016.7477715}{The ulg multimodality drowsiness database (called drozy) and examples of use}, Lake Placid, NY, United states, 2016, facial Expressions;Monitoring system;Multi-modality;Multiple modalities;Near Infrared;Passenger compartment;Range images;Road transportation;.
\newline\urlprefix\url{http://dx.doi.org/10.1109/WACV.2016.7477715}

\bibitem{67}
P.~Krishnan, S.~Yaacob, A.~P. Krishnan, M.~Rizon, C.~K. Ang, Eeg based drowsiness detection using relative band power and short-time fourier transform, Journal of Robotics Networks and Artificial Life 7~(3) (2020) 147--151.

\bibitem{68}
C.~M. S. R. M. R. N. R.~A. Martines, S.~Annaheim, Fatigue monitoring through wearables: A state-of-the-art review, Frontiers in Physiology 12 (2021) 790292.

\bibitem{69}
I.~Volker, C.~Kirchner, O.~L. Bock, \href{http://dx.doi.org/10.1080/00140139.2015.1110622}{On the relationship between subjective and objective measures of fatigue}, Ergonomics 59~(9) (2016) 1259 -- 1263, facial Expressions;Fatigue detection;Fatigue measures;multidimensional construct;Physical fatigues;Subjective and objective measures;Subjective methods;workplace;.
\newline\urlprefix\url{http://dx.doi.org/10.1080/00140139.2015.1110622}

\bibitem{70}
M.~E. McCauley, P.~McCauley, S.~M. Riedy, S.~Banks, A.~J. Ecker, L.~V. Kalachev, S.~Rangan, D.~F. Dinges, H.~P. Van~Dongen, \href{http://dx.doi.org/10.1016/j.trf.2021.04.006}{Fatigue risk management based on self-reported fatigue: Expanding a biomathematical model of fatigue-related performance deficits to also predict subjective sleepiness}, Transportation Research Part F: Traffic Psychology and Behaviour 79 (2021) 94 -- 106, aviation operations;Biomathematical modeling;Biomathematical models;Laboratory studies;Prediction accuracy;Psychomotor vigilance tests;Quantitative tool;Sleep deprivation;.
\newline\urlprefix\url{http://dx.doi.org/10.1016/j.trf.2021.04.006}

\bibitem{71}
M.~Hashemi, B.~Farahani, F.~Firouzi, \href{http://dx.doi.org/10.1109/COINS49042.2020.9191418}{Towards safer roads: A deep learning-based multimodal fatigue monitoring system}, Barcelona, Spain, 2020, driving conditions;Fatigue monitoring system;Learning technology;Multi-modal techniques;Physiological informations;Property damage;State Detection;Transportation industry;.
\newline\urlprefix\url{http://dx.doi.org/10.1109/COINS49042.2020.9191418}

\bibitem{72}
Y.~Jiao, Y.~Deng, Y.~Luo, B.-L. Lu, \href{http://dx.doi.org/10.1016/j.neucom.2019.05.108}{Driver sleepiness detection from eeg and eog signals using gan and lstm networks}, Neurocomputing 408 (2020) 100 -- 111, alpha blocking;Alpha waves;Classifier performance;Continuous Wavelet Transform;LSTM;Physiological signals;Sleepiness detection;Time and frequency domains;.
\newline\urlprefix\url{http://dx.doi.org/10.1016/j.neucom.2019.05.108}

\bibitem{73}
Z.~Zeng, Z.~Huang, K.~Leng, W.~Han, H.~Niu, Y.~Yu, Q.~Ling, J.~Liu, Z.~Wu, J.~Zang, \href{http://dx.doi.org/10.1021/acssensors.9b02451}{Nonintrusive monitoring of mental fatigue status using epidermal electronic systems and machine-learning algorithms}, ACS Sensors 5~(5) (2020) 1305 -- 1313, decision-tree algorithm;Fatigue classifications;Individual performance;Non-intrusive monitoring;Physiological features;Physiological signals;Predictive accuracy;Predictive modeling;.
\newline\urlprefix\url{http://dx.doi.org/10.1021/acssensors.9b02451}

\bibitem{74}
M.~Choi, G.~Koo, M.~Seo, S.~W. Kim, \href{http://dx.doi.org/10.1109/TIM.2017.2779329}{Wearable device-based system to monitor a driver's stress, fatigue, and drowsiness}, IEEE Transactions on Instrumentation and Measurement 67~(3) (2018) 634 -- 645, classification accuracy;Classification methods;Generalization performance;Optimal feature sets;Photoplethysmogram;Physiological informations;Physiological sensors;Wearable devices;.
\newline\urlprefix\url{http://dx.doi.org/10.1109/TIM.2017.2779329}

\bibitem{75}
V.~Vijayan, J.~Connolly, J.~Condell, N.~McKelvey, P.~Gardiner, \href{http://dx.doi.org/10.3390/s21165589}{Review of wearable devices and data collection considerations for connected health}, Sensors 21~(16), ambulatory movements;Disease diagnosis;Functional abilities;Human body movement;Human-activity detection;Self assessment;Sensor technologies;Wearable devices; (2021).
\newline\urlprefix\url{http://dx.doi.org/10.3390/s21165589}

\bibitem{76}
J.~Tan, X.-y. Shi, W.~Guo, \href{http://dx.doi.org/10.1007/978-981-15-0644-4_62}{Research on driving fatigue level using ecg signal from smart bracelet}, Vol. 617, Guilin, China, 2020, pp. 799 -- 810, bP neural networks;Comprehensive indices;Data compensation;Driving fatigue;Experimental analysis;Fatigue level;Fusion theory;Multi-index;.
\newline\urlprefix\url{http://dx.doi.org/10.1007/978-981-15-0644-4_62}

\bibitem{77}
Y.~Zhang, Y.~Hu, N.~Jiang, A.~K. Yetisen, \href{http://dx.doi.org/10.1016/j.bios.2022.114825}{Wearable artificial intelligence biosensor networks}, Biosensors and Bioelectronics 219, biosensing;Biosensor network;High quality;Machine-learning;Personalized medicines;Quality healthcare;Smart phones;Smartphone-based readout;Wearable biosensor;Well being; (2023).
\newline\urlprefix\url{http://dx.doi.org/10.1016/j.bios.2022.114825}

\bibitem{78}
Z.~Sedighi~Maman, Y.-J. Chen, A.~Baghdadi, S.~Lombardo, L.~A. Cavuoto, F.~M. Megahed, \href{http://dx.doi.org/10.1016/j.eswa.2020.113405}{A data analytic framework for physical fatigue management using wearable sensors}, Expert Systems with Applications 155, analytic approach;Fatigue detection;Fatigue management;Physical fatigues;Random forest modeling;Sensor technologies;Task assignment models;Workplace conditions; (2020).
\newline\urlprefix\url{http://dx.doi.org/10.1016/j.eswa.2020.113405}

\bibitem{79}
R.~Chattopadhyay, G.~Pradhan, S.~Panchanathan, \href{http://dx.doi.org/10.1109/IMTC.2010.5488258}{Towards fatigue and intensity measurement framework during continuous repetitive activities}, 2010, pp. 1341 -- 1346, continuous monitoring;Daily life activities;Intensity measurements;Machine learning techniques;Physiological state;Physiological status;Sensor technologies;Surface electromyogram;.
\newline\urlprefix\url{http://dx.doi.org/10.1109/IMTC.2010.5488258}

\bibitem{80}
A.~Baghdadi, L.~A. Cavuoto, A.~Jones-Farmer, S.~E. Rigdon, E.~T. Esfahani, F.~M. Megahed, \href{http://dx.doi.org/10.1080/00224065.2019.1640097}{Monitoring worker fatigue using wearable devices: A case study to detect changes in gait parameters}, Journal of Quality Technology 53~(1) (2021) 47 -- 71, clustering;Demographic characteristics;Exploratory data analysis;Gait parameters;Non-parametric;Research questions;Time series clustering;Wearable devices;.
\newline\urlprefix\url{http://dx.doi.org/10.1080/00224065.2019.1640097}

\bibitem{81}
S.~Huang, J.~Li, P.~Zhang, W.~Zhang, \href{http://dx.doi.org/10.1016/j.ijmedinf.2018.08.010}{Detection of mental fatigue state with wearable ecg devices}, International Journal of Medical Informatics 119 (2018) 39 -- 46, east Asian countries;Feature combination;Heart rate variability;K nearest neighbor (KNN);Mental fatigue;Public health issues;Public universities;Wearable devices;.
\newline\urlprefix\url{http://dx.doi.org/10.1016/j.ijmedinf.2018.08.010}

\bibitem{82}
L.~Zhao, M.~Li, Z.~He, S.~Ye, H.~Qin, X.~Zhu, Z.~Dai, \href{http://dx.doi.org/10.1016/j.measurement.2022.111648}{Data-driven learning fatigue detection system: A multimodal fusion approach of ecg (electrocardiogram) and video signals}, Measurement: Journal of the International Measurement Confederation 201, data driven;Deep learning;Detection system;Electrocardiogram signal;Fatigue detection;Features fusions;Learning analytic;Physiological signals;Video;Video signal; (2022).
\newline\urlprefix\url{http://dx.doi.org/10.1016/j.measurement.2022.111648}

\bibitem{83}
L.~Wang, J.~Li, Y.~Wang, \href{http://dx.doi.org/10.1109/ACCESS.2019.2956652}{Modeling and recognition of driving fatigue state based on r-r intervals of ecg data}, IEEE Access 7 (2019) 175584 -- 175593, conditional variance;Driving fatigue;Modeling and recognition;Multiple indicators;Prevention of accidents;Recognition algorithm;Recognition models;RR intervals;.
\newline\urlprefix\url{http://dx.doi.org/10.1109/ACCESS.2019.2956652}

\bibitem{84}
L.~B.~E. Butkeviciute, A.~Michalkovic, Ecg signal features classification for the mental fatigue recognition, MDPI (2022).
\newblock \href {https://doi.org/10.3390/math10183395} {\path{doi:10.3390/math10183395}}.

\bibitem{85}
M.~Babaeian, M.~Mozumdar, \href{http://dx.doi.org/10.1109/CCWC.2019.8666467}{Driver drowsiness detection algorithms using electrocardiogram data analysis}, Las Vegas, NV, United states, 2019, pp. 1 -- 6, biomedical signal analysis;Driver drowsiness;Heart rate variability;Heart rate variations;Innovative techniques;K nearest neighbor (KNN);K-nearest neighbor method;Short fourier transforms;.
\newline\urlprefix\url{http://dx.doi.org/10.1109/CCWC.2019.8666467}

\bibitem{86}
T.~Kundinger, N.~Sofra, A.~Riener, \href{http://dx.doi.org/10.3390/s20041029}{Assessment of the potential of wrist-worn wearable sensors for driver drowsiness detection}, Sensors (Switzerland) 20~(4), automated driving;Binary classification;Driver state;Driving behavior parameters;Drowsiness detection;Machine learning approaches;Physiological measurement;Physiological measures; (2020).
\newline\urlprefix\url{http://dx.doi.org/10.3390/s20041029}

\bibitem{87}
Z.~Ahmad, M.~N. Jamaludin, K.~Soeed, \href{http://dx.doi.org/10.1109/IECBES.2018.8626605}{Prediction of exhaustion threshold based on ecg features using the artificial neural network model}, Demak-Isthmus Bridge, Jalan Keruing, Sejingkat, Kuching, Malaysia, 2018, pp. 523 -- 528, artificial neural network modeling;Exercise-induced fatigues;Exhaustion threshold;Optimum parameters;Physical fatigues;Quantitative measurement;Real time monitoring;Specific information;.
\newline\urlprefix\url{http://dx.doi.org/10.1109/IECBES.2018.8626605}

\bibitem{88}
R.~Gao, H.~Yan, Study on the non-fatigue and fatigue states of orchard workers based on electrocardiogram signal analysis, Nature (2022).

\bibitem{89}
Y.~Bai, Y.~Guan, J.~Q. Shi, W.~Ng, \href{http://dx.doi.org/10.1145/3460421.3480429}{Towards automated fatigue assessment using wearable sensing and mixed-effects models}, Virtual, Online, United states, 2020, pp. 129 -- 131, demographic information;Fatigue assessments;Key factors;Mental energy;Mixed effects models;Patient health;Personalizations;Physical energy;Subjective perceptions;Wearable sensing;.
\newline\urlprefix\url{http://dx.doi.org/10.1145/3460421.3480429}

\bibitem{90}
J.~Wang, Y.~Sun, S.~Sun, \href{http://dx.doi.org/10.1109/ACCESS.2020.3038422}{Recognition of muscle fatigue status based on improved wavelet threshold and cnn-svm}, IEEE Access 8 (2020) 207914 -- 207922, classification accuracy;Mean power frequency;Statistical features;Support vector machine algorithm;Surface electromyography;Time and frequency domains;Training data sets;Wavelet threshold;.
\newline\urlprefix\url{http://dx.doi.org/10.1109/ACCESS.2020.3038422}

\bibitem{91}
J.~Wang, S.~Sun, Y.~Sun, \href{http://dx.doi.org/10.3390/s21196369}{A muscle fatigue classification model based on lstm and improved wavelet packet threshold}, Sensors 21~(19), anaerobic thresholds;Memory network;Model-based OPC;Muscle fatigues;Novel methods;Performance;Recognition models;Surface electromyography;Surface electromyography signals;Wavelet Packet; (2021).
\newline\urlprefix\url{http://dx.doi.org/10.3390/s21196369}

\bibitem{92}
L.~Boon-Leng, L.~Dae-Seok, L.~Boon-Giin, \href{http://dx.doi.org/10.1109/TENCON.2015.7372932}{Mobile-based wearable-type of driver fatigue detection by gsr and emg}, in: 35th IEEE Region 10 Conference (TENCON), 2015.
\newblock \href {https://doi.org/10.1109/TENCON.2015.7372932} {\path{doi:10.1109/TENCON.2015.7372932}}.
\newline\urlprefix\url{http://dx.doi.org/10.1109/TENCON.2015.7372932}

\bibitem{93}
Z.~Feng, Y.~Wang, L.~Liu, \href{http://dx.doi.org/10.1109/ICSP54964.2022.9778406}{Muscle fatigue detection method based on feature extraction and deep learning}, Xi'an, China, 2022, pp. 97 -- 100, deep learning;Detection methods;Fatigue detection;Frequency domains;Intelligent medical;Muscle fatigues;Signal-processing;Surface electromyography;Time domain;Time frequency;.
\newline\urlprefix\url{http://dx.doi.org/10.1109/ICSP54964.2022.9778406}

\bibitem{94}
J.~Sun, G.~Liu, Y.~Sun, K.~Lin, Z.~Zhou, J.~Cai, Application of surface electromyography in exercise fatigue: A review, Frontiers in Systems Neuroscience 16 (2022).
\newblock \href {https://doi.org/10.3389/fnsys.2022.893275} {\path{doi:10.3389/fnsys.2022.893275}}.

\bibitem{95}
P.~Karthick, D.~M. Ghosh, S.~Ramakrishnan, \href{http://dx.doi.org/10.1016/j.cmpb.2017.10.024}{Surface electromyography based muscle fatigue detection using high-resolution time-frequency methods and machine learning algorithms}, Computer Methods and Programs in Biomedicine 154 (2018) 45 -- 56, eMBD;Muscle fatigues;S transforms;Surface electromyography;Time frequency features;.
\newline\urlprefix\url{http://dx.doi.org/10.1016/j.cmpb.2017.10.024}

\bibitem{96}
M.~Papakostas, V.~Kanal, M.~Abujelala, K.~Tsiakas, F.~Makedon, \href{http://dx.doi.org/10.1145/3316782.3322772}{Physical fatigue detection through emg wearables and subjective user reports - a machine learning approach towards adaptive rehabilitation}, Rhodes, Greece, 2019, pp. 475 -- 481, daily lives;Dataset;EMG measurement;Machine learning approaches;Multiple sclerosis;Physical fatigues;Physiological monitoring;User Modeling;.
\newline\urlprefix\url{http://dx.doi.org/10.1145/3316782.3322772}

\bibitem{97}
S.~Shahmoradi, A.~Zare, S.~Behzadipour, \href{http://dx.doi.org/10.1109/ICBME.2017.8430264}{Fatigue status recognition in a post-stroke rehabilitation exercise with semg signal}, Tehran, Iran, 2017, maximum voluntary contraction;Post stroke patients;Post-stroke rehabilitation;Reaching task;Rehabilitation exercise;Semg signals;Stroke rehabilitation;Xbox kinect;.
\newline\urlprefix\url{http://dx.doi.org/10.1109/ICBME.2017.8430264}

\bibitem{98}
R.~Rejith, R.~Raj, K.~Sivanandan, \href{http://dx.doi.org/10.1109/INVENTIVE.2016.7823251}{Estimation of elbow kinematics under fatigue and non-fatigue conditions using mlpn network based on semg signal}, Vol.~1, Coimbatore, India, 2016, pp. IEEE; IEEE (EDS) Madras Section --, elbow kinematics;Integrated EMG;Multi-layered Perceptron;Surface EMG;Zero-crossings;.
\newline\urlprefix\url{http://dx.doi.org/10.1109/INVENTIVE.2016.7823251}

\bibitem{99}
R.~Foong, K.~K. Ang, Z.~Zhang, C.~Quek, \href{http://dx.doi.org/10.1088/1741-2552/ab255d}{An iterative cross-subject negative-unlabeled learning algorithm for quantifying passive fatigue}, Journal of Neural Engineering 16~(5), driving fatigue;Fatigue level;Fatigue scores;Manual labeling;Negative correlation;NU learning;Pu learning;Repeated measures; (2019).
\newline\urlprefix\url{http://dx.doi.org/10.1088/1741-2552/ab255d}

\bibitem{100}
U.~Budak, V.~Bajaj, Y.~Akbulut, O.~Atila, A.~Sengur, \href{http://dx.doi.org/10.1109/JSEN.2019.2917850}{An effective hybrid model for eeg-based drowsiness detection}, IEEE Sensors Journal 19~(17) (2019) 7624 -- 7631, cross-validation tests;Drowsiness detection;EEG signals;Energy distributions;Instantaneous frequency;Mean and standard deviations;Statistical features;Vehicular accidents;.
\newline\urlprefix\url{http://dx.doi.org/10.1109/JSEN.2019.2917850}

\bibitem{101}
T.~Nakamura, Y.~D. Alqurashi, M.~J. Morrell, D.~P. Mandic, \href{http://dx.doi.org/10.1109/IJCNN.2018.8489723}{Automatic detection of drowsiness using in-ear eeg}, Vol. 2018-July, Rio de Janeiro, Brazil, 2018, automatic Detection;Binary-class support vector machines;Classification accuracy;Complexity science;Proof of concept;Research communities;Sleep monitoring;Structural complexity;.
\newline\urlprefix\url{http://dx.doi.org/10.1109/IJCNN.2018.8489723}

\bibitem{102}
J.~Liu, C.~Zhang, C.~Zheng, \href{http://dx.doi.org/10.1016/j.bspc.2010.01.001}{Eeg-based estimation of mental fatigue by using kpca-hmm and complexity parameters}, Biomedical Signal Processing and Control 5~(2) (2010) 124 -- 130, approximate entropy;Electro-encephalogram (EEG);Kolmogorov complexity;KPCA-HMM;Mental fatigue;.
\newline\urlprefix\url{http://dx.doi.org/10.1016/j.bspc.2010.01.001}

\bibitem{103}
V.~Bajaj, S.~Taran, S.~K. Khare, A.~Sengur, \href{http://dx.doi.org/10.1016/j.apacoust.2020.107224}{Feature extraction method for classification of alertness and drowsiness states eeg signals}, Applied Acoustics 163, drowsiness detection;Electroencephalogram signals;Extreme learning machine;Feature extraction methods;Kruskal-Wallis tests;Mean and standard deviations;Non stationary characteristics;Q-factors; (2020).
\newline\urlprefix\url{http://dx.doi.org/10.1016/j.apacoust.2020.107224}

\bibitem{104}
J.~Kim, C.~Jeong, M.~Jung, J.~Park, D.~Jung, \href{http://dx.doi.org/10.1007/s12239-013-0106-z}{Highly reliable driving workload analysis using driver electroencephalogram (eeg) activities during driving}, International Journal of Automotive Technology 14~(6) (2013) 965 -- 970, driving workload;Electro-encephalogram (EEG);Engine vibration;Environmental factors;Human-vehicle interface;Statistical reliability;Statistical significance;Vibration properties;.
\newline\urlprefix\url{http://dx.doi.org/10.1007/s12239-013-0106-z}

\bibitem{105}
T.~G. Monteiro, C.~Skourup, H.~Zhang, \href{http://dx.doi.org/10.1109/THMS.2019.2938156}{Using eeg for mental fatigue assessment: A comprehensive look into the current state of the art}, IEEE Transactions on Human-Machine Systems 49~(6) (2019) 599 -- 610, classification algorithm;Electro-encephalogram (EEG);Human machine interaction;Kernel partial least squares;Machine learning techniques;Mental fatigue;Objective and subjective measures;Sensor fusion;.
\newline\urlprefix\url{http://dx.doi.org/10.1109/THMS.2019.2938156}

\bibitem{106}
A.~Lau-Zhu, M.~P. Lau, G.~McLoughlin, Mobile eeg in research on neurodevelopmental disorders: Opportunities and challenges, Developmental Cognitive Neuroscience 36 (2019) 100635.

\bibitem{107}
A.~Aryal, A.~Ghahramani, B.~Becerik-Gerber, \href{http://dx.doi.org/10.1016/j.autcon.2017.03.003}{Monitoring fatigue in construction workers using physiological measurements}, Automation in Construction 82 (2017) 154 -- 165, construction safety;Fatigue assessments;Physiological monitoring;Thermoregulation;Wearable sensing;.
\newline\urlprefix\url{http://dx.doi.org/10.1016/j.autcon.2017.03.003}

\bibitem{108}
S.-P. Cheon, S.-J. Kang, \href{http://dx.doi.org/10.1109/IVS.2017.7995924}{Sensor-based driver condition recognition using support vector machine for the detection of driver drowsiness}, Vol.~0, Redondo Beach, CA, United states, 2017, pp. 1517 -- 1522, classification accuracy;Condition recognition;Driver drowsiness;Fatal accidents;Feature vectors;Photoplethysmography (PPG);Precision and recall;Wearable devices;.
\newline\urlprefix\url{http://dx.doi.org/10.1109/IVS.2017.7995924}

\bibitem{109}
M.~Choi, G.~Koo, M.~Seo, S.~W. Kim, \href{http://dx.doi.org/10.1109/TIM.2017.2779329}{Wearable device-based system to monitor a driver's stress, fatigue, and drowsiness}, IEEE Transactions on Instrumentation and Measurement 67~(3) (2018) 634 -- 645, classification accuracy;Classification methods;Generalization performance;Optimal feature sets;Photoplethysmogram;Physiological informations;Physiological sensors;Wearable devices;.
\newline\urlprefix\url{http://dx.doi.org/10.1109/TIM.2017.2779329}

\bibitem{110}
V.~Chandel, D.~S. Jani, S.~Mukhopadhyay, S.~Khandelwal, D.~Jaiswal, A.~Pal, \href{http://dx.doi.org/10.1145/3144730.3144732}{C2p: An unobtrusive smartwatch-based platform for automatic background monitoring of fatigue}, Delft, Netherlands, 2018, pp. 19 -- 24, cardiopulmonary;Cardiopulmonary disease;Computationally efficient;Dyspnea;Functional capacity;Physiological measurement;Smartwatch;Wearable;.
\newline\urlprefix\url{http://dx.doi.org/10.1145/3144730.3144732}

\bibitem{111}
D.~Ichwana, R.~Z. Ikhlas, S.~Ekariani, \href{http://dx.doi.org/10.1109/ICITSI.2018.8696039}{Heart rate monitoring system during physical exercise for fatigue warning using non-invasive wearable sensor}, Bandung - Padang, Indonesia, 2018, pp. 497 -- 502, android;Android applications;Bluetooth communications;Heart-rate detection;Heart-rate monitoring;Photoplethysmograph;Physical activity;Physical exercise;.
\newline\urlprefix\url{http://dx.doi.org/10.1109/ICITSI.2018.8696039}

\bibitem{112}
M.~A. Ul~Alam, \href{http://dx.doi.org/10.1109/EMBC46164.2021.9630729}{Activity-aware deep cognitive fatigue assessment using wearables}, Virtual, Online, Mexico, 2021, pp. 7433 -- 7436.
\newline\urlprefix\url{http://dx.doi.org/10.1109/EMBC46164.2021.9630729}

\bibitem{113}
H.~F. Posada-Quintero, K.~H. Chon, \href{http://dx.doi.org/10.3390/s20020479}{Innovations in electrodermal activity data collection and signal processing: A systematic review}, Sensors (Switzerland) 20~(2), data collection;Data quality assessment;Electrodermal activity;New devices;Pathophysiological;Signal processing technique;Sympathetic function;Systematic Review; (2020).
\newline\urlprefix\url{http://dx.doi.org/10.3390/s20020479}

\bibitem{114}
R.~Zangroniz, A.~Martinez-Rodrigo, J.~M. Pastor, M.~T. Lopez, A.~Fernandez-Caballero, \href{http://dx.doi.org/10.3390/s17102324}{Electrodermal activity sensor for classification of calm/distress condition}, Sensors (Switzerland) 17~(10), arousal;Calmness;Distress;Electrodermal activity;Valence;Wearable; (2017).
\newline\urlprefix\url{http://dx.doi.org/10.3390/s17102324}

\bibitem{115}
M.-B. Hossain, H.~F. Posada-Quintero, Y.~Kong, R.~McNaboe, K.~H. Chon, \href{http://dx.doi.org/10.1016/j.bspc.2022.103483}{Automatic motion artifact detection in electrodermal activity data using machine learning}, Biomedical Signal Processing and Control 74, accurate analysis;Artifact detection;Diverse applications;Electrodermal activity;Emotion recognition;Features selection;LOSO validation;Motion artifact;Statistical frequency;Stress recognition; (2022).
\newline\urlprefix\url{http://dx.doi.org/10.1016/j.bspc.2022.103483}

\bibitem{116}
A.~Jaiswal, M.~Z. Zadeh, A.~Hebri, F.~Makedon, \href{http://dx.doi.org/10.48550/arXiv.2205.00287}{Assessing fatigue with multimodal wearable sensors and machine learning}, cognitive fatigue;Cognitive performance;Condition;Machine-learning;Multi-modal;Multimodal sensor;Physical fatigues;Physical performance;Physiological factors;Work hours; (2022).
\newline\urlprefix\url{http://dx.doi.org/10.48550/arXiv.2205.00287}

\bibitem{117}
R.~Varandas, R.~Lima, S.~B.~I. Badia, H.~Silva, H.~Gamboa, \href{http://dx.doi.org/10.3390/s22114010}{Automatic cognitive fatigue detection using wearable fnirs and machine learning}, Sensors 22~(11), cognitive fatigue;Cognitive state;Fatigue detection;Functional near infrared spectroscopy;Impact performance;Learning settings;Machine learning algorithms;Machine-learning;Multiple domains;Wearable devices; (2022).
\newline\urlprefix\url{http://dx.doi.org/10.3390/s22114010}

\bibitem{118}
V.~Camomilla, E.~Bergamini, S.~Fantozzi, G.~Vannozzi, \href{http://dx.doi.org/10.3390/s18030873}{Trends supporting the in-field use of wearable inertial sensors for sport performance evaluation: A systematic review}, Sensors (Switzerland) 18~(3), activity classifications;Athlete;Capacity assessment;Device characteristics;Inertial sensor;Performance assessment;Performance indicators;Technological development; (2018).
\newline\urlprefix\url{http://dx.doi.org/10.3390/s18030873}

\bibitem{119}
M.~J. Pinto-Bernal, C.~A. Cifuentes, O.~Perdomo, M.~Rincon-Roncancio, M.~Munera, \href{http://dx.doi.org/10.3390/s21196401}{A data-driven approach to physical fatigue management using wearable sensors to classify four diagnostic fatigue states}, Sensors 21~(19), classification models;Data-driven approach;EMG;Fatigue diagnose;Fatigue management;Inertial measurements units;Performance;Physical exercise;Physical fatigues;Walking rehabilitation; (2021).
\newline\urlprefix\url{http://dx.doi.org/10.3390/s21196401}

\bibitem{120}
P.~Li, R.~Meziane, M.~J.-D. Otis, H.~Ezzaidi, P.~Cardou, \href{http://dx.doi.org/10.1109/ROSE.2014.6952983}{A smart safety helmet using imu and eeg sensors for worker fatigue detection}, Timisoara, Romania, 2014, pp. 55 -- 60, accident avoidance;Anomalous behavior;Fatigue detection;Head motion;Human machine interaction;Inertial measurement unit;Risk of accidents;Vision based system;.
\newline\urlprefix\url{http://dx.doi.org/10.1109/ROSE.2014.6952983}

\bibitem{121}
S.~S. Bangaru, C.~Wang, F.~Aghazadeh, \href{http://dx.doi.org/10.3390/s22249729}{Automated and continuous fatigue monitoring in construction workers using forearm emg and imu wearable sensors and recurrent neural network}, Sensors 22~(24), activity recognition;Aerobic fatigue threshold;Construction labor shortage;Construction labours;Fatigue monitoring;Fatigue threshold;Labor shortages;Muscle activities;Musculoskeletal disorders;Oxygen prediction;Scaffold building;Work-related;Work-related musculoskeletal disorder; (2022).
\newline\urlprefix\url{http://dx.doi.org/10.3390/s22249729}

\bibitem{122}
A.~Biro, A.~I. Cuesta-Vargas, L.~Szilagyi, \href{http://dx.doi.org/10.3390/s24010132}{Ai-assisted fatigue and stamina control for performance sports on imu-generated multivariate times series datasets}, Sensors 24~(1), assessment;Deep learning;Fatigue control;IMU;LSTM;Machine-learning;Multivariate time series;Performance;Stamen;Time-series data; (2024).
\newline\urlprefix\url{http://dx.doi.org/10.3390/s24010132}

\bibitem{123}
Y.~Dong, H.~Zhu, Z.~Zhou, W.~Fu, \href{http://dx.doi.org/10.1109/ICTech58362.2023.00044}{Fatigue monitoring and awakening system based on eye electrical and head movement parameters monitoring}, Wuhan, China, 2023, pp. 174 -- 178, awaken;EOG signal;Eye electricity;Fatigue monitoring;Head movements;Head posture;Human bodies;Parameter monitoring;Three-axis gyroscopes;Wake up;.
\newline\urlprefix\url{http://dx.doi.org/10.1109/ICTech58362.2023.00044}

\bibitem{124}
S.~Murugan, P.~K. Sivakumar, C.~Kavitha, A.~Harichandran, W.-C. Lai, \href{http://dx.doi.org/10.3390/s23062944}{An electro-oculogram (eog) sensor’s ability to detect driver hypovigilance using machine learning}, Sensors 23~(6), behavioural measurements;Drowsiness;Drowsiness detection;Electro-oculogram;Financial loss;Machine-learning;Maximum accuracies;Physical state;Signal;Visual inattention; (2023).
\newline\urlprefix\url{http://dx.doi.org/10.3390/s23062944}

\bibitem{125}
Y.~Tian, J.~Cao, \href{http://dx.doi.org/10.1186/s13640-021-00575-1}{Fatigue driving detection based on electrooculography: a review}, Eurasip Journal on Image and Video Processing 2021~(1), behaviour monitoring;Driving behavior monitoring;Driving behaviour;Driving state;Electro-oculogram;Electrooculogram signal;Fatigue driving state monitoring;Literature reviews;Monitoring system;State monitoring; (2021).
\newline\urlprefix\url{http://dx.doi.org/10.1186/s13640-021-00575-1}

\bibitem{126}
M.~Kolodziej, P.~Tarnowski, D.~J. Sawicki, A.~Majkowski, R.~J. Rak, A.~Bala, A.~Pluta, \href{http://dx.doi.org/10.1109/JSEN.2020.3012404}{Fatigue detection caused by office work with the use of eog signal}, IEEE Sensors Journal 20~(24) (2020) 15213 -- 15223, accuracy of classifications;Automatic method;Classification accuracy;Fatigue detection;Repetitive works;Research results;User independents;Work efficiency;.
\newline\urlprefix\url{http://dx.doi.org/10.1109/JSEN.2020.3012404}

\bibitem{127}
B.~Tag, A.~W. Vargo, A.~Gupta, G.~Chernyshov, K.~Kunze, T.~Dingler, \href{http://dx.doi.org/10.1145/3290605.3300694}{Continuous alertness assessments: Using eog glasses to unobtrusively monitor fatigue levels in-the-wild}, Glasgow, United kingdom, 2019, pp. ACM SIGCHI --, circadian computing;Cognitive functions;Continuous monitoring;Eye blink;Homeostatic process;Lab conditions;Robust modeling;Self assessment;.
\newline\urlprefix\url{http://dx.doi.org/10.1145/3290605.3300694}

\bibitem{128}
J.~Shi, K.~Wang, \href{http://dx.doi.org/10.1016/j.bspc.2023.104744}{Fatigue driving detection method based on time-space-frequency features of multimodal signals}, Biomedical Signal Processing and Control 84, correlation coefficient;Deep learning;Detection methods;Fatigue detection;Features fusions;Multi-modality;Multimodality fatigue detection;Performance;Root mean square errors;Signal features; (2023).
\newline\urlprefix\url{http://dx.doi.org/10.1016/j.bspc.2023.104744}

\bibitem{129}
C.~Zhou, J.~Li, \href{http://dx.doi.org/10.1109/CCDC52312.2021.9602058}{A real-time driver fatigue monitoring system based on lightweight convolutional neural network}, Kunming, China, 2021, pp. 1548 -- 1553, accident rate;Convolutional neural network;Driver fatigue monitoring;Driver monitoring system;Facial feature;Fatigue detection;Fatigue monitoring system;Jetson TX2;Mobilenetv3;Real- time;.
\newline\urlprefix\url{http://dx.doi.org/10.1109/CCDC52312.2021.9602058}

\bibitem{130}
A.~Jaiswal, M.~Z. Zadeh, A.~Hebri, F.~Makedon, \href{http://dx.doi.org/10.48550/arXiv.2205.00287}{Assessing fatigue with multimodal wearable sensors and machine learning}, cognitive fatigue;Cognitive performance;Condition;Machine-learning;Multi-modal;Multimodal sensor;Physical fatigues;Physical performance;Physiological factors;Work hours; (2022).
\newline\urlprefix\url{http://dx.doi.org/10.48550/arXiv.2205.00287}

\bibitem{131}
S.~Chen, K.~Xu, X.~Yao, J.~Ge, L.~Li, S.~Zhu, Z.~Li, \href{http://dx.doi.org/10.1016/j.cmpb.2021.106451}{Information fusion and multi-classifier system for miner fatigue recognition in plateau environments based on electrocardiography and electromyography signals}, Computer Methods and Programs in Biomedicine 211, electrocardiography;Electromyography;Electromyography signals;Fatigue recognition;Field test;Multimodal information fusion;Performance;Physiological indices;Physiological signals;Principal-component analysis; (2021).
\newline\urlprefix\url{http://dx.doi.org/10.1016/j.cmpb.2021.106451}

\bibitem{132}
S.~Chen, K.~Xu, X.~Yao, S.~Zhu, B.~Zhang, H.~Zhou, X.~Guo, B.~Zhao, \href{http://dx.doi.org/10.1016/j.compbiomed.2021.104413}{Psychophysiological data-driven multi-feature information fusion and recognition of miner fatigue in high-altitude and cold areas}, Computers in Biology and Medicine 133, cold area;Electrocardiography;Electromyography;Fatigue recognition;Feature information;Field test;Machine-learning;Multi features;Psychophysiological;Support vectors machine; (2021).
\newline\urlprefix\url{http://dx.doi.org/10.1016/j.compbiomed.2021.104413}

\bibitem{143}
K.~K. Singh, V.~K. Jha, \href{http://dx.doi.org/10.1002/cpe.6926}{An optimized novel public cloud system to secure the medical record from third parties}, Vol.~35, 2023, african buffalo optimization;Cloud systems;Cloud-computing;Encryption and decryption;Medical cloud computing;Medical record;Optimisations;Public cloud system;Public clouds;Third parties;.
\newline\urlprefix\url{http://dx.doi.org/10.1002/cpe.6926}

\bibitem{144}
D.~Lin, Y.~Tang, \href{http://dx.doi.org/10.1109/JSYST.2019.2923991}{Edge computing-based mobile health system: Network architecture and resource allocation}, IEEE Systems Journal 14~(2) (2020) 1716 -- 1727, cellular network;Computational capability;ITS architecture;ITS data;Medical data;Medical treatment;Mobile health application;Mobile health systems;.
\newline\urlprefix\url{http://dx.doi.org/10.1109/JSYST.2019.2923991}

\bibitem{145}
R.~Zanetti, A.~Arza, A.~Aminifar, D.~Atienza, \href{http://dx.doi.org/10.1109/TBME.2021.3092206}{Real-time eeg-based cognitive workload monitoring on wearable devices}, IEEE Transactions on Biomedical Engineering 69~(1) (2022) 265 -- 277, cognitive workloads;Human machine interaction;Model size reductions;Optimization strategy;Processing strategies;Real-time data processing;Resourceconstrained devices;Sensitivity and specificity;.
\newline\urlprefix\url{http://dx.doi.org/10.1109/TBME.2021.3092206}

\bibitem{135}
M.~K. K.~A. Subasi, Muscle fatigue detection in emg using time–frequency methods, ica and neural networks, Springer Link (2010).

\bibitem{136}
L.~Boon-Leng, L.~Dae-Seok, L.~Boon-Giin, \href{http://dx.doi.org/10.1109/TENCON.2015.7372932}{Mobile-based wearable-type of driver fatigue detection by gsr and emg}, Vol. 2016-January, Macau, China, 2015, bluetooth low energies (BTLE);Fatigue monitoring;Frequency Analysis;Frequency domains;Galvanic skin response;mobile;Mobile applications;Time domain features;.
\newline\urlprefix\url{http://dx.doi.org/10.1109/TENCON.2015.7372932}

\bibitem{137}
Z.~Feng, Y.~Wang, L.~Liu, \href{http://dx.doi.org/10.1109/ICSP54964.2022.9778406}{Muscle fatigue detection method based on feature extraction and deep learning}, Xi'an, China, 2022, pp. 97 -- 100, deep learning;Detection methods;Fatigue detection;Frequency domains;Intelligent medical;Muscle fatigues;Signal-processing;Surface electromyography;Time domain;Time frequency;.
\newline\urlprefix\url{http://dx.doi.org/10.1109/ICSP54964.2022.9778406}

\bibitem{140}
M.~S.~H. Lee, J.~Lee, \href{http://dx.doi.org/10.3390/electronics8020192}{Using wearable ecg/ppg sensors for driver drowsiness detection based on distinguishable pattern of recurrence plots}, MDPI (2019).
\newblock \href {https://doi.org/10.3390/electronics8020192} {\path{doi:10.3390/electronics8020192}}.
\newline\urlprefix\url{http://dx.doi.org/10.3390/electronics8020192}

\bibitem{141}
R.~Khushaba, S.~Kodagoda, S.~Lal, G.~Dissanayake, \href{http://dx.doi.org/10.1109/TBME.2010.2077291}{Driver drowsiness classification using fuzzy wavelet-packet-based feature-extraction algorithm}, IEEE Transactions on Biomedical Engineering 58~(1) (2011) 121 -- 131, bio-signal processing;Classification accuracy;Driver drowsiness;Electro-encephalogram (EEG);Electrocardiogram signal;Feature extraction algorithms;Feature extraction methods;Wavelet packet transforms;.
\newline\urlprefix\url{http://dx.doi.org/10.1109/TBME.2010.2077291}

\bibitem{170}
S.~Leonhardt, L.~Leicht, D.~Teichmann, \href{http://dx.doi.org/10.3390/s18093080}{Unobtrusive vital sign monitoring in automotive environments—a review}, Sensors (Switzerland) 18~(9), ballistocardiography;Car seats;Magnetic impedances;State monitoring;Steering wheel;Unobtrusive monitoring; (2018).
\newline\urlprefix\url{http://dx.doi.org/10.3390/s18093080}

\bibitem{142}
J.-M. A.~J. Gielen, Feature extraction and evaluation for driver drowsiness detection based on thermoregulation, Applied Science (2019).

\bibitem{171}
J.~Vicente, P.~Laguna, A.~Bartra, R.~Bailon, \href{http://dx.doi.org/10.1007/s11517-015-1448-7}{Drowsiness detection using heart rate variability}, Medical and Biological Engineering and Computing 54~(6) (2016) 927 -- 937, autonomic nervous system;Heart rate variability;Impaired driving;Linear discriminant analysis;Sleep debt;.
\newline\urlprefix\url{http://dx.doi.org/10.1007/s11517-015-1448-7}

\bibitem{146}
T.~Kalia, \href{https://outdesign.medium.com/6-key-challenges-of-wearable-product-development-49717d88c684}{6 key challenges of wearable product development}, in: Medium, 2017.
\newline\urlprefix\url{https://outdesign.medium.com/6-key-challenges-of-wearable-product-development-49717d88c684}

\bibitem{147}
J.~Amor, C.~James, \href{http://dx.doi.org/10.4018/978-1-5225-5484-4.ch025}{Unobtrusive wearable technology for health monitoring}, 2018, pp. 562 -- 577, health and wellness;Health monitoring;Long term monitoring;Short term;Unobtrusive monitoring;Wearable devices;.
\newline\urlprefix\url{http://dx.doi.org/10.4018/978-1-5225-5484-4.ch025}

\bibitem{148}
H.~C. Ates, P.~Q. Nguyen, L.~Gonzalez-Macia, E.~Morales-Narvaez, F.~Guder, J.~J. Collins, C.~Dincer, \href{http://dx.doi.org/10.1038/s41578-022-00460-x}{End-to-end design of wearable sensors}, Nature Reviews Materials 7~(11) (2022) 887 -- 907, biochemical information;Chemical and biological sensors;Clinical diagnostics;End-to-end design;Face masks;Minimally invasive;Parkinson's disease;Physical sensors;Real- time;Wearable devices;.
\newline\urlprefix\url{http://dx.doi.org/10.1038/s41578-022-00460-x}

\bibitem{149}
A.~Kiaghadi, M.~Baima, J.~Gummeson, T.~Andrew, D.~Ganesan, \href{http://dx.doi.org/10.1145/3274783.3274845}{Fabric as a sensor: Towards unobtrusive sensing of human behavior with triboelectric textiles}, Shenzhen, China, 2018, pp. 199 -- 210, activity classifications;Athletic performance;Computing machinery;Discriminative features;Layered architecture;Real-world performance;Triboelectric signals;Ubiquitous sensing;.
\newline\urlprefix\url{http://dx.doi.org/10.1145/3274783.3274845}

\bibitem{150}
S.~Costantini, F.~Persia, L.~De~Lauretis, An intelligent ecosystem to improve patient monitoring using wearables and artificial intelligence, Vol. 3204, Bologna, Italy, 2022, pp. 141 -- 153, complex event processing;Complex events;Creative Commons;Creatives;Event Processing;Health condition;Images processing;Patient health;Smart applications;Wearables;.

\bibitem{151}
N.~S. E.~M. Chang, E.~F. Gillespie, Truthfulness in patient-reported outcomes: Factors affecting patients’ responses and impact on data quality (2019).

\bibitem{152}
M.~S. G. C. D. K. C.~J. Kelly, A.~Karthikesalingam, Key challenges for delivering clinical impact with artificial intelligence, BMC Medicine (2019).
\newblock \href {https://doi.org/10.1186/s12916-019-1426-2} {\path{doi:10.1186/s12916-019-1426-2}}.

\bibitem{153}
X.~Qiu, F.~Tian, Q.~Shi, Q.~Zhao, B.~Hu, \href{http://dx.doi.org/10.1109/BIBM49941.2020.9313129}{Designing and application of wearable fatigue detection system based on multimodal physiological signals}, Virtual, Seoul, Korea, Republic of, 2020, pp. 716 -- 722, electro-encephalogram (EEG);Environmental noise;Fatigue assessments;Pervasive applications;Physiological data;Physiological signal acquisitions;Physiological signals;Portable applications;.
\newline\urlprefix\url{http://dx.doi.org/10.1109/BIBM49941.2020.9313129}

\bibitem{154}
P.~S. Game, V.~Vaze, M.~Emmanuel, \href{http://dx.doi.org/10.1007/s12065-019-00267-w}{Optimized decision tree rules using divergence based grey wolf optimization for big data classification in health care}, Evolutionary Intelligence 15~(2) (2022) 971 -- 987, artificial bee colony algorithms;Data classification;Data classification models;Decision tree classifiers;Medical data;Optimizers;Particle swarm optimization algorithm;Principle component analysis;.
\newline\urlprefix\url{http://dx.doi.org/10.1007/s12065-019-00267-w}

\bibitem{155}
S.~Kundu, Ai in medicine must be explainable, World View (2021).
\newblock \href {https://doi.org/10.1038/s41591-021-01461-z} {\path{doi:10.1038/s41591-021-01461-z}}.

\bibitem{156}
\href{https://towardsdatascience.com/}{Towards data science} (2022).
\newline\urlprefix\url{https://towardsdatascience.com/}

\bibitem{157}
F.~Zhou, A.~Alsaid, M.~Blommer, R.~Curry, R.~Swaminathan, D.~Kochhar, W.~Talamonti, L.~Tijerina, \href{http://dx.doi.org/10.1080/10447318.2021.1965774}{Predicting driver fatigue in monotonous automated driving with explanation using gpboost and shap}, International Journal of Human-Computer Interaction 38~(8) (2022) 719 -- 729, behavioral measures;Gaussian Processes;Individual prediction;Machine learning models;Machine learning techniques;Mean absolute error;Predictor variables;Root mean squared errors;.
\newline\urlprefix\url{http://dx.doi.org/10.1080/10447318.2021.1965774}

\bibitem{158}
M.~Abdar, F.~Pourpanah, S.~Hussain, D.~Rezazadegan, L.~Liu, M.~Ghavamzadeh, P.~Fieguth, X.~Cao, A.~Khosravi, U.~R. Acharya, V.~Makarenkov, S.~Nahavandi, \href{http://dx.doi.org/10.1016/j.inffus.2021.05.008}{A review of uncertainty quantification in deep learning: Techniques, applications and challenges}, Information Fusion 76 (2021) 243 -- 297, ensemble learning;Fundamental research;NAtural language processing;Optimization and decision makings;Real-world problem;Science and engineering;Text classification;Uncertainty quantifications;.
\newline\urlprefix\url{http://dx.doi.org/10.1016/j.inffus.2021.05.008}

\bibitem{159}
P.~Tabarisaadi, A.~Khosravi, S.~Nahavandi, M.~Shafie-Khah, J.~P.~S. Catalao, \href{http://dx.doi.org/10.1109/TNNLS.2022.3213315}{An optimized uncertainty-aware training framework for neural networks}, IEEE Transactions on Neural Networks and Learning Systems 35~(5) (2024) 6928 -- 6935, bayes method;Deep neural network;Prediction algorithms;Predictive models;Training framework;Uncertainty;Uncertainty accuracy;Uncertainty quantification;Uncertainty quantifications;.
\newline\urlprefix\url{http://dx.doi.org/10.1109/TNNLS.2022.3213315}

\bibitem{160}
J.~M. Dolezal, A.~Srisuwananukorn, D.~Karpeyev, S.~Ramesh, S.~Kochanny, B.~Cody, A.~Mansfield, S.~Rakshit, R.~Bansa, M.~Bois, A.~O. Bungum, J.~J. Schulte, E.~E. Vokes, M.~C. Garassino, A.~N. Husain, A.~T. Pearson, Uncertainty-informed deep learning models enable high-confidence predictions for digital histopathology, clinical users;Confidence predictions;High confidence;Learning models;Medical settings;Predictive uncertainty;Real-world;Squamous cell carcinoma;Uncertainty;Uncertainty quantifications; (2022).

\bibitem{161}
E.~F. M. A.~R. Lara, R.~Echeveste, Addressing fairness in artificial intelligence for medical imaging, Nature Communications (2022).

\bibitem{162}
X.~Jin, L.~Li, F.~Dang, X.~Chen, Y.~Liu, \href{http://dx.doi.org/10.1016/j.dsp.2021.103146}{A survey on edge computing for wearable technology}, Digital Signal Processing: A Review Journal 125, compact size;Computation scheduling;Computing technology;Device resources;Future research directions;Information perception;Smart wearables;Wearable devices; (2022).
\newline\urlprefix\url{http://dx.doi.org/10.1016/j.dsp.2021.103146}

\bibitem{163}
H.~J. Eoh, M.~K. Chung, S.-H. Kim, \href{http://dx.doi.org/10.1016/j.ergon.2004.09.006}{Electroencephalographic study of drowsiness in simulated driving with sleep deprivation}, International Journal of Industrial Ergonomics 35~(4) (2005) 307 -- 320, car movements;Drowsiness;Mental alertness;Sleep deprivation;.
\newline\urlprefix\url{http://dx.doi.org/10.1016/j.ergon.2004.09.006}

\bibitem{164}
R.~Paradiso, C.~Belloc, G.~Loriga, N.~Taccini, Wearable healthcare systems, new frontiers of e-textile, Vol. 117, Belfast, United kingdom, 2005, pp. 9 -- 16, environmental conditions;Fabric sensors;Integrated electronics;Physiological parameters;Physiological reactions;Piezoresistive materials;Wearable;Wearable healthcare systems;.

\bibitem{165}
T.~Chalder, G.~Berelowitz, T.~Pawlikowska, L.~Watts, S.~Wessely, D.~Wright, E.~Wallace, \href{https://www.sciencedirect.com/science/article/pii/002239999390081P}{Development of a fatigue scale}, Journal of Psychosomatic Research 37~(2) (1993) 147--153.
\newblock \href {https://doi.org/10.1016/0022-3999(93)90081-P} {\path{doi:10.1016/0022-3999(93)90081-P}}.
\newline\urlprefix\url{https://www.sciencedirect.com/science/article/pii/002239999390081P}

\bibitem{166}
H.~Al-Libawy, A.~Al-Ataby, W.~Al-Nuaimy, M.~A. Al-Taee, \href{http://dx.doi.org/10.1109/SSD.2016.7473750}{Hrv-based operator fatigue analysis and classification using wearable sensors}, Leipzig, Germany, 2016, pp. 268 -- 273, alertness;Classification methods;Fatigue assessments;Heart rate variability;Prediction accuracy;Skin temperatures;Supervised machine learning;Wearable devices;.
\newline\urlprefix\url{http://dx.doi.org/10.1109/SSD.2016.7473750}

\bibitem{167}
D.~Sandberg, \href{http://dx.doi.org/10.1109/ITSC.2011.6082939}{The performance of driver sleepiness indicators as a function of interval length}, Washington, DC, United states, 2011, pp. 1735 -- 1740, blink duration;Lateral positions;Sleepiness detection;Standard deviation;.
\newline\urlprefix\url{http://dx.doi.org/10.1109/ITSC.2011.6082939}

\bibitem{168}
S.~Shirmohammadi, K.~Barbe, D.~Grimaldi, S.~Rapuano, S.~Grassini, \href{http://dx.doi.org/10.1109/MIM.2016.7579063}{Instrumentation and measurement in medical, biomedical, and healthcare systems}, IEEE Instrumentation and Measurement Magazine 19~(5) (2016) 6 -- 12, customer dissatisfaction;Health-care system;Incorrect measurements;Instrumentation and measurements;Measurement methods;Medical measurement;Medical practitioner;Medical professionals;.
\newline\urlprefix\url{http://dx.doi.org/10.1109/MIM.2016.7579063}

\bibitem{169}
K.~Pal, B.~V. Patel, \href{http://dx.doi.org/10.1109/ICCMC48092.2020.ICCMC-00016}{Data classification with k-fold cross validation and holdout accuracy estimation methods with 5 different machine learning techniques}, Erode, India, 2020, pp. 83 -- 87, accuracy estimation;Decision-tree algorithm;Feature extraction techniques;K fold cross validations;Machine learning methods;Machine learning techniques;NAtural language processing;Random forest methods;.
\newline\urlprefix\url{http://dx.doi.org/10.1109/ICCMC48092.2020.ICCMC-00016}

\bibitem{172}
F.~Qu, N.~Dang, B.~Furht, M.~Nojoumian, \href{http://dx.doi.org/10.1186/s40537-024-00890-0}{Comprehensive study of driver behavior monitoring systems using computer vision and machine learning techniques}, Journal of Big Data 11~(1), autonomous Vehicles;Behavior monitoring system;Behaviour classification;Driver behavior classification;Driver behavior monitoring system;Driver's behavior;Machine-learning;Neural-networks;Vision learning; (2024).
\newline\urlprefix\url{http://dx.doi.org/10.1186/s40537-024-00890-0}

\bibitem{173}
N.~Chiapello, I.~Gerlero, V.~Gatteschi, F.~Lamberti, \href{http://dx.doi.org/10.1109/ICCE56470.2023.10043418}{Driver state monitoring through driving style estimation}, Vol. 2023-January, Las Vegas, NV, United states, 2023, 'current;Aggressive driving;Driving behaviour;Driving situations;Driving styles;Estimation techniques;Impaired driving;Monitoring system;Road safety;State monitoring;.
\newline\urlprefix\url{http://dx.doi.org/10.1109/ICCE56470.2023.10043418}

\bibitem{174}
L.~Zhao, M.~Li, Z.~He, S.~Ye, H.~Qin, X.~Zhu, Z.~Dai, \href{http://dx.doi.org/10.1016/j.measurement.2022.111648}{Data-driven learning fatigue detection system: A multimodal fusion approach of ecg (electrocardiogram) and video signals}, Measurement: Journal of the International Measurement Confederation 201, data driven;Deep learning;Detection system;Electrocardiogram signal;Fatigue detection;Features fusions;Learning analytic;Physiological signals;Video;Video signal; (2022).
\newline\urlprefix\url{http://dx.doi.org/10.1016/j.measurement.2022.111648}

\bibitem{175}
H.~Lee, J.~Lee, M.~Shin, \href{https://api.semanticscholar.org/CorpusID:115439049}{Using wearable ecg/ppg sensors for driver drowsiness detection based on distinguishable pattern of recurrence plots}, Electronics (2019).
\newline\urlprefix\url{https://api.semanticscholar.org/CorpusID:115439049}

\bibitem{176}
T.~G. Monteiro, C.~Skourup, H.~Zhang, \href{http://dx.doi.org/10.1109/ACCESS.2020.2976601}{Optimizing cnn hyperparameters for mental fatigue assessment in demanding maritime operations}, IEEE Access 8 (2020) 40402 -- 40412, bayesian optimization;Classification tasks;Gaussian Processes;Generalization performance;Maritime operation;Mental fatigue;Network structures;Tree-like structures;.
\newline\urlprefix\url{http://dx.doi.org/10.1109/ACCESS.2020.2976601}

\bibitem{177}
Z.~Zhou, V.~Tam, K.~Lui, E.~Lam, X.~Hu, A.~Yuen, N.~Law, \href{http://dx.doi.org/10.1109/ICALT49669.2020.00097}{A sophisticated platform for learning analytics with wearable devices}, Virtual, Online, Estonia, 2020, pp. 300 -- 304, learning Activity;Learning effectiveness;Learning models;Mobile applications;Platform for learning;Real-time learning;Sleeping monitoring;Wearable devices;.
\newline\urlprefix\url{http://dx.doi.org/10.1109/ICALT49669.2020.00097}

\bibitem{178}
X.~Zhu, S.~Ye, L.~Zhao, Z.~Dai, \href{http://dx.doi.org/10.3390/s21062003}{Hybrid attention cascade network for facial expression recognition}, Sensors 21~(6) (2021) 1 -- 16, competitive performance;Facial expression recognition;Facial feature points;Natural environments;Position information;Recognition accuracy;State-of-the-art methods;Uneven illuminations;.
\newline\urlprefix\url{http://dx.doi.org/10.3390/s21062003}

\bibitem{179}
Q.~Massoz, T.~Langohr, C.~Francois, J.~G. Verly, \href{http://dx.doi.org/10.1109/WACV.2016.7477715}{The ulg multimodality drowsiness database (called drozy) and examples of use}, Lake Placid, NY, United states, 2016, facial Expressions;Monitoring system;Multi-modality;Multiple modalities;Near Infrared;Passenger compartment;Range images;Road transportation;.
\newline\urlprefix\url{http://dx.doi.org/10.1109/WACV.2016.7477715}

\bibitem{180}
C.~B.~S. Maior, M.~J. d.~C. Moura, J.~M.~M. Santana, I.~D. Lins, \href{http://dx.doi.org/10.1016/j.eswa.2020.113505}{Real-time classification for autonomous drowsiness detection using eye aspect ratio}, Expert Systems with Applications 158, automated systems;Biological approach;Complex environments;Drowsiness detection;Electro-oculogram;Human supervision;Neuroscience literature;Performance indicators; (2020).
\newline\urlprefix\url{http://dx.doi.org/10.1016/j.eswa.2020.113505}

\bibitem{181}
A.~Lambay, Y.~Liu, P.~L. Morgan, Z.~Ji, Machine learning assisted human fatigue detection, monitoring, and recovery, Digital Engineering (2024) 100004.

\end{thebibliography}

\appendix
\section{}

\onecolumn
\begin{landscape}
\small 
\begin{longtable}[c]{@{}cccccccc@{}}
\caption{Summary of Proposed Method and Overall Performance of Various Modalities for Fatigue Monitoring.}
\label{tab:Apendix}\\
\hline
\textbf{Reference} & \textbf{Application} & \textbf{Methods Used} & \textbf{Key Findings} & \textbf{Challenges/Limitations} \\ \hline
\endfirsthead

\hline
\textbf{Reference} & \textbf{Application} & \textbf{Methods Used} & \textbf{Key Findings} & \textbf{Challenges/Limitations} \\ \hline
\endhead

\hline
\hline
\endfoot

\hline
\endlastfoot

Taylor et al. {\cite{9}} & Healthcare & \begin{tabular}[c]{@{}c@{}}Wearable bracelet,   \\ RF sensing, \\ Random Forest, \\ ResNet\end{tabular} & \begin{tabular}[c]{@{}c@{}}100\% accuracy in \\ detecting fatigue\end{tabular} & Limited to simulated fatigue \\\hline
Zakharov et al. {\cite{6}} & Aviation & \begin{tabular}[c]{@{}c@{}}AI, \\ FRMS, \\ EEG, \\ Smartwatches\end{tabular} & \begin{tabular}[c]{@{}c@{}}Enhanced real-time detection \\ and prediction of fatigue\end{tabular} & Data reliability and privacy concerns \\\hline
Biro et al. {\cite{10}} & Sports & \begin{tabular}[c]{@{}c@{}}IMUs, \\ Random Forest, \\ Gradient Boosting,   \\ LSTM\end{tabular} & \begin{tabular}[c]{@{}c@{}}High predictive accuracy \\ for fatigue and stamina\end{tabular} & Real-time model adaptation challenges \\\hline
Lambay et al. {\cite{181}} & Manufacturing & ML Techniques & \begin{tabular}[c]{@{}c@{}}Complexity in holistic approaches \\ for fatigue detection\end{tabular} & Varying task-specific fatigue patterns \\\hline
Kong et al. {\cite{8}} & Transportation & \begin{tabular}[c]{@{}c@{}}RGB cameras, \\ CNN, \\ BiLSTM\end{tabular} & \begin{tabular}[c]{@{}c@{}}98.2\% accuracy in \\ driver fatigue detection\end{tabular} & Sensitivity to lighting and motion conditions \\\hline
Mu et al. {\cite{16}} & Exercise & \begin{tabular}[c]{@{}c@{}}ECG, \\ HRV, \\ Transformer framework\end{tabular} & \begin{tabular}[c]{@{}c@{}}94\% accuracy in exercise-\\ induced fatigue detection\end{tabular} & High computational cost \\\hline
Zong et al. {\cite{4}} & Construction & Systematic review & \begin{tabular}[c]{@{}c@{}}Identification of causes and   \\ interventions for fatigue\end{tabular} & Lack of practical implementation \\\hline
Zhang et al. {\cite{11}} & Healthcare & Wearable biosensors & \begin{tabular}[c]{@{}c@{}}Advances in non-invasive \\ fatigue diagnosis\end{tabular} & Performance limits of biosensors \\\hline
Alam et al. {\cite{12}} & Cognitive Work & \begin{tabular}[c]{@{}c@{}}Wearables, \\ RNN\end{tabular} & \begin{tabular}[c]{@{}c@{}}19\% improvement in \\ cognitive fatigue estimation\end{tabular} & Limited sample size \\\hline
Goumpougosoulos et al. {\cite{13}} & Workplace & \begin{tabular}[c]{@{}c@{}}Wearables, \\ HRV analysis, \\ SVM\end{tabular} & \begin{tabular}[c]{@{}c@{}}High accuracy in mental \\ fatigue detection\end{tabular} & Limited to HRV features \\\hline
Russell et al. {\cite{17}} & Sports & \begin{tabular}[c]{@{}c@{}}Single sensor, \\ CNN\end{tabular} & \begin{tabular}[c]{@{}c@{}}Practical applications \\ in field fatigue testing\end{tabular} & Single participant study \\\hline
Nartey et al. {\cite{19}} & Occupational Safety & \begin{tabular}[c]{@{}c@{}}XAI, \\ LLM, \\ DT, \\ NN, \\ SVM, \\ XGBoost\end{tabular} & \begin{tabular}[c]{@{}c@{}}Explainable AI for \\ fatigue prediction\end{tabular} & Balancing explainability and accuracy \\\hline
Giorgi et al. {\cite{20}} & Automotive & \begin{tabular}[c]{@{}c@{}}EEG, \\ EOG, \\ PPG, \\ EDA\end{tabular} & \begin{tabular}[c]{@{}c@{}}Real-time mental \\ fatigue detection\end{tabular} & Delay in ocular parameter sensitivity \\\hline
Li et al. {\cite{5}} & Construction & \begin{tabular}[c]{@{}c@{}}Decentralized deep learning, \\ Facial images\end{tabular} & \begin{tabular}[c]{@{}c@{}}Improved fatigue monitoring \\ without privacy risks\end{tabular} & Technical efficiency vs. data privacy \\\hline
Yaacob et al. {\cite{15}} & Cognitive Science & \begin{tabular}[c]{@{}c@{}}BCI, \\ AI techniques\end{tabular} & \begin{tabular}[c]{@{}c@{}}Systematic review of \\ mental fatigue detection\end{tabular} & Research gaps in neurofeedback \\\hline
Jiao et al. {\cite{21}} & Transportation & \begin{tabular}[c]{@{}c@{}}HRV, \\ EDA, \\ ML techniques\end{tabular} & \begin{tabular}[c]{@{}c@{}}High accuracy in driver \\ fatigue detection\end{tabular} & Subject-independent classification bias \\\hline
Lambert et al. {\cite{14}} & Cognitive Science & AI models & \begin{tabular}[c]{@{}c@{}}Guidelines for accurate \\ mental fatigue modeling\end{tabular} & Experimental validation \\\hline
Zhao et al. {\cite{23}} & Education & \begin{tabular}[c]{@{}c@{}}RPG, \\ CNN\end{tabular} & \begin{tabular}[c]{@{}c@{}}High accuracy in early \\ learning fatigue detection\end{tabular} & Limited application in educational field \\\hline
El et al. {\cite{22}} & Transportation & \begin{tabular}[c]{@{}c@{}}ML, \\ DL techniques\end{tabular} & \begin{tabular}[c]{@{}c@{}}Comprehensive review of \\ driver fatigue detection methods\end{tabular} & Applicability and reliability issues \\\hline
Bosch et al. {\cite{24}} & Manufacturing & \begin{tabular}[c]{@{}c@{}}Wearable sensors, \\ ML techniques\end{tabular} & \begin{tabular}[c]{@{}c@{}}High accuracy in repetitive \\ physical task fatigue detection\end{tabular} & Task-specific fatigue dependency \\\hline
This work & Physiological Healthcare & Wearable Technology and AI & \multicolumn{2}{c}{\begin{tabular}[c]{@{}c@{}}Identified advancements in real-time   \\ fatigue detection using AI and wearables, \\ highlighted the importance of multi-modal \\ data fusion for accuracy, and explored the \\ potential of hidden technology in safety-critical \\ environments.\end{tabular}}
\end{longtable}
\end{landscape}
\twocolumn

\end{document}